# Lecture notes on the design of low-pass digital filters with wireless-communication applications

Hugh L. Kennedy

*DEWC Services (hugh.kennedy@dewc.com)*

*UniSA STEM Unit, University of South Australia (hugh.kennedy@unisa.edu.au)*

*Mawson Lakes SA 5095*

## Executive summary

The low-pass filter is a fundamental building block from which digital signal-processing systems (e.g. radio and radar) are built. Signals in the electromagnetic spectrum extend over all timescales/frequencies and are used to transmit and receive very long or very short pulses of very narrow or very wide bandwidth. Time/Frequency agility is the key for optimal spectrum utilization (i.e. to suppress interference and enhance propagation) and low-pass filtering is the low-level digital mechanism for manoeuvre in this domain. By increasing and decreasing the bandwidth of a low-pass filter, thus decreasing and increasing its pulse duration, the engineer may shift energy concentration between frequency and time. Simple processes for engineering such components are described and explained below.

These lecture notes are part of a short course that is intended to help recent engineering graduates design low-pass digital filters for this purpose, who have had some exposure to the topic during their studies, and who are now interested in the sending and receiving signals over the electromagnetic spectrum, in wireless communication (i.e. radio) and remote sensing (e.g. radar) applications, for instance. The best way to understand the material is to interact with the spectrum using receivers and or transmitters and software-defined radio development-kits provide a convenient platform for experimentation. Fortunately, wireless communication in the radio-frequency spectrum is an ideal application for the illustration of waveform agility in the electromagnetic spectrum.

In Parts I and II, the theoretical foundations of digital low-pass filters are presented, i.e. signals-and-systems theory, then in Part III they are applied to the problem of radio communication and used to concentrate energy in time or frequency. Expressions to predict link quality (i.e. symbol resolvability) from the parameters of the pulse-shaping filters in a noisy channel are also derived.

# Contents

# Introduction

Low-pass filters are essential elements of multi-rate digital systems that rely on up-conversion and down-conversion to transmit and receive pulses for the transfer of information (e.g. wireless communication) or for remote sensing (e.g. radar) via radio frequencies. Simple procedures for the design and analysis of digital finite-impulse-response (FIR) and infinite-impulse-response (IIR) filters are provided here. FIR filters are discussed in Part I, followed by IIR filters in Part II; software-defined radio applications are considered in Part III. These lecture notes are intended to be bridging material for engineering graduates who are considering working on wireless communication or radar systems.

Software-defined radios are intended to be rapidly re-configurable, for instance, to support a variety of modulation and coding schemes; furthermore, they are inexpensive and widely available for civilian and academic use. Wireless communication via software-defined radio is therefore used in the lab work that accompanies these lecture notes to illustrate the fundamental concept of waveform agility and the important role that the humble low-pass digital filter plays in agile spectrum utilization more generally.

## FIR and IIR filters

Finite Impulse-Response (FIR) filters are a special limiting-case of a linear discrete-time system and as such, many of the more daunting mathematical formalisms (outlined in Part II) of the more general Infinite Impulse Response (IIR) filters may be avoided. The behaviour of a digital (FIR or IIR) filter is represented in the frequency domain by its frequency response $H(\omega)$ and in the (discrete) time domain by its impulse response $h[n]$.

The frequency response $H(\omega)$ defines the *steady-state* output of the filter for a never-ending (complex) sinusoid $e^{\omega i n}$ where $i$ is the imaginary unit and $n$ is the sample index. The relative angular frequency $\omega$, is a continuous real variable in radians per sample. It is a 'circular' quantity, i.e. $\omega = \omega + 2k\pi$, where $k$ is an integer, due to information loss from finite sampling, at a rate of $F_{\mathrm{smp}}$ (samples per second or Hz).

The impulse response $h[n]$ defines the *transient* output of the filter, for a unit impulse input at $n = 0$ with a zero input for all $n$ thereafter.

In digital-electronic systems $H(\omega)$ & $h[n]$ are the complement of each other, i.e. compression in one domain is expansion in the other. In radio-frequency applications, e.g. wireless communication and radar, it may be advantageous to concentrate signal power in either frequency or time, thus both response definitions and filter types are important.

The output of an IIR filter at the $n$th sample $y[n]$, for $n \geq M$, is evaluated using

$$y[n] = \sum_{m=0}^{M-1} b[m]x[n-m] - \sum_{m=1}^{M-1} a[m]y[n-m] \qquad (1)$$

where $M$ is the filter length, $m$ is the delay index, $b[m]$ are the filter coefficients that are applied to prior (and current) inputs $x[n-m]$, and $a[m]$ are the filter coefficients that are applied to prior outputs $y[n-m]$. For $n < M$, i.e. during the filter start-up transient, the upper limit of the summations in (1) is $n$. Feedback is utilized in these digital systems thus the magnitude of the impulse response, i.e. $|h[n]|$, either decays over an infinite duration (stable) remains constant forever (marginally stable) or grows without bound (unstable). Feedback allows only a few non-zero $a[m]$ coefficients to do the work of many $b[m]$ coefficients.

If it were feasible to store the infinite impulse response of such a filter in computer memory, then its output could also be evaluated using

$$y[n] = \sum_{m=0}^{n} h[m]x[n-m] \text{ for } n = 0 \dots \infty. \qquad (2)$$

An FIR filter does not employ feedback and its output $y[n]$ is simply evaluated by convolving the filter coefficients $b[m]$ with the input sequence $x[n]$, using

$$y[n] = \sum_{m=0}^{M-1} b[m]x[n-m]. \qquad (3)$$

For a unit impulse input $y[n] = b[n]$ for the first $M$ samples (i.e. $n = 0 \dots M-1$); with $y[n] = 0$ thereafter (i.e. $n = M \dots \infty$). Thus $h[n] = b[n]$ (for $n = 0 \dots M-1$) or $h[m] = b[m]$ (for $m = 0 \dots M-1$) which allows $h[n]$, $h[m]$ and $b[m]$ to be used interchangeably in FIR filters.

## Signals-and-systems definitions used in Parts I & II

DFT: Discrete Fourier Transform.

IDFT: Inverse Discrete Fourier Transform.

DTFT: Discrete-Time Fourier Transform.

FFT: Fast Fourier Transform.

IFFT: Inverse Fast Fourier Transform.

FIR: Finite Impulse Response.

IIR: Infinite Impulse Response.

tx: The transmitting node.

rx: The receiving node.

dc: Direct current, i.e. zero Hz.

$(\blacksquare)$: Denotes a function of continuous argument e.g. in continuous time or frequency.

$[\blacksquare]$: Denotes a sampled or discrete function of integer argument. Indexed from zero, with the zeroth element at the initial sample (odd or even length) or at the centre sample (for odd length only).

$|\blacksquare|$: Magnitude of a complex variable.

$\angle \blacksquare$: Angle of a complex variable.

$\text{Re}\{\blacksquare\}$: Real part of a complex variable.

$\text{Im}\{\blacksquare\}$: Imaginary part of a complex variable.

$\blacksquare^{-1}$: Matrix inverse.

$\blacksquare^{\text{T}}$: Transpose of a real matrix or vector.

$\blacksquare^{*}$: Complex conjugation.

$f(x)|_{x=a}$: Evaluation of a function, e.g. $f(x)$ at $x = a$, i.e. $f(a)$.

$i = \sqrt{-1}$: Complex unit.

$\Omega$: Angular frequency (radians per second).

$F = \Omega/2\pi$: Frequency (cycles per second or Hz).

$F_{\text{smp}}$: Sampling frequency (cycles per second or Hz) i.e. the sampling rate of the digital-to-analogue and the analogue-to-digital converters.

$T_{\text{smp}} = 1/F_{\text{smp}}$: Sampling period (seconds).

$t$: Time (seconds).

$\omega = \Omega/F_{\text{smp}}$: Normalized angular frequency (radians per sample).

$f = \omega/2\pi = F/F_{\text{smp}}$: Normalized frequency (cycles per sample).

$M$: Number of delays.

$N$: Number of samples.

$K$: Number of bin pairs $M = 2K + 1$.

$n$: Time index, into a sampled sequence (samples, $0 \le n < N$, $t = nT_{\text{smp}}$).

$m$: Delay index, into digital filter coefficients (samples, $0 \leq m < M$, $t = nT_{\text{smp}} - mT_{\text{smp}}$).

$q$: Group delay (samples, non-integer, $-\infty < q < \infty$, $t = nT_{\text{smp}} - qT_{\text{smp}}$).

$k$: Bin index ($-K \leq k \leq K$).

$H(\omega)$: Frequency response.

$h[m]$ or $h[n]$: Impulse response.

$b[m]$: Input filter coefficients of an FIR or IIR filter. Is equal to $h[m]$ for an FIR filter.

$a[m]$: Output filter coefficients, for feedback, in an IIR filter.

$x[n]$: Filter input.

$y[n]$: Filter output.

$t_{\text{pls}}$: Pulse half-duration, pulse duration is $2t_{\text{pls}}$ (seconds).

$\Omega_{\text{pls}}$: Pulse half-bandwidth, pulse bandwidth is $2\Omega_{\text{pls}}$ (radians per second).

$\mathcal{S}(t)$ and $\mathcal{S}(\Omega)$: Sinc function in time and frequency.

$t_{\Delta}$: Time shift, a delay (seconds).

$\Omega_{\Delta}$: Frequency shift (radians per second).

$\psi(\blacksquare)$: A complex sinusoid or oscillator in frequency or continuous time.

$\psi_{\Delta}(\Omega)$: A sinusoid in frequency from a shift in time.

$\psi_{\Delta}(t)$: A sinusoid in time from a shift in frequency.

$\mathcal{D}(t)$ and $\mathcal{D}(\omega)$: Dirichlet kernel in time and frequency.

$\Omega_k$: Angular frequency of the $k$th bin (radians per second).

$\varphi$: Argument of the Dirichlet kernel, an angle (radians).

$k_{\Delta}$: Frequency shift (bins).

$f_{\Delta} = k_{\Delta}/M$: Frequency shift (cycles per sample).

$h_{\text{rec}}[m]$: Impulse response of FIR low-pass filter with a rectangular time window.

$f_{\text{rec}}$: Peak-to-node main-lobe half-width (cycles per sample) of a rectangular filter, $f_{\text{rec}} = 1/M$.

$\overleftarrow{H}(\omega)$: Frequency response of a non-causal filter.

$\overrightarrow{H}(\omega)$: Frequency response of a causal filter.

$\psi_{\text{osc}}[m]$: A sampled oscillator, i.e. discrete-time sinusoid.

$\phi_{\text{osc}}$: Phase of an oscillator (radians).

$\omega_{\text{osc}}$: Relative angular frequency of an oscillator (radians per sample).

$f_{\text{osc}}$: Relative frequency of an oscillator (cycles per sample).

$k_{\text{osc}}$: Frequency bin of an oscillator.

$\omega_c$ or $f_c$: Cut-off or critical frequency (radians per cycle or samples per cycle) of a low-pass filter.

$P_{\omega_c}$: Pass-band power.

$P_{\pi}$: Pass-band plus stop-band (i.e. total) power.

$\boldsymbol{h}$: an $M \times 1$ (column) vector containing the coefficients, i.e. $h[m]$, of an FIR filter.

$\boldsymbol{S}_{\omega_c}$: An $M \times M$ (Toeplitz) matrix with definite integral elements, for pass-band power evaluation.

$S_{\omega_c}[m, n]$: The element in the $m$th row and $n$th column of $\boldsymbol{S}_{\omega_c}$.

$\boldsymbol{S}_{\pi}$: An $M \times M$ diagonal matrix with definite integral elements, for total power evaluation.

$S_{\pi}[m, n]$: The element in the $m$th row and $n$th column of $\boldsymbol{S}_{\pi}$.

$\bar{P}$: Pass-band power concentration (a Rayleigh quotient).

$\boldsymbol{I}$: Identity matrix.

$\boldsymbol{h}_{\max}$: Raw coefficients of an un-normalized Slepian low-pass filter with unity power.

$h_{\max}[m]$: Elements of $\boldsymbol{h}_{\max}$.

$c_{\mathrm{dc}}$: dc magnitude, for normalization.

$\boldsymbol{h}_{\mathrm{eig}}$: Coefficients of a Slepian low-pass filter with unity dc gain.

$D(\omega)$: Desired frequency response of a digital low-pass (FIR or IIR) filter.

$E(\omega)$: Error response of a digital low-pass (FIR or IIR) filter.

$2\omega_{\mathrm{snc}}$: Bandwidth of a sinc function, half-bandwidth or cut-off frequency is $\omega_{\mathrm{snc}}$ (radians per sample).

$f_{\mathrm{snc}}$: Cut-off frequency (cycles per sample) of a sinc function.

WISE: Weighted Integral of the Squared Error.

$\tilde{w}$: Squared error weight in the pass band.

$\bar{w}$: Squared error weight in the stop band.

$\widetilde{\mathrm{ISE}}$: Squared error integral over the pass band.

$\overline{\mathrm{ISE}}$: Squared error integral over the stop band.

$\omega_{\mathrm{lo}}$ and $\omega_{\mathrm{hi}}$: Lower and upper cut-off frequencies (radians per sample) for a transition band with $\omega_{\mathrm{lo}} \leq \omega_{\mathrm{c}} \leq \omega_{\mathrm{hi}}$.

$f_{\mathrm{lo}}$ and $f_{\mathrm{hi}}$: Lower and upper cut-off frequencies (cycles per sample) for a transition band with $f_{\mathrm{lo}} \leq f_{\mathrm{c}} \leq f_{\mathrm{hi}}$.

$\bar{D}(\omega)$: Desired pass-band frequency response.

$\boldsymbol{S}_{xx}, \tilde{\boldsymbol{S}}_{xx}, \bar{\boldsymbol{S}}_{xx}, \boldsymbol{s}_{xy}, \tilde{\boldsymbol{s}}_{xy}, s_{yy}$ and $\tilde{s}_{xy}$: Used to evaluate the WISE.

$\boldsymbol{h}_{\mathrm{isq}}$: Filter coefficient vector of a low-pass FIR filter, designed by minimizing the WISE.

$s$: Complex $s$-plane coordinate, reached via the Laplace transform.

$\mathcal{L}\{\blacksquare\}$: Laplace transform ($t \to s$).

$\mathcal{H}(s)$: Continuous-time transfer function.

$\mathcal{B}(s)$: Numerator polynomial of $\mathcal{H}(s)$.

$\mathcal{A}(s)$: Denominator polynomial $\mathcal{H}(s)$.

$z = e^{i\omega}$: Complex $z$-plane coordinate, reached via the $\mathcal{Z}$ transform, note that $\omega = \angle z$.

$\mathcal{Z}\{\blacksquare\}$: $\mathcal{Z}$ transform ($n \to z$).

$\mathcal{Z}^{-1}\{\blacksquare\}$: Inverse $\mathcal{Z}$ transform ($z \to n$).

$\mathcal{H}(z)$: Discrete-time transfer function.

$\mathcal{B}(z)$: Numerator polynomial of $\mathcal{H}(z)$.

$\mathcal{A}(z)$: Denominator polynomial of $\mathcal{H}(z)$.

$\beta_m$: The $m$th root (i.e. a 'zero') of $\mathcal{B}(z)$.

$M_b$: Degree of $\mathcal{B}(z)$.

$b_0 \ldots b_m \ldots b_{M_b-1}$: Coefficients of $\mathcal{B}(z)$.

$\boldsymbol{b}$: Vector $(1 \times M_b)$ of $\mathcal{B}(z)$ coefficients $b[m]$.

$M_a$: Degree of $\mathcal{A}(z)$.

$\alpha_m$: The $m$th root (i.e. a 'pole') of $\mathcal{A}(z)$.

$a_0 \ldots a_m \ldots a_{M_b-1}$: Coefficients of $\mathcal{A}(z)$, with $a_0 = 1$.

$\boldsymbol{a}$: Vector $(1 \times M_a)$ of $\mathcal{A}(z)$ coefficients $a[m]$.

$\overset{\leftrightarrow}{h}[m]$: Non-causal discrete-time impulse-response.

$\vec{h}[m]$: Causal discrete-time impulse-response. Generated in the forward time direction.

$\overleftarrow{h}[m]$: Anti-causal discrete-time impulse-response. Generated in the backward (i.e. reverse) time direction.

$\overset{\leftrightarrow}{\mathcal{H}}(z)$: Non-causal discrete-time transfer-function.

$\vec{\mathcal{H}}(z)$: Causal discrete-time transfer-function.

$\overleftarrow{\mathcal{H}}(z)$: Anti-causal discrete-time transfer-function.

$\phi(\omega)$: Phase response of a digital filter.

$\mathcal{H}_{\mathrm{dc}}(z)$ and $\mathcal{H}'_{\mathrm{dc}}(z)$: Used to determine $q$ from $\mathcal{H}(z)$.

# Part I: Finite Impulse-Response (FIR) filters

Most people with an interest in science and technology are familiar with the discrete Fourier transform (DFT) for the frequency-domain analysis of sampled signals in the time domain. And its inverse (the IDFT) for the synthesis of signals at samples in time from bins in a spectrum. The discrete-time Fourier transform (DTFT) is closely related but less well known and it is essential for the design of FIR filters. The DFT (or more commonly, its fast implementation, the FFT) is a *system component* that transforms a discrete-time signal at $M$ samples into the frequency domain at $M$ discrete 'bins'; whereas the DTFT is an *analytical tool* that is used to generate the frequency response $H(\omega)$ from the $M$ coefficients of an FIR filter $h[m]$, where $H(\omega)$ is a (complex-valued) function of a continuous frequency variable $\omega$ (the relative angular frequency in radians per sample) and $m$ are delay indices (for $m = 0 \dots M-1$) into a vector of FIR filter coefficients.

The frequency response of an FIR filter is determined using

$$H(\omega) = \sum_{m=0}^{M-1} h[m]e^{-im\omega}. \tag{4}$$

The magnitude-scaling and phase-shift experienced by a (real or complex) sinusoid of angular frequency $\omega$ as it passes through the filter is then computed using $|H(\omega)|$ and $\angle H(\omega)$, where $|arg|$ and $\angle arg$ are the magnitude and angle operators of the complex argument, respectively. The form of (4) reveals that each filter coefficient yields a sinusoid in the frequency domain, where the frequency of each sinusoid is proportional to the delay applied by the corresponding filter coefficient. The frequency response $H(\omega)$ of a digital filter is generated via the DTFT of its impulse response $h[m]$. The idea of waves in the time domain is a relatively easy concept to grasp. The idea of complex sinusoids in the frequency domain requires more effort to appreciate.

The console of a musician's homemade audio synthesizer, that generates a sound by summing $M$ sinusoidal components in the time domain, is a familiar and useful analogy to help us understand the inverse DFT. The console is a frequency-domain representation of the audio signal with a slider and dial available to independently set the magnitude and phase of each sinusoidal component in the time domain. Conversely, a similar analogy may be used to describe the way in which the DTFT is used by an engineer – this 'console' is a time-domain representation of an FIR filter's coefficients $h[m]$, and the frequency response $H(\omega)$, is shaped by adjusting the magnitude (and sign or phase) of each filter coefficient during the design process until the desired response is reached. Compromise is inevitable for an FIR filter with a maximum of only $M$ degrees of freedom thus some ripple in the pass band and/or stop band is unavoidable and a sharp step-like transition between the bands is impossible. Simple procedures for the setting of filter coefficients to satisfy various design requirements are discussed in the subsections below.

## Seeking refuge from the diabolical Colonel Dirichlet in the Sleepy Inn

The plots and discussion below are intended to show the severe shortcomings of frequency-domain analysis of discrete time-domain signals using an un-tapered rectangular window in FIR filters (of length $M$). Rectangular time windows 'resonate' at spurious frequencies, according to the Dirichlet kernel. Ways of suppressing this effect in digital systems using Slepians will now be described, beginning with the sinc function, which is its analogue-system equivalent. Slepians are again used in Part III to shape the envelope of pulsed waveforms for radio communication.

The sinc function plays a pivotal role in the Fourier analysis of continuous-time signals. The Fourier transform of a rectangular pulse in time is a sinc function in frequency ($\Omega$). And the inverse Fourier transform of a perfectly band-limited signal with a rectangular frequency-response is a sinc function in time ($t$). In these complementary domains, $\Omega$ is the angular frequency in radians per second and $t$ is in seconds. In the former case the rectangular pulse is centred on $t = 0$ ($-\infty < t < \infty$) with a duration of $2t_{\text{pls}}$ seconds. In the latter case the rectangular pulse is centred on $\Omega = 0$ ($-\infty < \Omega < \infty$) with a bandwidth of $2\Omega_{\text{pls}}$ radians per second. The sinc function ($\mathcal{S}$) in the frequency and time domains is defined as follows:

$$\mathcal{S}(\Omega) = \frac{1}{2t_{\text{pls}}} \int_{-t_{\text{pls}}}^{t_{\text{pls}}} e^{-i\Omega t} dt = \frac{\sin(\Omega t_{\text{pls}})}{\Omega t_{\text{pls}}} \tag{5}$$

and

$$\mathcal{S}(t) = \frac{1}{2\Omega_{\text{pls}}} \int_{-\Omega_{\text{pls}}}^{\Omega_{\text{pls}}} e^{i\Omega t} d\Omega = \frac{\sin(\Omega_{\text{pls}} t)}{\Omega_{\text{pls}} t}. \tag{6}$$

When the continuous rectangular function is shifted in time (by $t_\Delta$ seconds) or in frequency (by $\Omega_\Delta$ radians per second) the sinc functions above are multiplied by the complex sinusoids $\psi_\Delta(\Omega) = e^{i\Omega t_\Delta}$ or $\psi_\Delta(t) = e^{i\Omega_\Delta t}$, respectively.

The Dirichlet kernel (also known as the periodic sinc function) plays an analogous role in the Fourier analysis of discrete-time signals, where definite integrals over a continuum of time or frequency are replaced by finite summations over discrete samples in time (e.g. $m$ samples) or discrete bins in frequency (i.e. $k$). For an odd number of samples and bins, with $M = 2K + 1$ (i.e. the window length in samples) indexed using $-K \leq m \leq K$ and $-K \leq k \leq K$, respectively, the Dirichlet kernel ($\mathcal{D}$) is defined in the frequency and time domains as follows:

$$\mathcal{D}(\omega) = \frac{1}{M} \sum_{m=-K}^{K} e^{-im\omega} = \frac{\sin(M\varphi/2)}{\sin(\varphi/2)} \text{ with } \varphi = \omega \tag{7}$$

and

$$\mathcal{D}(t) = \frac{1}{M} \sum_{k=-K}^{K} e^{i\Omega_k t} = \frac{\sin(M\varphi/2)}{\sin(\varphi/2)} \text{ with } \varphi = 2\pi t/MT_{\text{smp}} \text{ and } \Omega_k = 2\pi k/MT_{\text{smp}} \tag{8}$$

where $T_{\text{smp}}$ is the sampling period (seconds per sample, with $T_{\text{smp}} = 1/F_{\text{smp}}$).

Similarly, when the discrete rectangular function is shifted in time by $m_\Delta$ samples or in frequency by $k_\Delta$ bins, the Dirichlet kernels above are (respectively) multiplied by $\psi_\Delta(\omega) = e^{im_\Delta \omega}$ or $\psi_\Delta(t) = e^{i\Omega_\Delta t}$ (with $\Omega_k = 2\pi k_\Delta /MT_{\text{smp}}$).

The expression for $\mathcal{D}(t)$ is not essential for discrete-time analysis; however, it is included here for completeness and because the summation over bins in frequency is arguably more intuitive than the summation over samples in time (as discussed in the 'console' analogy above). Furthermore, the units of the parameters that define $\mathcal{D}(t)$ are somewhat awkward whereas a simpler definition of $\mathcal{D}(\omega)$ is chosen because it forms the foundation of what follows.

The construction of $\mathcal{D}(t)$ and $\mathcal{D}(\omega)$ from a sum of sinusoids is illustrated in Figure 1 and Figure 2 for $K = 8$. The real and imaginary parts of complex quantities are plotted in blue and red (respectively) with magnitudes in green. In both figures, the upper subplot in the right column shows the $m$th or $k$th sinusoid being added to the sum and the lower subplot shows the cumulative sum; thus the (real) Dirichlet kernel is shown in the lower subplot of the final row in the right column, in the other rows its magnitude is also shown using a dotted green line. The frequency axis has been extended in Figure 2 to illustrate the periodicity of the Dirichlet kernel, due to sampling at a finite rate. In these plots and throughout these lecture notes, relative frequency $f$ (cycles per sample) is used with $f = F/F_{\mathrm{smp}} = \Omega/2\pi F_{\mathrm{smp}} = \omega/2\pi$ where $F$ is the frequency in samples per second (i.e. Hz) and $F_{\mathrm{smp}}$ is the sampling rate (also in Hz, with $F_{\mathrm{smp}} = 1/T_{\mathrm{smp}}$). The delay index $m$ may also be regarded as relative time. It is a dimensionless quantity (in samples) with $m = t/T_{\mathrm{smp}}$. The Dirichlet kernel is evaluated over continuous time in Figure 1.

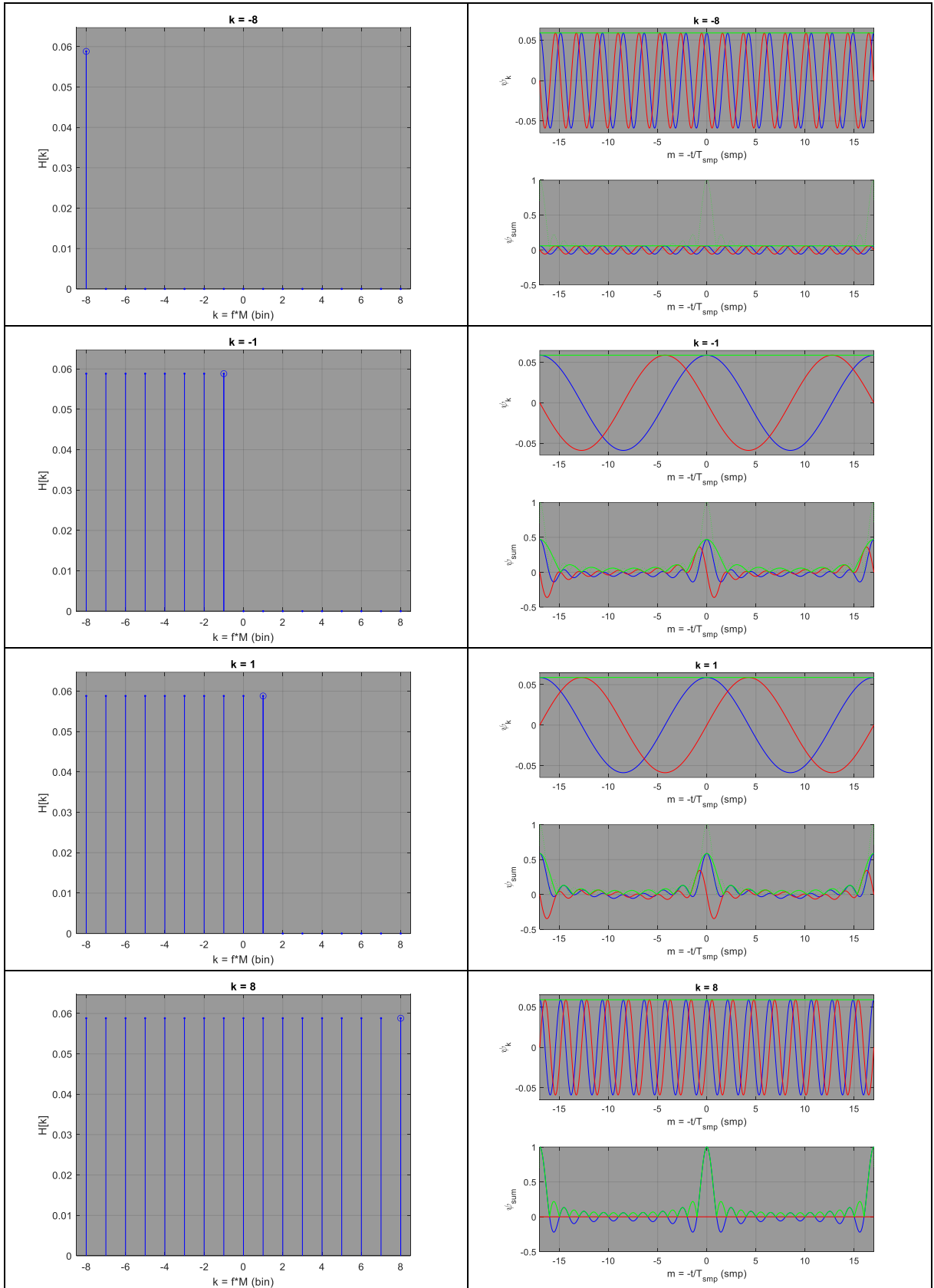

*Figure 1. A summation in the frequency domain yields the Dirichlet kernel in the time domain*

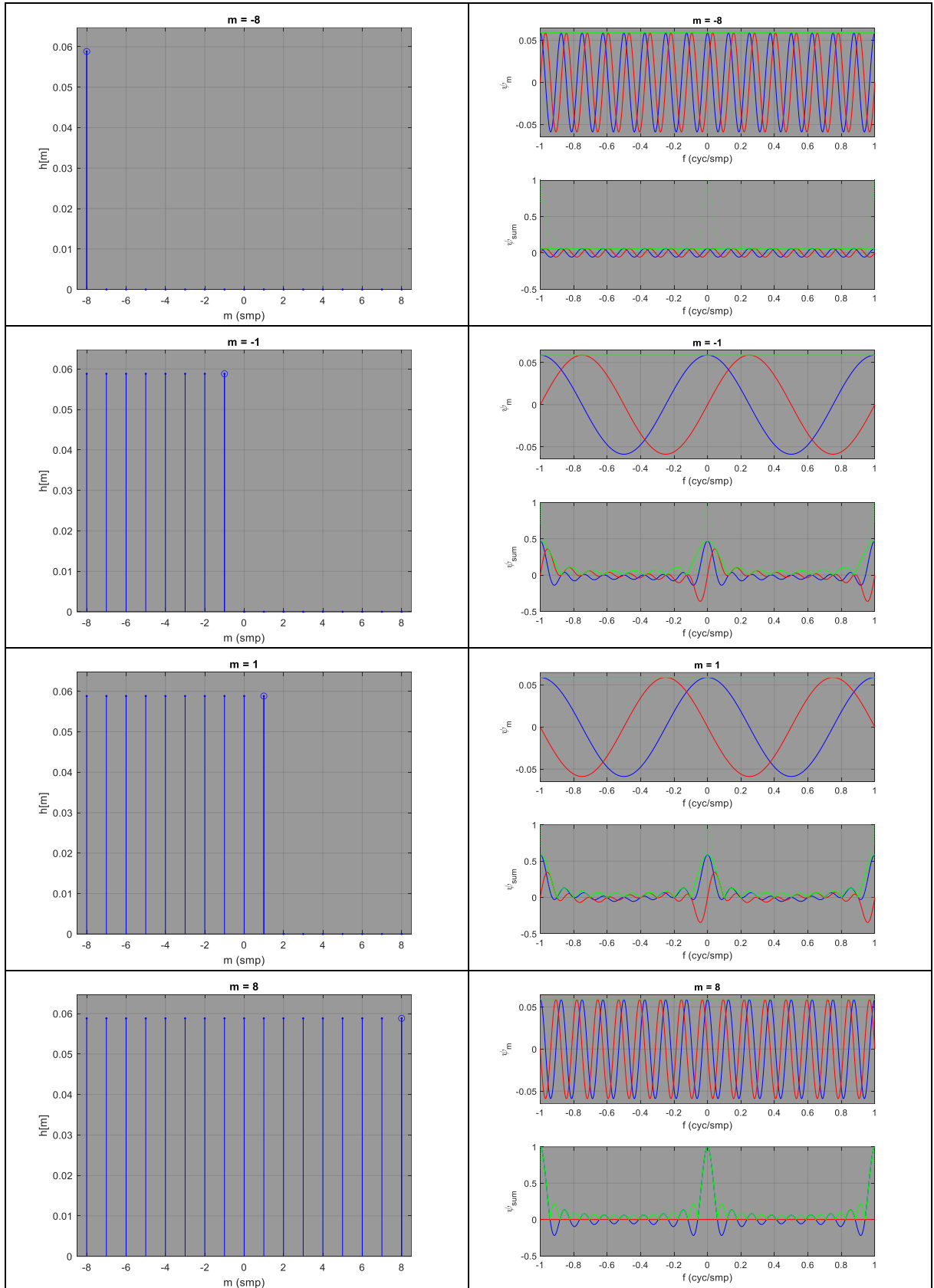

*Figure 2. A summation in the time domain yields the Dirichlet kernel in the frequency domain*

The nodal and 'lobal' structure of the Dirichlet kernel in the frequency domain is clearer in Figure 4, for a rectangular FIR filter with $K = 16$ (see Figure 3). To determine the response of this rectangular FIR filter (upper subplot) to a sinusoidal input of a given frequency, we simply evaluate the Dirichlet kernel (lower subplot) at that frequency. The lower subplot indicates that 'direct-current' (i.e. dc where $\omega = 0$) components of an input signal pass through this filter unchanged, as the magnitude and phase at $f = 0$ are unity and zero, respectively. Furthermore, the frequency response indicates there is zero transmission when the length of this rectangular filter is a whole multiple of a component's wavelength (commensurate) i.e. at $f_k = k/M$ for $-K \leq k < 0$ and $0 < k \leq K$. The response at these frequencies is shown using dots in Figure 4. The (main and side) lobes of the Dirichlet kernel indicate that other components at intervening frequencies 'slip through the cracks', with very little attenuation if they are near dc. As the frequency response of this filter (i.e. Dirichlet kernel) is real at all frequencies, no components experience a phase shift. The main lobe contracts and the sidelobes are lowered as the length (i.e. $M$) of the FIR filter increases. This rectangular FIR filter $h_{\text{rec}}[m]$ is a simple moving-average low-pass filter and $M h_{\text{rec}}[m]$ evaluates the $k = 0$ bin of a sliding (non-recursive) $M$-point DFT.

The design and analysis of FIR filters with $M$ odd is simplified by advancing the filter by $K$ samples so that the impulse response is centred on $m = 0$. This results in a non-causal filter because one sample is used from the present time ($m = 0$), $K$ samples are used from the past ($m > 0$) and $K$ samples are used from the future ($m < 0$). Furthermore, if the impulse response is symmetric or anti-symmetric around $m = 0$ then the frequency response of a non-causal filter is real which means that the phase response is zero for all frequencies. However, the conversion to a causal filter with the same magnitude response is trivial. It is reached by applying a delay of $K$ samples to the non-causal filter. The now complex frequency-response of the causal filter is therefore $\vec{H}(\omega) = e^{-iK\omega} \overleftrightarrow{H}(\omega)$, where $\vec{H}(\omega)$ and $\overleftrightarrow{H}(\omega)$ are the frequency response of the causal and non-causal filters, respectively.

In Figure 5 and Figure 6 a new FIR filter with complex coefficients is created by multiplying the rectangular FIR filter above (with $M = 33$) by a complex sinusoid $\psi_{\text{osc}}[m] = e^{i\omega_{\text{osc}}m}$ where $\omega_{\text{osc}} = 2\pi f_{\text{osc}}$ with $f_{\text{osc}} = k_{\text{osc}}/M$ where $k_{\text{osc}} = 3$. The rectangular 'window' (normalized for unity dc gain) is shown in green; with the real and imaginary parts of $\psi_{\text{osc}}$ in blue and red, respectively. Note that $\psi_{\text{osc}}$ is scaled by a factor of $1/M$ in the plot to assist visualization so that it has the same magnitude as $h_{\text{rec}}[m]$. The real and imaginary parts of the coefficients of the resulting FIR filter are shown in cyan and magenta, respectively. The frequency response of this filter is evaluated in the usual way using (4).

Multiplication in the (discrete) time domain is (circular) convolution in the frequency domain. Since the frequency response of the sinusoid is a unit impulse at $\omega_{\text{osc}}$ this convolution reduces to a simple shift of the centre frequency of the Dirichlet kernel using $\varphi = \omega - \omega_{\text{osc}}$ in (7). The frequency response is now unity at $f_k = k/M$ for $k = 3$ and zero at all other bins. The output of this FIR filter (multiplied by a factor of $M$) evaluates the $k$th bin of a (non-recursive) sliding DFT. The phase centre of $\psi_{\text{osc}}$ is at $m = 0$ in Figure 5 and Figure 6. Shifting the phase of $\psi_{\text{osc}}$, i.e. using $\psi_{\text{osc}}[m] = e^{i\omega_{\text{osc}}m + i\phi_{\text{osc}}} = e^{i\phi_{\text{osc}}} e^{i\omega_{\text{osc}}m}$, results in a complex frequency response because the impulse response and the frequency response are both multiplied by a complex (scalar) factor of $e^{i\phi_{\text{osc}}}$.

In Figure 7 and Figure 8 the rectangular window modulates a complex sinusoid $\psi_{\text{osc}}[m] = e^{i\omega_{\text{osc}}m}$ with $\omega_{\text{osc}} = 2\pi f_{\text{osc}}$ with $f_{\text{osc}} = 0.1079$, which falls in between bins $k = 3$ & $k = 4$. When the Dirichlet kernel $\mathcal{D}(\omega)$ is shifted in the frequency domain so that it is centred on $\omega_{\text{osc}}$ the peak and nodes of $\mathcal{D}(\omega)$ no longer coincide with the DFT frequency bins.

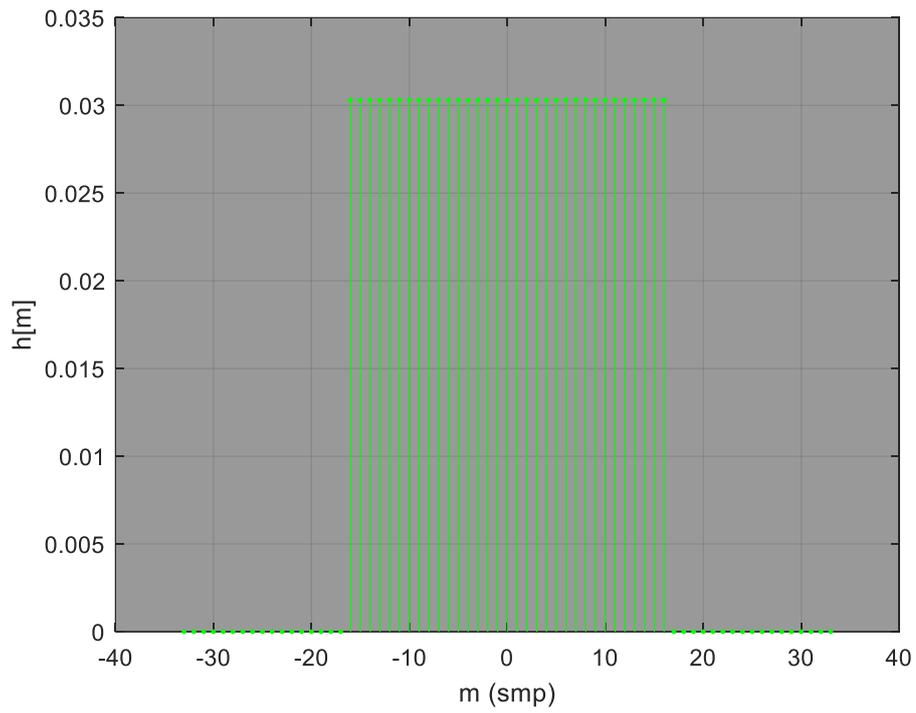

*Figure 3. Impulse response of a rectangular low-pass filter*

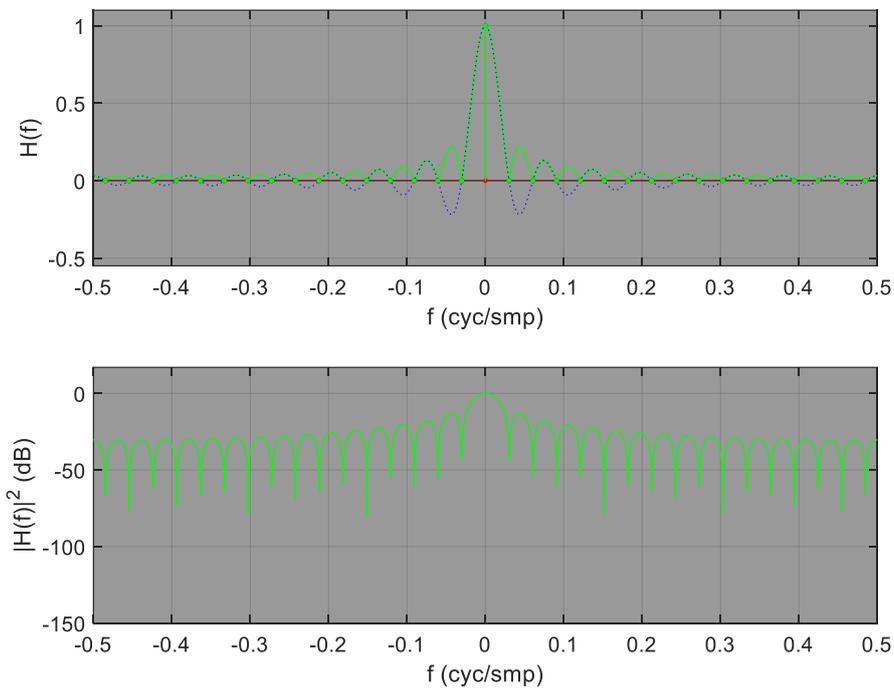

*Figure 4. Frequency response of a rectangular low-pass filter*

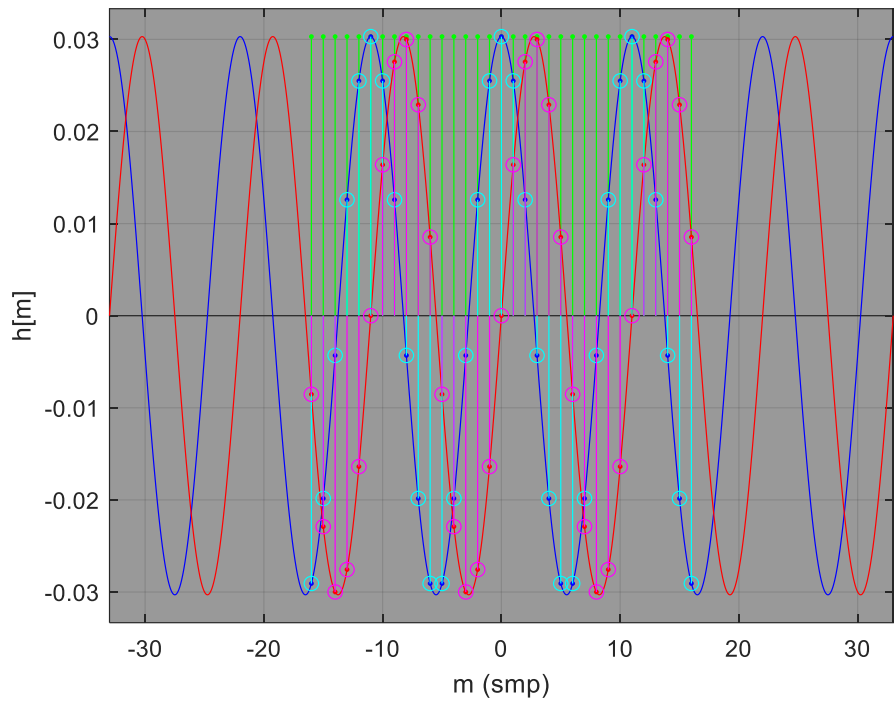

*Figure 5. Impulse response of a rectangular low-pass filter that is multiplied by a complex sinusoid with a commensurate wavelength*

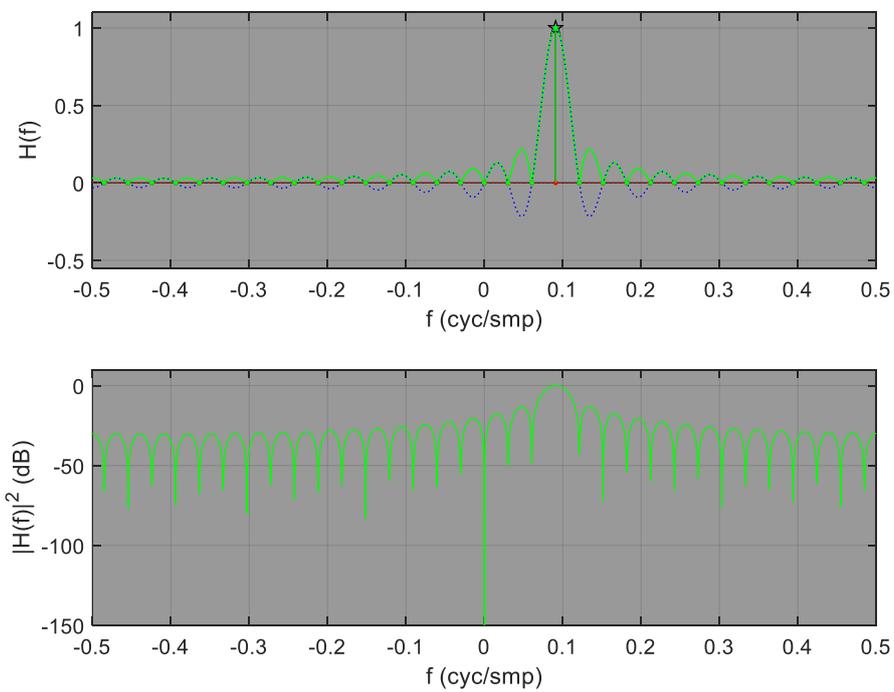

*Figure 6. Frequency response of a rectangular low-pass filter that is multiplied by a complex sinusoid with a commensurate wavelength*

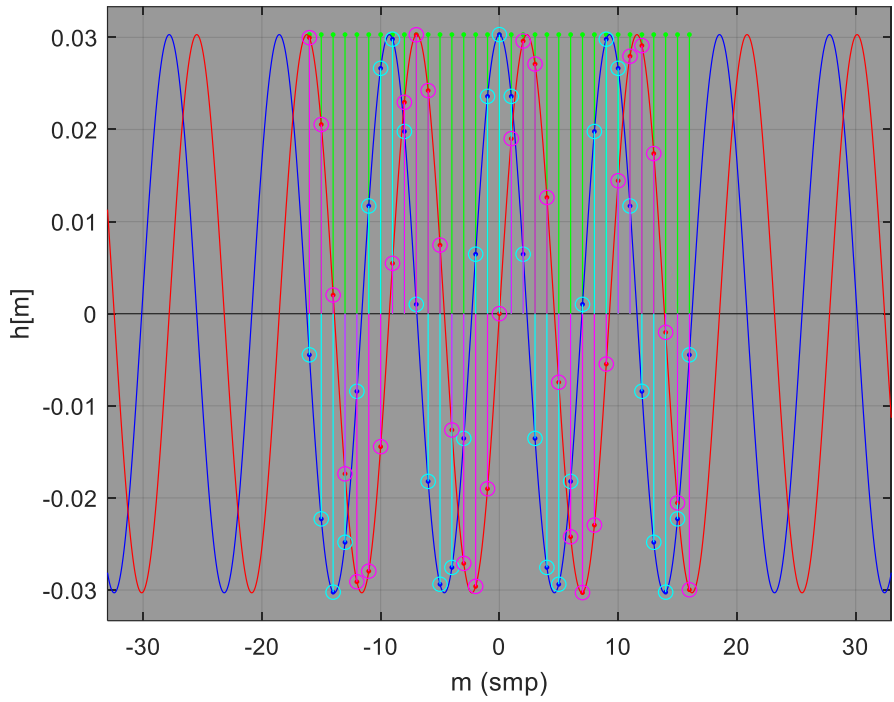

*Figure 7. Impulse response of a rectangular low-pass filter that is multiplied by a complex sinusoid with an incommensurate wavelength*

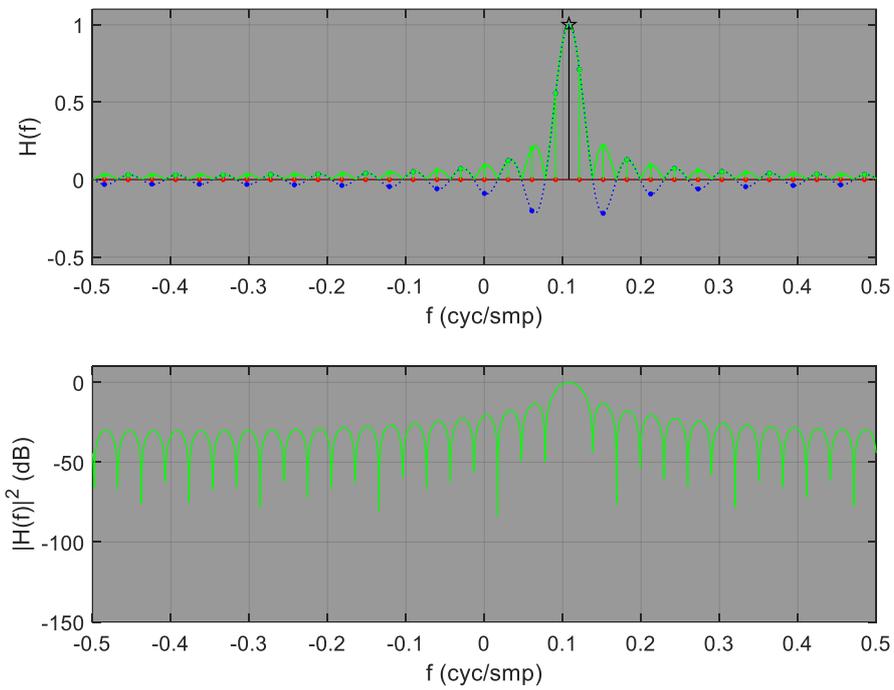

*Figure 8. Frequency response of a rectangular low-pass filter that is multiplied by a complex sinusoid with an incommensurate wavelength*

The 'hard' truncation of the rectangular FIR filter in the time domain (see Figure 3) yields high side-lobes in the frequency domain (see Figure 4). A procedure popularized by David Slepian at Bell Laboratories (and published in 1961) that seeks to maximize the pass-band power concentration (thus minimize stop-band power) provides a simple means of designing low-pass filters with a more gradual taper in the time domain for lower side-lobes beyond the cut-off frequency in the frequency domain. The power contained in the pass band of a low-pass filter with a cut-off frequency of $\omega_c$ is evaluated by integrating the squared magnitude over $\pm\omega_c$. For an FIR filter this is readily done using

$$P_{\omega_c} = \boldsymbol{h}^\mathrm{T} \boldsymbol{S}_{\omega_c} \boldsymbol{h} \tag{9}$$

where

$\boldsymbol{h}$ is an $M \times 1$ (column) vector containing the FIR filter coefficients

$\blacksquare^\mathrm{T}$ is the transpose operator and

$\boldsymbol{S}_{\omega_c}$ is an $M \times M$ (Toeplitz) matrix with elements (in the $m$th row and $n$th column)

$$S_{\omega_c}[m,n] = \int_{-\omega_c}^{\omega_c} e^{i(m-n)\omega} = \begin{cases} 2\sin\{(m-n)\omega_c\}/(m-n), & m-n \neq 0 \\ 2\omega_c, & m-n = 0 \end{cases}. \tag{10}$$

Similarly, the total power over $\pm\pi$ is determined using

$$P_\pi = \boldsymbol{h}^\mathrm{T} \boldsymbol{S}_\pi \boldsymbol{h} \tag{11}$$

where

$\boldsymbol{S}_\pi$ is an $M \times M$ diagonal matrix with elements

$$S_\pi[m,n] = \int_{-\pi}^{\pi} e^{i(m-n)\omega} = \begin{cases} 0, & m-n \neq 0 \\ 2\pi, & m-n = 0 \end{cases}. \tag{12}$$

The pass-band power concentration is therefore the Rayleigh quotient

$$\tilde{P} = P_{\omega_c}/P_\pi = \boldsymbol{h}^\mathrm{T} \boldsymbol{S}_{\omega_c} \boldsymbol{h} / \boldsymbol{h}^\mathrm{T} \boldsymbol{S}_\pi \boldsymbol{h}. \tag{13}$$

The set of $M$ orthonormal vectors $\boldsymbol{h}$ that minimize this quantity are found by solving the general eigenvalue problem $\mathrm{eig}(\boldsymbol{S}_{\omega_c}, \boldsymbol{S}_\pi)$ or simply the eigenvalue problem $\mathrm{eig}(\boldsymbol{S}_{\omega_c}/2\pi)$ since $\boldsymbol{S}_\pi = 2\pi\boldsymbol{I}$ (where $\boldsymbol{I}$ is the identity matrix). The eigenvalues of each eigenvector represent the power concentration ($\tilde{P}$) of each possible solution. Each eigenvector is normalized for unity total power i.e.

$$\sum_{m=-K}^{K} |h[m]|^2 = 1. \tag{14}$$

The low-pass FIR filter coefficients $\boldsymbol{h}_\mathrm{eig}$ are set equal to the eigenvector with the greatest eigenvalue $\boldsymbol{h}_\mathrm{max}$, i.e. the one with maximal pass-band power concentration, multiplied by a scaling factor $1/c_\mathrm{dc}$ for unity dc gain, where $c_\mathrm{dc}$ is the sum of the elements of $\boldsymbol{h}_\mathrm{max}$, thus $\boldsymbol{h}_\mathrm{eig} = \boldsymbol{h}_\mathrm{max}/c_\mathrm{dc}$ with

$$c_\mathrm{dc} = \sum_{m=-K}^{K} h_\mathrm{max}[m]. \tag{15}$$

The impulse response and magnitude response of a Slepian low-pass filter designed using $\omega_c = 2\pi f_c$ with $f_c = 0.1$ and $K = 16$ (for $M = 33$) is shown in Figure 9 and Figure 10, respectively. Note that for an FIR filter the impulse response (i.e. the filter output for a unit impulse input) is simply equal to the coefficient vector of the filter. The sidelobes of this Slepian low-pass filter are no longer visible on a linear scale (see the upper subplot of Figure 10). The Slepian frequency response is (circularly) shifted and centred at $\omega_\mathrm{osc}$ when the impulse response is multiplied by a complex sinusoid (see Figure 11 and Figure 12). Relative to the Dirichlet kernel, the power that spills into other frequencies via the side lobes is now negligible. The pass-band power concentration is near unity and the stop-band power concentration is $1.5820 \times 10^{-8}$.

Note that without a specified timescale (i.e. the sampling period in seconds) what is simply referred to as 'power' in this document is (more correctly) the energy density in the time domain or energy spectral density in the frequency domain. Furthermore, if the units of measurement are (milli-) Volts then an impedance must also be specified for the actual power to be computed.

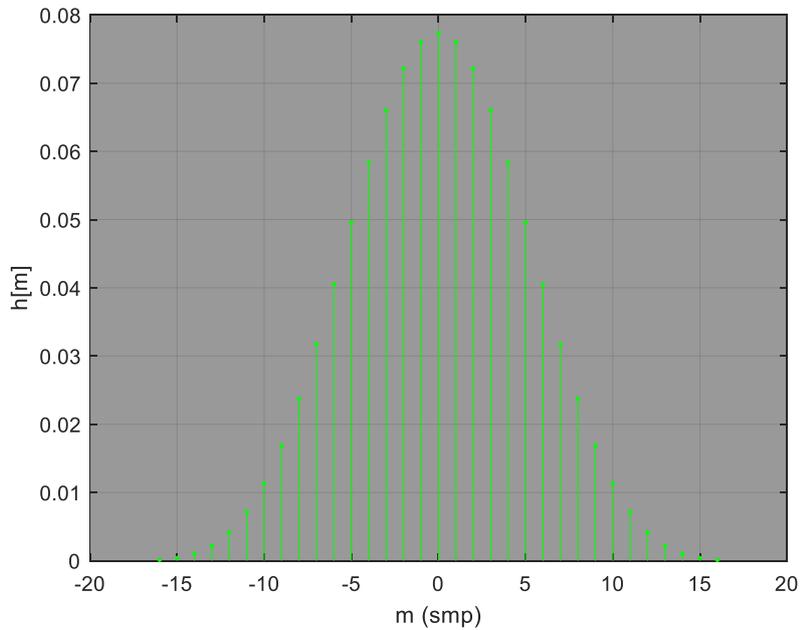

*Figure 9. Impulse response of a Slepian low-pass filter*

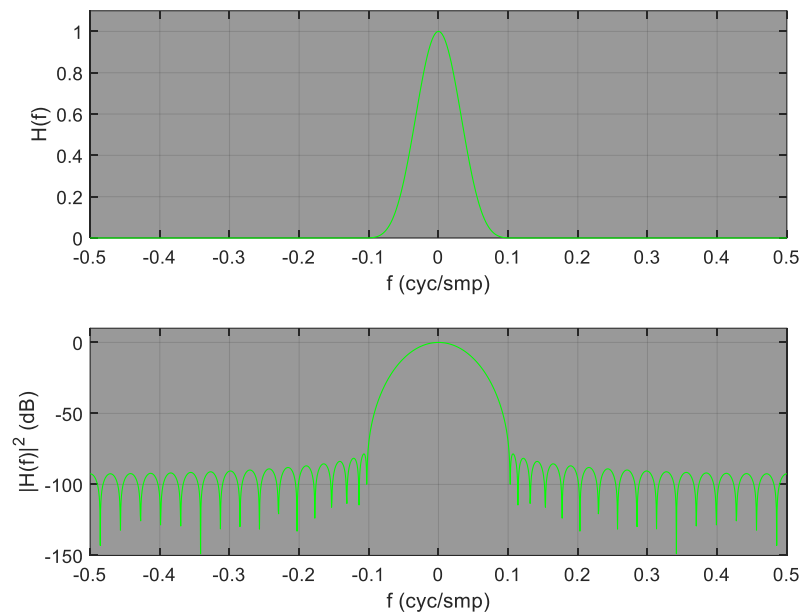

*Figure 10. Frequency response of a Slepian low-pass filter*

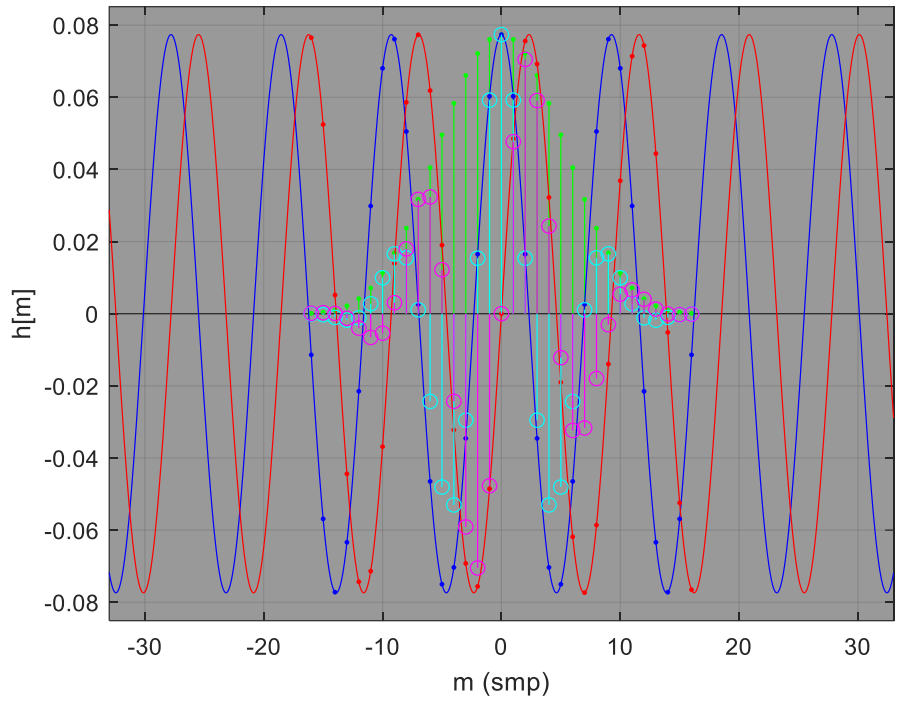

*Figure 11. Impulse response of a Slepian low-pass filter that is multiplied by a complex sinusoid with an incommensurate wavelength*

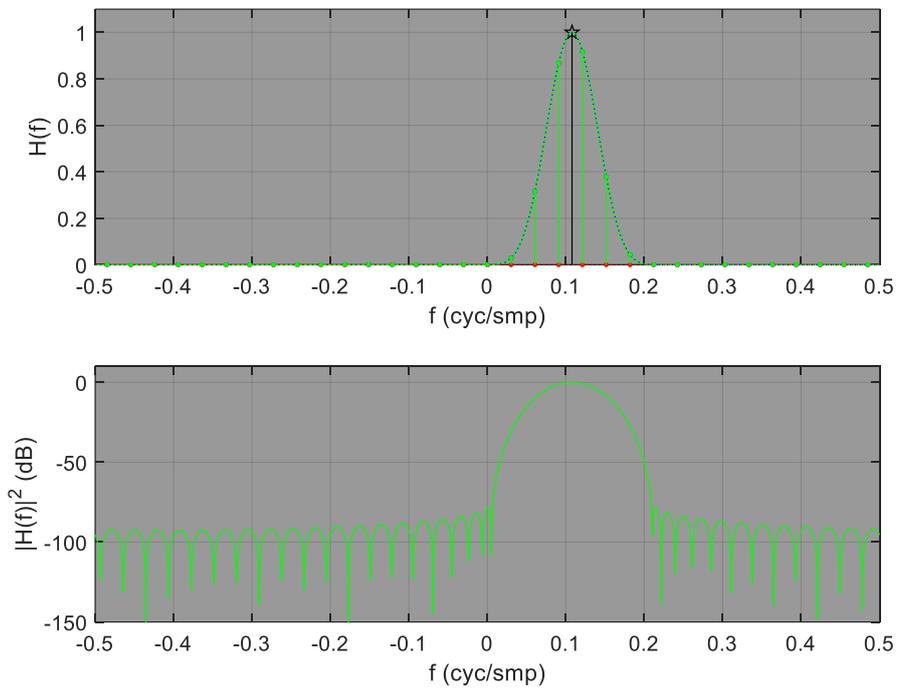

*Figure 12. Frequency response of a Slepian low-pass filter that is multiplied by a complex sinusoid with an incommensurate wavelength*

## Slepian-windowed-sinc and least-squared-error designs

Wide-band low-pass filters of high order (e.g. with $f_c = 0.3$ and $M = 33$) cannot be designed using the Slepian method described above. Dilation of the frequency response leads to a contraction of the impulse response and if the optimal taper in the time domain is too severe so that many of the coefficients at the filter's extremes are negligible (i.e. near $\pm K$) then the eigen-problem is likely to be ill conditioned because there are too many degrees of freedom and erroneous solutions should be expected (e.g. oscillatory and asymmetric). Furthermore, the passband of the Slepian filter is by no measure flat and there is a gradual monotonic roll-off from dc to the passband cut-off frequency over the pass band. The absence of a sharp step-like transition between the pass and stop bands in the frequency domain does however eliminate 'ripple' or 'ringing' in the time domain, i.e. the Gibbs phenomenon. This yields an impulse response $h[m]$ that is positive for $-K \leq m < K$, which allows the Slepian to be interpreted as a weight or tapered window.

Two ways of overcoming the aforementioned shortcomings of Slepian low-pass filters are described below and used to design a wide-band low-pass filter with $f_c = 0.3$ and $M = 33$. In the first approach a Slepian window is used instead of a rectangular window to lower the sidelobes of a discretised sinc function. In the second approach the filter coefficients are chosen so that the (weighted) integral of the squared response error is minimized, where the response error $E(\omega)$ is the difference between the desired (ideal) response $D(\omega)$ and the realized (actual) response $H(\omega)$. These methods offer simplicity and flexibility, respectively.

A continuous-time signal with $H(\Omega) = 1/2\omega_{\mathrm{snc}}$ over $\pm\omega_{\mathrm{snc}}$ and zero elsewhere (where $\omega_{\mathrm{snc}} = \Omega_c/F_{\mathrm{smp}} = 2\pi f_c$ with $f_c = 0.2$) is a sinc function $\mathcal{S}(t)$ in the time-domain (blue lines in Figure 13 and Figure 14). For a realizable discrete-time FIR filter of reasonable complexity, the sinc function is sampled at $t = mT_{\mathrm{smp}}$ and truncated at $m = \pm K$ in the time domain then normalized for unity gain at dc (cyan dots in Figure 13). This discretization operation is equivalent to the multiplication of $\mathcal{S}(t)$ by $h_{\mathrm{rec}}[m]$ (blue line and green dots, respectively, in Figure 13).

Multiplication (or modulation) in the time domain is convolution in the frequency domain (see Figure 14). Thus the ideal rectangular response (blue line) of the sinc function, in continuous time over an infinite extent, is convolved with the Dirichlet kernel $\mathcal{D}(\omega)$ (green line) yielding the realized response $H(\omega)$ of the FIR filter (cyan line). Unfortunately, the high side-lobes of $\mathcal{D}(\omega)$ are conferred to $H(\omega)$. The side lobes may be (slowly) lowered by increasing $K$ at the expense of increased computational complexity for the deployed filter.

Alternatively, a Slepian window $h_{\mathrm{eig}}[m]$ with a cut-off frequency of $f_{\mathrm{eig}} = 0.1$ and a length of $M = 33$, which has much lower sidelobes in the frequency, domain may be used instead of $h_{\mathrm{rec}}[m]$ to discretize (and taper) the sinc function $\mathcal{S}(t)$ in the time domain (see Figure 15). Convolution of the ideal rectangular low-pass response of the sinc function (blue line) with the frequency response of the Slepian window $H_{\mathrm{eig}}(\omega)$ (green line) is a frequency-domain smoothing operation, that yields an FIR low-pass filter with much lower side lobes (cyan line) and a broader passband with the desired cut-off frequency of $f_c = f_{\mathrm{snc}} + f_{\mathrm{eig}} = 0.3$ (see Figure 16). Note that in Figure 13 and Figure 15 the impulse responses of the sinc function and the Slepian have been scaled so they have the same magnitude as the impulse response of the low-pass filter at $m = 0$, to facilitate visualization and comparison. For both rectangular and Slepian windows, the coefficients of the resulting low-pass filter are scaled to ensure unity dc gain, i.e. $|H(\omega)|_{\omega=0}| = 1$.

Any tapered window-function (e.g. Gaussian, Hann, Hamming or Blackman) could be used instead of the Slepian window to manage the trade-off between side-lobe lowering and main-lobe broadening;

however, the Slepian is ideal because it yields the lowest sidelobes possible for the amount of main lobe broadening that the designer considers to be acceptable, as specified using its $f_c$ (i.e. $f_{eig}$ above). The peak-to-first-null half-bandwidth of a rectangular window is equal to the frequency resolution of an $M$-sample DFT, i.e. $f_{rec} = 1/M$. If $f_{eig} = f_{rec}$ then only a slight taper is applied in the time domain. More severe tapers (thus wider main-lobes and lower side-lobes) are produced for $f_{eig}/f_{rec} \gg 1$.

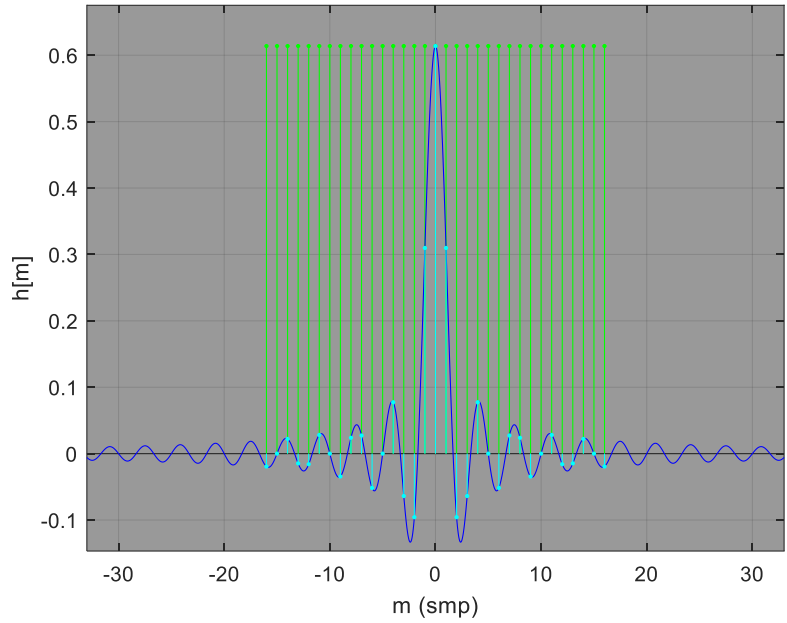

*Figure 13. Impulse response of a discretised (sampled and truncated) sinc function*

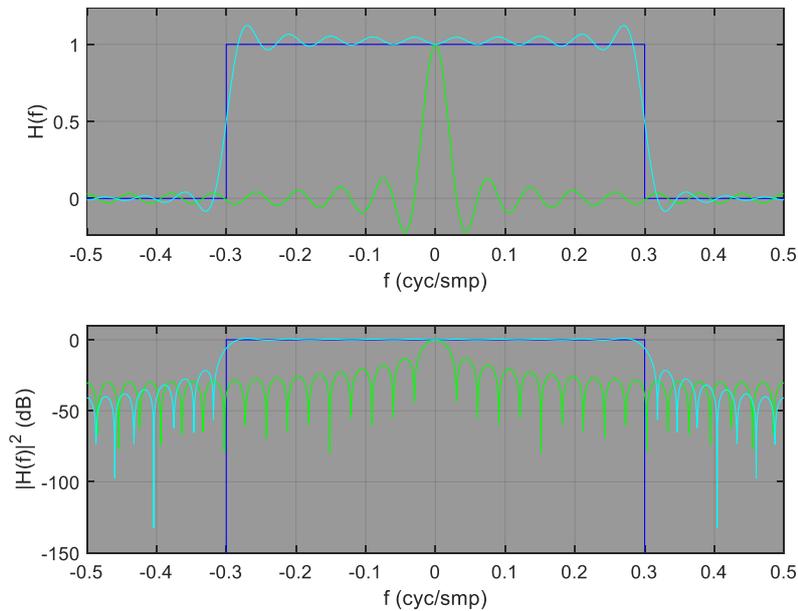

*Figure 14. Frequency response of a discretised (sampled and truncated) sinc function*

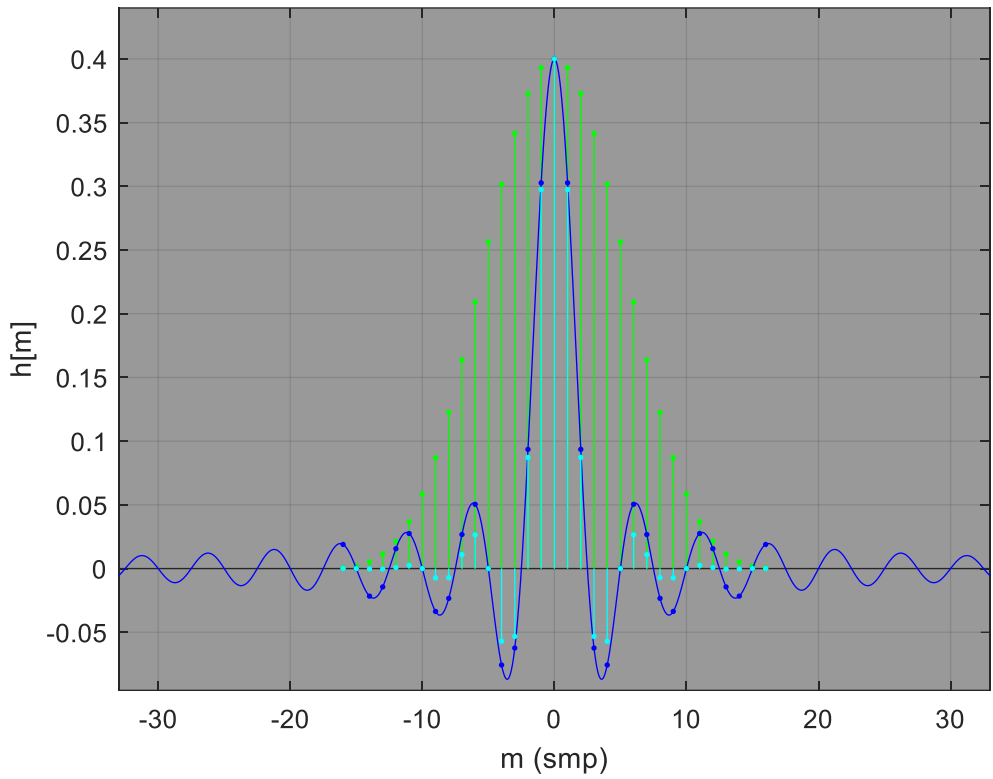

*Figure 15. Impulse response of a discretised (sampled and truncated) sinc function with a Slepian taper*

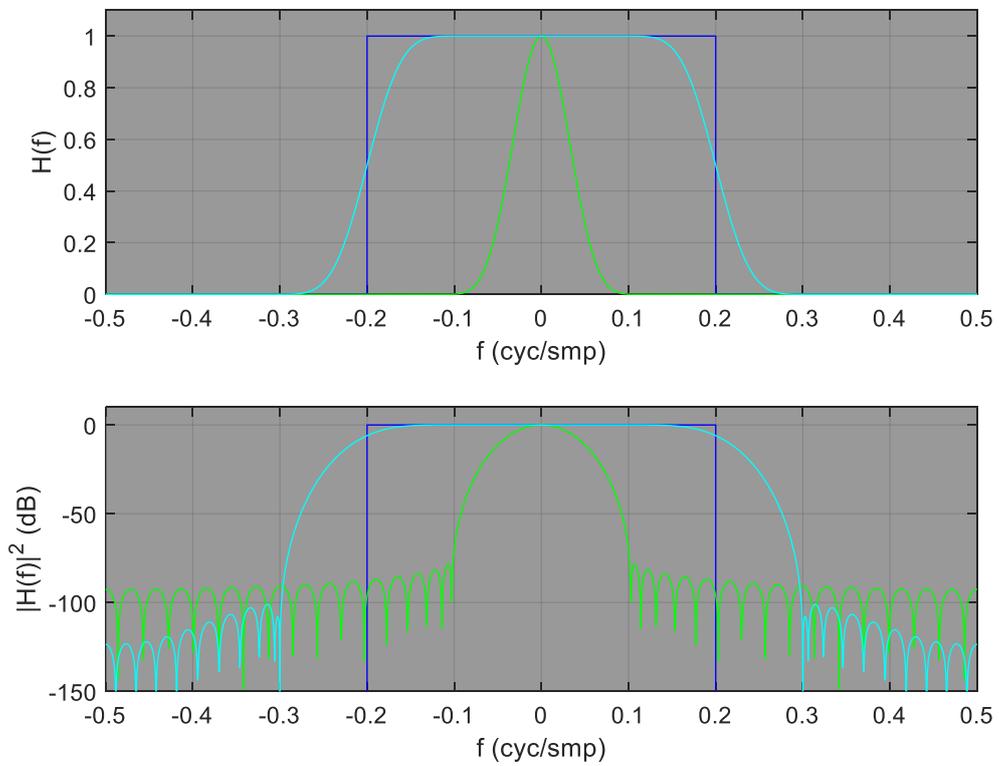

*Figure 16. Frequency response of a discretised (sampled and truncated) sinc function with a Slepian taper*

An alternative FIR design procedure, that does not rely on window functions, minimizes the Weighted Integral of the Squared Error (WISE), with

$$\text{WISE} = \widetilde{w}\widehat{\text{ISE}} + \overline{w}\,\overline{\text{ISE}} \;. \tag{16}$$

In the above equation, $\widehat{\text{ISE}}$ & $\overline{\text{ISE}}$ are squared error integrals over the pass band ($0 \leq |\omega| \leq \omega_{\text{lo}}$) and stop band ($\omega_{\text{hi}} \leq |\omega| \leq \pi$) respectively; and $\widetilde{w}$ & $\overline{w}$ are the corresponding weights (positive real scalars) that may be used to set the relative importance of a flat pass-band and low stop-band power. Using a gap between $\omega_{\text{lo}}$ and $\omega_{\text{hi}}$ introduces a transition band or a so-called 'don't-care' region that reduces band-edge error peaks. It also encourages a gradual roll-off between the pass band and the stop band for reduced ringing in the time domain. Indeed, an FIR filter with a Slepian-like response (with $\omega_{\text{eig}} = \omega_{\text{hi}}$) results as $\omega_{\text{lo}}$ approaches zero. The squared-error integrals are defined as follows:

$$\widehat{\text{ISE}} = \int_{-\omega_{\text{lo}}}^{\omega_{\text{lo}}} |E(\omega)|^2 \, d\omega \quad \text{and} \tag{17}$$

$$\overline{\text{ISE}} = \int_{-\pi}^{\pi} |E(\omega)|^2 \, d\omega - \int_{-\omega_{\text{hi}}}^{\omega_{\text{hi}}} |E(\omega)|^2 \, d\omega \tag{18}$$

where $E(\omega)$ is the difference between the desired (or ideal) frequency response and the realized (or actual) response of the low-pass FIR filter i.e. $E(\omega) = D(\omega) - H(\omega)$. For a low-pass filter $|\overline{D}(\omega)| = 1$ and $\overline{D}(\omega) = 0$ over the pass band and stop band, respectively. Better approximations are possible, in principle, when more filter coefficients are used, albeit at the expense of increased computational complexity for the deployed filter. This procedure is sufficiently flexible to accommodate filters of odd and even length (i.e. $M$) with an arbitrary pass-band group-delay of $q$ samples (this quantity is discussed in more detail in Part II). When $q = (M-1)/2$, the impulse response is symmetric for an FIR filter with a perfectly linear-phase response over $-\pi \leq \omega \leq \pi$. If the group delay is reduced for an asymmetric impulse-response, then the phase response is only approximately linear over the pass band. The response over the pass band is determined by the desired group delay with $\overline{D}(\omega) = e^{-iq\omega}$. As the group delay is now considered explicitly in the design process from the outset, $H(\omega)$ is complex and the filter coefficients are indexed using $m = 0 \ldots M - 1$.

Substitution of $E(\omega) = D(\omega) - H(\omega)$ into (17) & (18) then (17) & (18) into (16) and expanding yields

$$\text{WISE} = \boldsymbol{h}^{\mathrm{T}} \boldsymbol{S}_{xx} \boldsymbol{h} - 2\boldsymbol{h}^{\mathrm{T}} \boldsymbol{s}_{xy} + s_{yy}^2 \tag{19}$$

where

$\boldsymbol{h}$ is an $M \times 1$ (column) vector containing the FIR filter coefficients

$\boldsymbol{S}_{xx} = \widetilde{w}\overline{\boldsymbol{S}}_{xx} + \overline{w}\overline{\boldsymbol{S}}_{xx}$ with the elements of $\overline{\boldsymbol{S}}_{xx}$ and $\overline{\boldsymbol{S}}_{xx}$ (both $M \times M$ and Toeplitz) evaluated using (10) i.e.

$\overline{S}_{xx}[m,n] = S_{\omega_{lo}}[m,n]$ and

$\overline{S}_{xx}[m,n] = 2\pi - S_{\omega_{hi}}[m,n]$

$\boldsymbol{s}_{xy} = \widetilde{w}\overline{\boldsymbol{s}}_{xy}$ is an $M \times 1$ (column) vector with elements

$$\overline{s}_{xy}[m] = \int_{-\omega_{\text{lo}}}^{\omega_{\text{lo}}} e^{i(m-q)\omega} = \begin{cases} 2\sin\{(m-q)\omega_{\text{lo}}\}/(m-q), & m - q \neq 0 \\ 2\omega_{\text{lo}}, & m - q = 0 \end{cases} \quad \text{and} \tag{20}$$

$s_{yy}^2 = 2\widetilde{w}^2 \omega_{\text{lo}}.$

Differentiating the WISE with respect to the filter coefficients, setting the derivative to zero, then rearranging yields

$$\boldsymbol{s}_{xy} = \boldsymbol{S}_{xx}\boldsymbol{h} \tag{21}$$

which is solved for the filter coefficients using

$$\boldsymbol{h} = \boldsymbol{S}_{xx}^{-1}\boldsymbol{s}_{xy}. \tag{22}$$

Even for relatively large pass-band weights, the dc gain may not be unity (due to ripple in the pass band) therefore normalization is recommended using $\boldsymbol{h}_{\text{lsq}} = \boldsymbol{h}/c_{\text{dc}}$ where

$$c_{\text{dc}} = \sum_{m=0}^{M-1} h[m]. \tag{23}$$

The impulse and frequency responses of two low-pass FIR filters designed by the least squared-error procedure are shown below (see Figure 17-Figure 20). Both are designed using with $M = 33$, $f_{\text{c}} = 0.3$, $f_{\text{lo}} = f_{\text{c}}/2$ & $f_{\text{hi}} = f_{\text{c}}$ (for a wide transition band) and $\tilde{w} = 1.0$ & $\bar{w} = 1000.0$ (for low side-lobes). As $H(\omega)$ is complex, $Re\{H(\omega)\}$, $Im\{H(\omega)\}$ and $|H(\omega)|$ are shown in blue, red and green (respectively) in the upper subplots of Figure 18 and Figure 20. The desired phase-response (as set using $q$) is shown with a dashed black line in the lower subplots. The response of a linear-phase FIR filter with $q = 16$ is shown in Figure 17 and Figure 18. The response of a non-linear-phase FIR filter with a reduced group delay of $q = 8$ is shown in Figure 19 and Figure 20. The WISE of these filters is $9.6194 \times 10^{-8}$ and $1.9193 \times 10^{-6}$, respectively. The response over the transition band is not considered in these metrics and in the latter case, there are erroneous magnitude peaks near $\pm f_{\text{lo}}$ and the phase error gradually grow large as is $\pm f_{\text{hi}}$ is approached, otherwise, the response is reasonable.

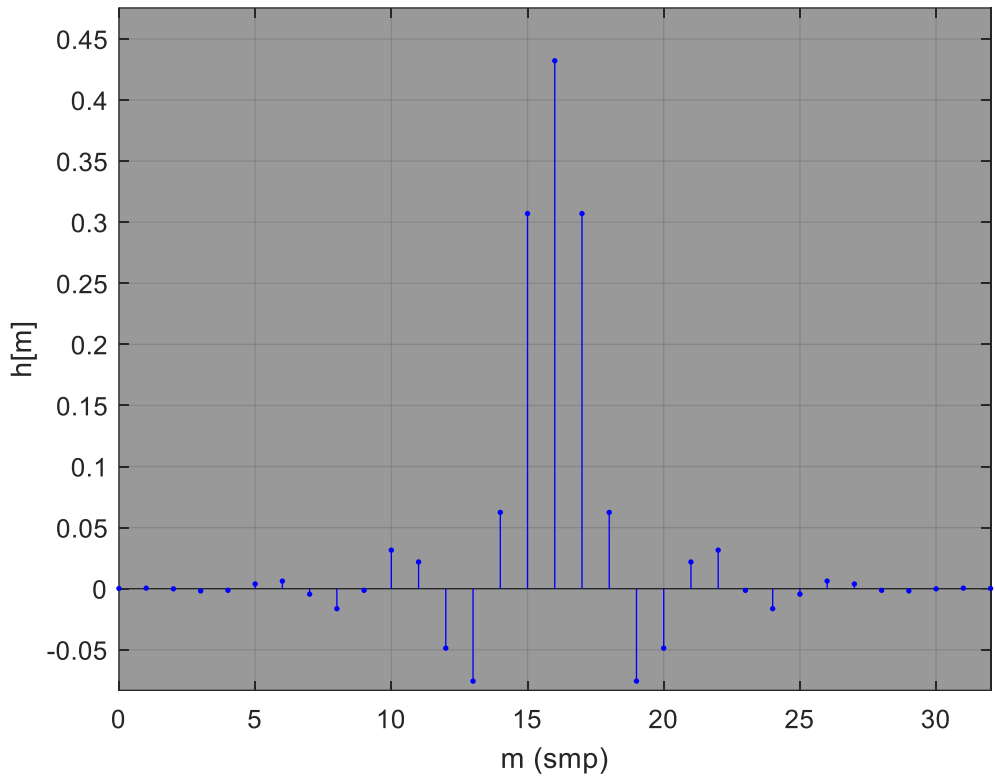

*Figure 17. Impulse response of least-squared-error low-pass filter with linear phase*

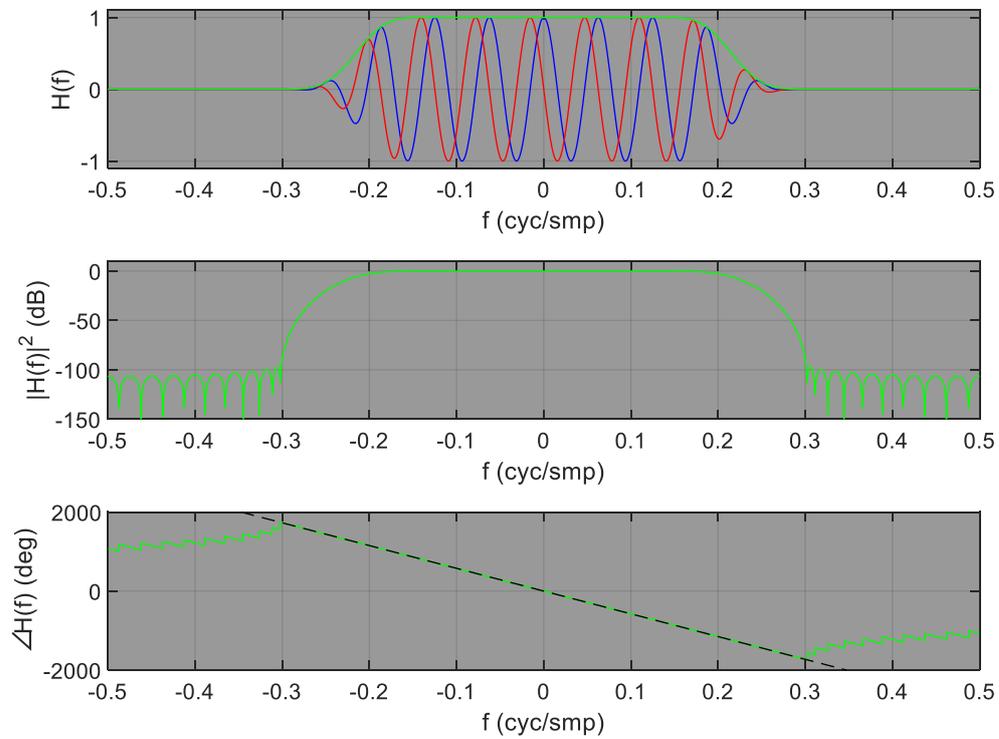

*Figure 18. Frequency response of least-squared-error low-pass filter with linear phase*

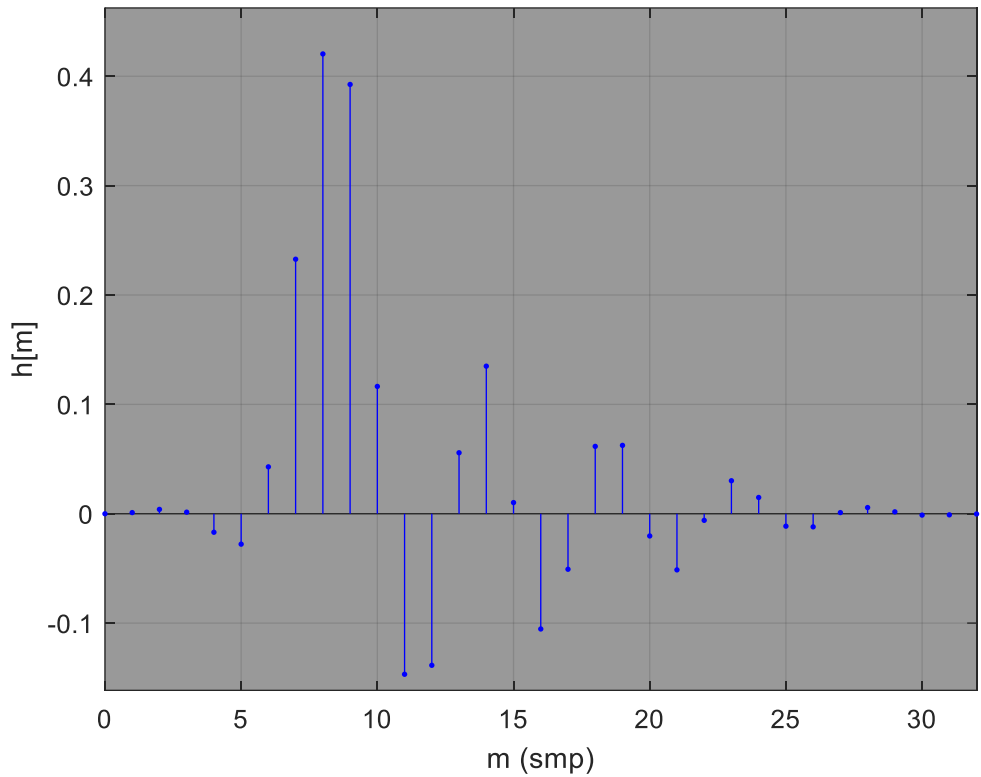

*Figure 19. Impulse response of least-squared-error low-pass filter with reduced group-delay*

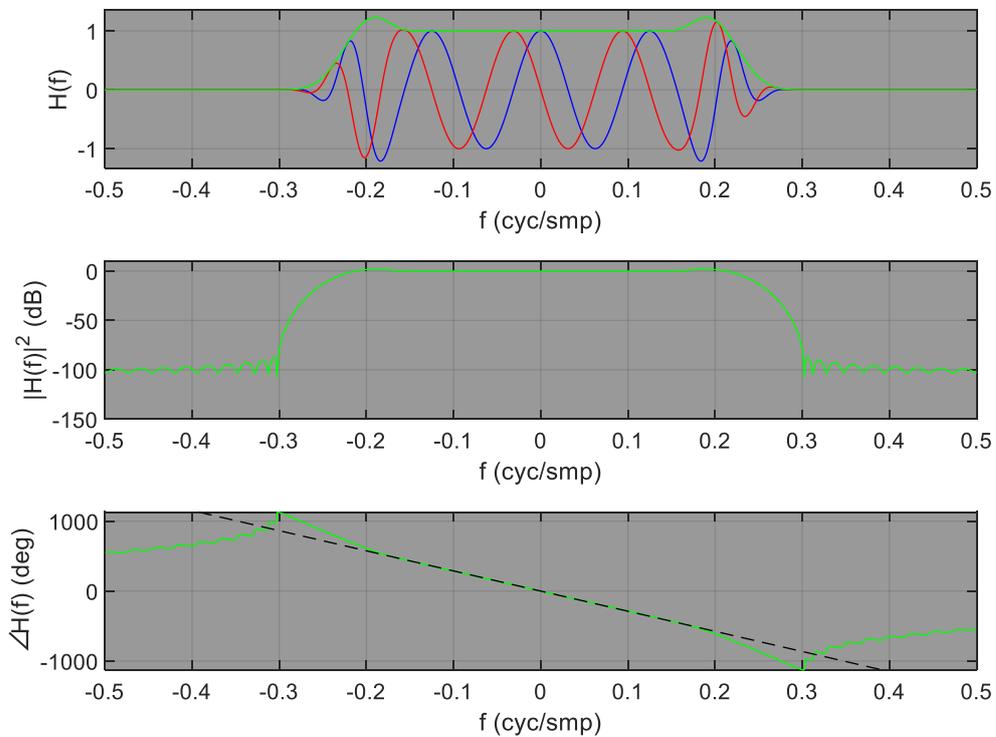

*Figure 20. Frequency response of least-squared-error low-pass filter with reduced group-delay, yielding a non-linear phase response*

# Part II: Infinite Impulse-Response (IIR) filters

In Part I of this document, non-recursive FIR filters were designed in the $n$ and $\omega$ domains using only the (discrete-time) Fourier transform (i.e. $n \to \omega$). For the IIR filters considered here, the $\mathcal{Z}$ transform (and its inverse) can no longer be avoided because the complex $z$-domain is a 'magical gateway' that provides the link (i.e. $n \leftrightarrow z \to \omega$) between the linear-difference equation, impulse response, and frequency response, of a recursive digital filter that utilizes feedback. Unfortunately, the behaviour of even simple feedback processes defies intuition; however, the otherwise abstract field of complex analysis, as explored by pioneers such as Cauchy and Riemann, provides a path. A brief conceptual overview is therefore provided here, with as little mathematics and theory as possible.

The $\mathcal{L}$ transform is a standard analytical tool that is used in most branches of science and engineering; thus, it is a core component of the undergraduate curriculum. However, use of the $\mathcal{Z}$ transform is largely confined to digital electronic engineering thus it is more commonly encountered in elective courses. Their roles are however analogous, and an appreciation of the former transform may help students to understand the latter. Moreover, the $\mathcal{Z}$ transform is arguably easier to understand than the $\mathcal{L}$ transform because concepts and applications may be explored with a digital computer, using only delay, multiply, and add operations, without the need for calculus (i.e. solving differential equations), a soldering iron and supplies of analogue circuit components.

The $\mathcal{Z}$ transform ($n \to z$) is to discrete-time systems what the $\mathcal{L}$ (i.e. Laplace) transform ($t \to s$) is to continuous-time systems.

In the $s$-domain, differentiation and integration operations (with respect to time) are replaced by $s$ and $1/s$ operators, respectively. This allows the ordinary differential equations used to model natural processes or synthetic systems (e.g. analogue filters in electronic devices) to be represented using a continuous-time transfer function $\mathcal{H}(s)$. It is the ratio of two polynomials (of numerator degree $M_b$ and denominator degree $M_a$, with $M_a \geq M_b$) in $s$, i.e.

$$\mathcal{H}(s) = \mathcal{B}(s)/\mathcal{A}(s) \tag{24}$$

$$\mathcal{H}(s) = \frac{b_0 s^{M_b-1} + b_1 s^{M_b-2} + b_2 s^{M_b-3} + \ldots + b_{M_b-1}}{a_0 s^{M_a-1} + a_1 s^{M_a-2} + a_2 s^{M_a-3} + \ldots + a_{M_a-1}}. \tag{25}$$

Analogue filters are designed using networks of integrating elements, each with an ideal transfer function of $1/s$. Thus, dividing the numerator and denominator polynomials by $s^{M-1}$ yields the following form that is better aligned with the system realization:

$$\mathcal{H}(s) = \frac{b_0 + b_1 s^{-1} + b_2 s^{-2} + \ldots + b_{M-1} s^{-M+1}}{a_0 + a_1 s^{-1} + a_2 s^{-2} + \ldots + a_{M-1} s^{-M+1}}. \tag{26}$$

Note that in (26) it is assumed that the degrees of polynomials $\mathcal{B}(s)$ and $\mathcal{A}(s)$ are the same and both equal to $M - 1$.

In the $z$-domain, advance and delay operations (by one sample) in the discrete-time domain are replaced by $z$ and $1/z$ operators, respectively. This allows the finite-difference equations used to define synthetic discrete-time systems (e.g. digital filters in electronic devices) to be represented using a discrete-time transfer function $\mathcal{H}(z)$. It is the ratio of two polynomials in $z$, i.e.

$$\mathcal{H}(z) = \mathcal{B}(z)/\mathcal{A}(z) \tag{27}$$

$$\mathcal{H}(z) = \frac{b_0 z^{M_b-1} + b_1 z^{M_b-2} + b_2 z^{M_b-3} + \dots + b_{M_b-1}}{z^{M_a-1} + a_1 z^{M_a-2} + a_2 z^{M_a-3} + \dots + a_{M_a-1}}. \tag{28}$$

Digital electronic devices are designed using networks of delay operations, each with a transfer function of $1/z$. Thus, dividing the numerator and denominator polynomials by $z^{M-1}$ yields the following form that is better aligned with the system realization:

$$\mathcal{H}(z) = \frac{b_0 + b_1 z^{-1} + b_2 z^{-2} + \dots + b_{M-1} z^{-M+1}}{1 + a_1 z^{-1} + a_2 z^{-2} + \dots + a_{M-1} z^{-M+1}}. \tag{29}$$

Note that in (29) it is assumed that the degrees of polynomials $\mathcal{B}(z)$ and $\mathcal{A}(z)$ are the same (both equal to $M-1$) and that $a_0 = 1$. Or if the polynomial coefficients are held in the $(1 \times M)$ vectors $\boldsymbol{b}$ and $\boldsymbol{a}$

$$\mathcal{H}(z) = \frac{b[0] + b[1] z^{-1} + b[2] z^{-2} + \dots + b[M-1] z^{-M+1}}{1 + a[1] z^{-1} + a[2] z^{-2} + \dots + a[M-1] z^{-M+1}}. \tag{30}$$

The discrete-time transfer-function $H(z)$ is a complex function of complex argument, with 'poles' at the roots of $\mathcal{A}(z)$ and 'zeros' at the roots of $\mathcal{B}(z)$, i.e. points in the complex $z$-plane, where $|\mathcal{H}(z)| = \infty$ and $|\mathcal{H}(z)| = 0$, respectively. The 'order' of a digital (FIR or IIR) filter is equal to the number of poles or the degree of the $\mathcal{A}(z)$ polynomial, i.e. $M-1$. In the FIR case, all poles are at the origin of the complex $z$-plane, i.e. at $z = 0$. In the more general IIR case, let $\alpha_m$ and $\beta_m$ be the $m$th roots of $\mathcal{A}(z)$ and $\mathcal{B}(z)$ in (27), respectively.

For causal discrete-time systems, e.g. digital filters that operate on the oldest sample first and the newest sample last, i.e. the 'forward' direction, all poles must be inside the unit circle, i.e. $|\alpha_m| < 1$, for $\mathcal{H}(z)$ to have a stable impulse response, i.e. $\lim\limits_{n \to \infty} |h[n]| = 0$.

For anti-causal discrete-time systems, e.g. digital filters that operate on the newest sample first and the oldest sample last, i.e. the 'backward' direction, all poles must be outside the unit circle, i.e. $|\alpha_m| > 1$, for $\mathcal{H}(z)$ to have a stable impulse response, i.e. $\lim\limits_{n \to -\infty} |h[n]| = 0$.

In both cases, the rate of impulse-response decay decreases as the poles approach the unit circle. For a real impulse response, all complex poles (and zeros) occur in conjugate pairs.

It is convenient to design causal and anti-causal IIR filters in the complex $z$-domain so that constraints may be placed on the pole positions to ensure that the above properties are satisfied. The filter coefficients (i.e. $\boldsymbol{b}$ and $\boldsymbol{a}$) are set, subject to these constraints, so that the frequency response of $\mathcal{H}(z)$, i.e. $H(\omega)$, approximates the desired frequency response $D(\omega)$, where

$$H(\omega) = \mathcal{H}(z)|_{z=e^{i\omega}} \tag{31}$$

i.e. by evaluating $\mathcal{H}(z)$ on the unit circle where $|z| = 1$.

By definition, $Y(z) = \mathcal{H}(z)X(z)$, or $\mathcal{A}(z)Y(z) = \mathcal{B}(z)X(z)$, after substituting $\mathcal{H}(z) = \mathcal{B}(z)/\mathcal{A}(z)$ and rearranging, where $X(z)$ and $Y(z)$ are the $\mathcal{Z}$-transforms of the input sequence and output sequence respectively, i.e. $X(z) = \mathcal{Z}\{x[n]\}$ and $Y(z) = \mathcal{Z}\{y[n]\}$. The linear-difference equation that is used to realize the filter in the discrete-time domain is therefore reached by taking the inverse $\mathcal{Z}$-transforms i.e.

$$\mathcal{Z}^{-1}\{\mathcal{A}(z)Y(z)\} = \mathcal{Z}^{-1}\{\mathcal{B}(z)X(z)\}. \tag{32}$$

After substituting the $\mathcal{A}(z)$ and $\mathcal{B}(z)$ polynomials:

$$\mathcal{Z}^{-1}\{(1 + a[1] z^{-1} + a[2] z^{-2} + \dots + a[M-1] z^{-M+1})Y(z)\} =$$

$$\mathcal{Z}^{-1}\{(b[0] + b[1]z^{-1} + b[2]z^{-2} + \ ... \ + b[M-1]z^{-M+1})X(z)\}. \tag{33}$$

After expanding the arguments of the inverse $\mathcal{Z}$-transform:

$$\mathcal{Z}^{-1}\{Y(z) + a[1]z^{-1}Y(z) + a[2]z^{-2}Y(z) + \ ... \ + a[M-1]z^{-M+1}Y(z)\} =$$

$$\mathcal{Z}^{-1}\{b[0]X(z) + b[1]z^{-1}X(z) + b[2]z^{-2}X(z) + \ ... \ + b[M-1]z^{-M+1}X(z)\}. \tag{34}$$

After applying the inverse $\mathcal{Z}$-transform, and using the fact that, $z^{-m}$ is a shift by $m$ samples backwards in time, e.g. $\mathcal{Z}^{-1}\{z^{-m}X(z)\} = x[n-m]$, the products of polynomials in $z$ above become the convolutions in $n$ below, i.e.

$$y[n] + a[1]y[n-1] + a[2]y[n-2] + \ ... \ + a[M-1]y[n-M+1] =$$

$$b[0]x[n] + b[1]x[n-1] + b[2]x[n-2] + \ ... \ + b[M-1]x[n-M+1]. \tag{35}$$

After rearranging the above, the linear-difference equation is reached:

$$y[n] = b[0]x[n] + b[1]x[n-1] + b[2]x[n-2] + \ ... \ + b[M-1]x[n-M+1]$$

$$- a[1]y[n-1] - a[2]y[n-2] - \ ... - a[M-1]y[n-M+1] \tag{36}$$

which is used to realize the (recursive) IIR filter in the discrete-time domain.

The *ab-initio* design of low-pass IIR filters is somewhat more complicated than the design of low-pass FIR filters and for this reason the discretisation of analogue prototypes is a commonly used design procedure. For instance, the transfer function of a classical Butterworth filter in continuous relative time (i.e. $t/T_{smp}$) with an order of $2M$ and a cut-off frequency of $\omega_c$ in relative angular frequency ($\omega_c = 2\pi F_c/F_{smp}$, where $F_c$ is the cut-off frequency in Hz) has a simple analytical expression:

$$\mathcal{H}(s) = \mathcal{B}(s)/\mathcal{A}(s) \text{ where } \mathcal{B}(s) = 1 \text{ and } \mathcal{A}(s) = \left(\frac{-1}{\omega_c^2}\right)^M s^{2M} + 1 \ . \tag{37}$$

This transfer function is discretised, i.e. mapped from the $s$ to the $z$ domains, using the bilinear transformation

$$\mathcal{H}(z) = \mathcal{H}(s)\big|_{s=2\frac{(z-1)}{(z+1)}} \ . \tag{38}$$

Performing this substitution, expanding, then simplifying to get $\mathcal{B}(z)$ & $\mathcal{A}(z)$, is not particularly onerous, due to simplicity of the $\mathcal{B}(s)$ & $\mathcal{A}(s)$ polynomials for the Butterworth filter. This mapping preserves the important properties of the Butterworth filter, e.g. the monotonic roll-off from the passband to the stop band and a ripple-free frequency response. Furthermore, it preserves the flatness of the first $2M$ derivatives of the frequency response at dc, i.e.

$$d^k H(\omega)/d\,\omega^k\big|_{\omega=0} = d^k D(\omega)/d\,\omega^k\big|_{\omega=0} = \begin{cases} 1, & k=0 \\ 0, & 0 < k < 2M \end{cases} . \tag{39}$$

Moreover, $2M$ zeros are placed at $z = -1$ (or very near that point, due to finite numerical precision) for a deep and wide null at $\omega = \pi$, i.e.

$$d^k H(\omega)/d\,\omega^k\big|_{\omega=\pi} = d^k D(\omega)/d\,\omega^k\big|_{\omega=\pi} = 0 \text{ for } 0 \le k < 2M \ . \tag{40}$$

However, the cut-off frequency is not maintained precisely, i.e. $|H(\omega_c)| \ne 1/2$. In most cases, the match is reasonable; however, for wide-band low-pass filters the bandwidth is narrower than it should be, i.e. $|H(\omega_c)| < 1/2$. Other mappings that maintain the cut-off frequency exactly may be applied; however, they are not discussed here.

The frequency response of a digital 8ᵗʰ-order Butterworth filter (i.e. $M = 4$) with $\omega_c = 2\pi f_c$ and $f_c = 0.3$ is plotted in Figure 22 (solid green line). The frequency response is real because the impulse response is symmetric about $m = 0$. As this filter has $M$ poles inside the unit circle and $M$ poles outside the unit circle, a stable linear-difference equation is not reached via the inverse $\mathcal{Z}$-transform of $\mathcal{H}(z)$; however, it is readily obtained by factoring $\mathcal{H}(z)$, which is non-causal, into a product of discrete-time transfer functions. That is, let $\overleftrightarrow{\mathcal{H}}(z) = \mathcal{H}(z)$ thus $\overleftrightarrow{\mathcal{H}}(z) = \overleftarrow{\mathcal{H}}(z)\overrightarrow{\mathcal{H}}(z)$, where $\overleftarrow{\mathcal{H}}(z)$ is an anti-causal transfer function with $M$ poles outside the unit circle and $\overrightarrow{\mathcal{H}}(z)$ is a causal transfer-function with all $M$ poles inside the unit circle. Their respective linear-difference equations are obtained via the inverse $\mathcal{Z}$-transform of $\overleftarrow{\mathcal{H}}(z)$ and $\overrightarrow{\mathcal{H}}(z)$. The impulse response of $\overleftrightarrow{\mathcal{H}}(z)$ may now be obtained by filtering a unit impulse input in the backward direction (until steady-state is reached) using the linear-difference equation of $\overleftarrow{\mathcal{H}}(z)$ to yield $\overleftarrow{h}[n]$, followed by the filtering of $\overleftarrow{h}[n]$ in the forward direction (until steady-state is reached) using the linear-difference equation of $\overrightarrow{\mathcal{H}}(z)$ to yield $\overleftrightarrow{h}[n]$.

The resulting impulse response of this non-causal recursive IIR filter, is plotted in Figure 21. In this plot, the discrete-time index is $m$ (instead of $n$) and the impulse response is truncated at $m = \pm K$ (i.e. $m = -K \dots K$, where $K = 16$ so that it may be compared with the impulses responses of the wide-band FIR filters considered previously. The coefficients of an FIR may be derived from this truncated impulse response by setting $b[m] = \overleftrightarrow{h}[m]$ for $m = -K \dots K$. The corresponding frequency response of this approximately equivalent FIR filter is also plotted in Figure 22 (dotted green line). The side-lobes of the FIR filter are lowered, thus the approximation improves, as $K \to \infty$. For this wide-band low-pass filter the reduction in the computational complexity of the recursive IIR filter (of order $2M$) relative to a non-recursive FIR filter (of order $2K$, that is realized in the time domain) is not significant. However, the computational complexity of an IIR filter is independent of its impulse-response duration; therefore, the relative computational savings become more significant as the bandwidth of the digital filter decreases.

The impulse response and frequency response of the causal filter, i.e. $\overrightarrow{h}[m]$ and $\overrightarrow{H}(\omega)$, are shown in Figure 23 and Figure 24, respectively. As the impulse response of this filter is clearly asymmetric, the phase response is not perfectly linear over $\omega = \pm\pi$ (see lower subplot of Figure 24); however, the magnitude and phase response are both reasonably flat and linear (respectively) over the pass band until the band-edge is approached.

After factoring the non-causal transfer function, reached via the bilinear transform, into causal and non-causal parts, only the first three derivatives of $\overrightarrow{H}(\omega)$ evaluated at $\omega = 0$ match the desired values i.e. for $k = 0 \dots 2$, which are now complex due to the delay that has been introduced, i.e. $D(\omega) = e^{-iq\omega}$ in the pass band. The $M$ zero derivatives at $\omega = \pi$ are maintained after the factorization. Further investigation suggests these important features of $\overrightarrow{H}(\omega)$ for this bilinear mapping, are independent of the filter order, i.e. for $k = 0 \dots \min(2, M-1)$ at $\omega = 0$ and for $k = 0 \dots M-1$ at $\omega = \pi$.

The pass-band group-delay $q$ (in samples) of a low-pass digital filter is the time shift that a band-limited input pulse experiences as it passes through the filter. It is proportional to the derivative of the phase response with respect to frequency, i.e.

$$q = -\frac{d\phi(\omega)}{d\omega}. \tag{41}$$

For low-pass filters with a non-linear phase response, the pass-band group-delay at the dc limit (i.e. as $\omega \to 0$) is evaluated using

$$q = Im\left\{\frac{\mathcal{H}'_{dc}(z)}{\mathcal{H}_{dc}(z)}\right\} \text{ where} \tag{42}$$

$$\mathcal{H}_{dc}(z) = \mathcal{H}(z)|_{z=1} \text{ and} \tag{43}$$

$$\mathcal{H}'_{dc}(z) = \frac{d\mathcal{H}(z)}{dz}\bigg|_{z=1}. \tag{44}$$

The group delay of the non-causal filter in Figure 21 and Figure 22 is of course zero. The group delay of the causal filter in Figure 23 and Figure 24 is 1.3863 samples. The ideal linear-phase response (in radians) is computed from this value using $\phi(\omega) = -q\omega$ and shown for comparison in the phase response plot (dashed black line in the lower subplot of Figure 24). Note that a non-linear phase response in the stop band is of no consequence if the stop-band gain is negligible. Note also that the pass-band phase-error is not as bad as the asymmetry of the impulse response might suggest. The adequacy of the pass-band phase response is illustrated (see Figure 25) where a band-limited pulse is processed (i.e. a Slepian input of unit power, with $K = 8$ and $f_c = 0.3$) instead of an all-band pulse (i.e. an impulse input of unit power with $f_c = 0.5$). For a symmetric input pulse (cyan) the band-edge phase-error does not appear to significantly distort the filter output (blue) and the pulse passes through the filter (largely) unaffected. Coarse 'visual-interpolation' of the response suggests that the output is delayed by around 1.5 samples, relative to the input, in accordance with the computed group delay of 1.4 samples.

Non-causal IIR filters may be applied when the data samples land at the processor together in frames or batches, for example in an image from a digital camera, or time series data that are processed off-line. In online systems where samples arrive sequentially, non-causal filters are only feasible, if large latencies are acceptable, while the system waits for a data buffer to fill before batch processing is commenced. And in such cases, an FIR filter realized via an FFT is also an attractive processing alternative. However, in communication or radar systems, a hybrid approach may be used, whereby an FIR filter is applied on transmit (tx) and a causal IIR filter is used on receive (rx) where the coefficients of the FIR filter are set equal to truncated impulse response of the conjugate anti-causal IIR filter, as discussed above and illustrated in Figure 23 and Figure 24. The response of the proposed filter (tx & rx combined) is shown in Figure 26 and Figure 27. The impulse response (see Figure 26) is only truncated on the left-hand (anti-causal) side; the right-hand (causal) side of the impulse response extends to infinity, and it is only truncated for display purposes. In the frequency-response plot (i.e. Figure 27) its magnitude and phase response is plotted (solid blue, red, and green, lines) along with the magnitude response of the corresponding non-causal IIR filter (dotted green line). The magnitude response of the FIR filter (dash-dot green line) and the causal IIR filter (dashed green line) are also shown.

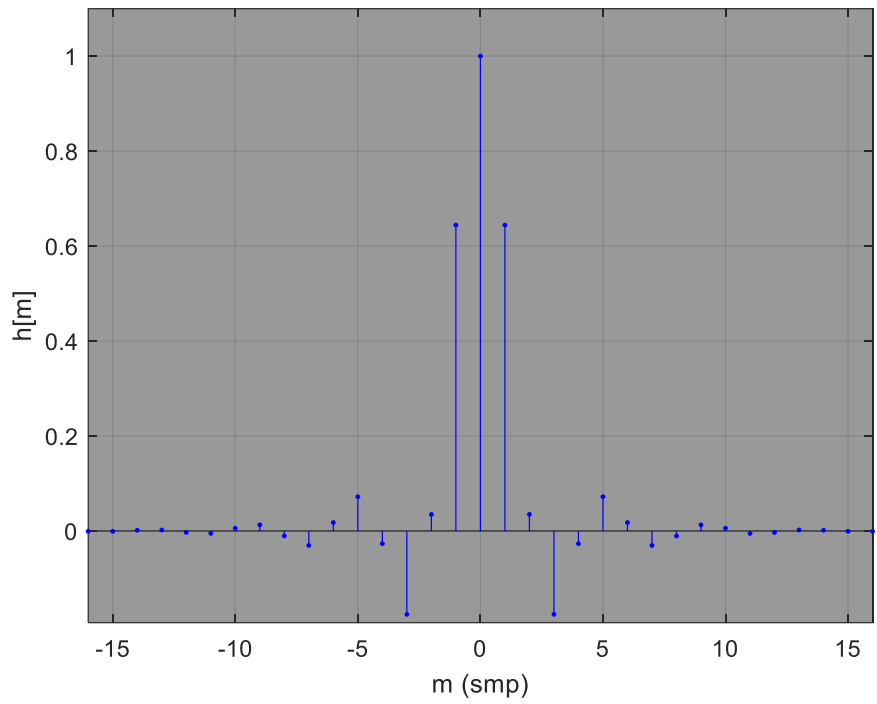

*Figure 21. Impulse response of a non-causal discretised Butterworth filter*

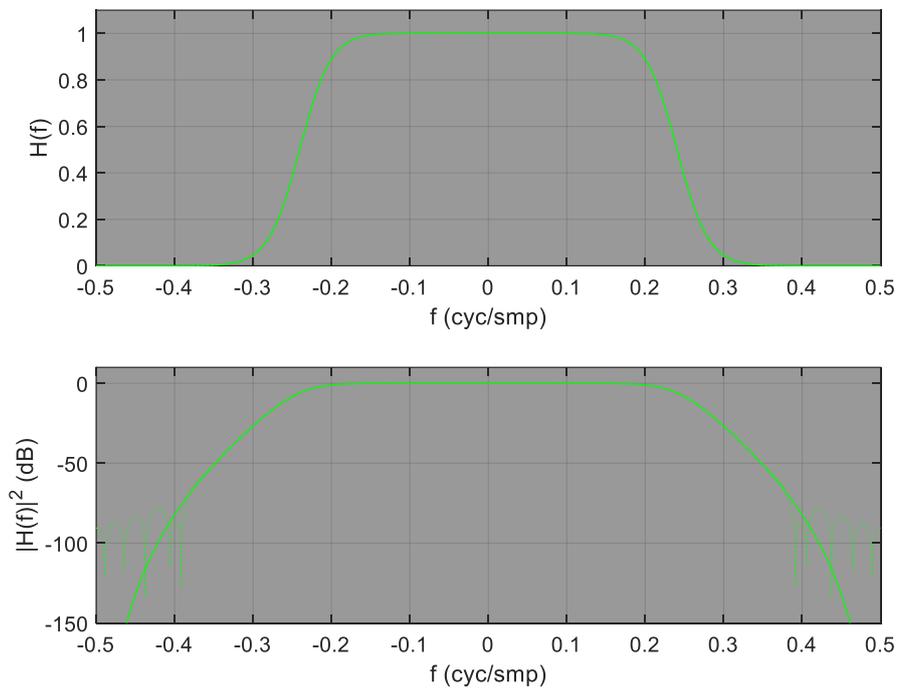

*Figure 22. Frequency response of a non-causal discretised Butterworth filter*

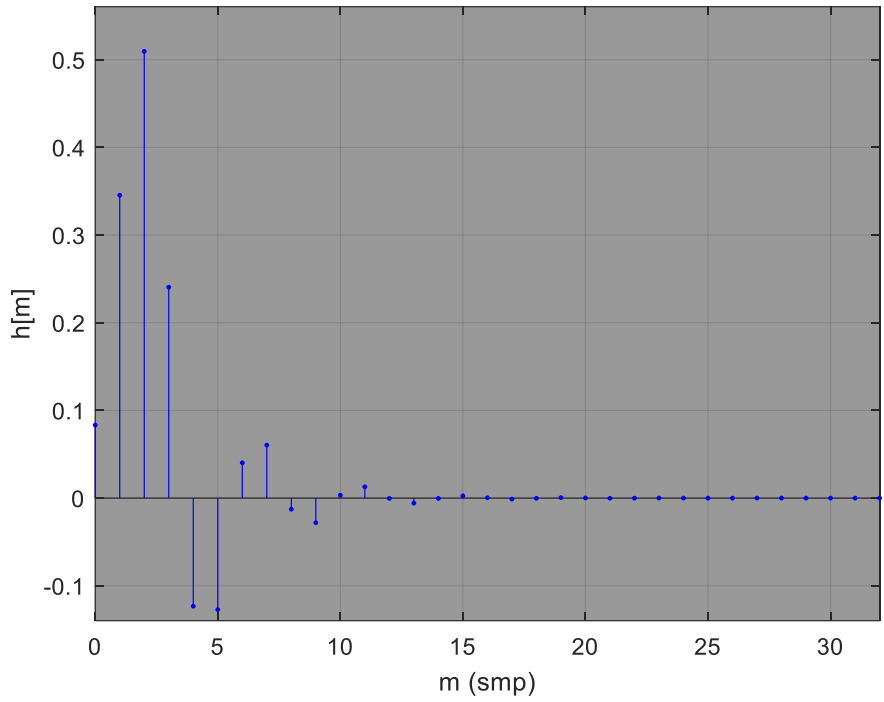

*Figure 23. Impulse response of a causal discretised Butterworth filter*

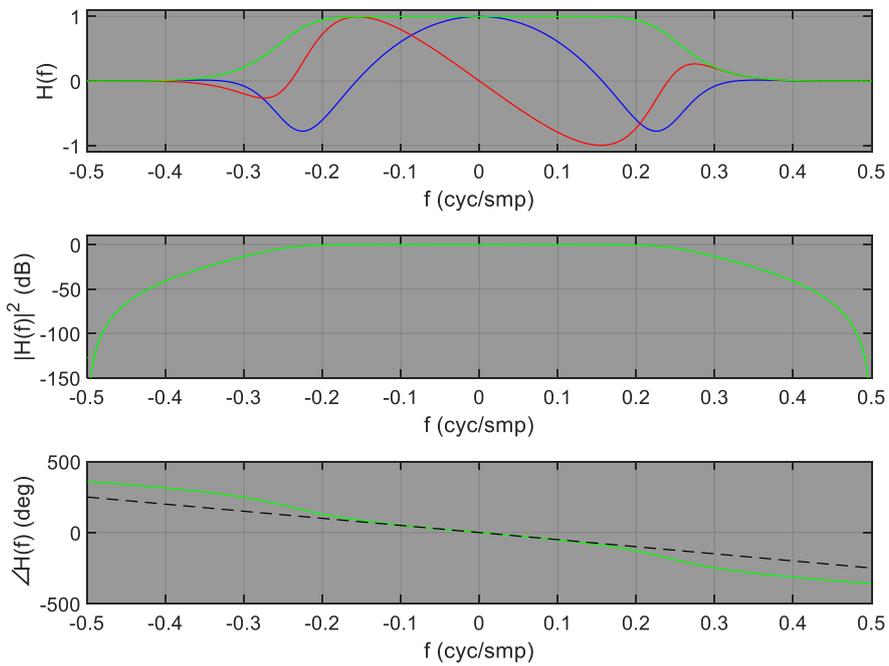

*Figure 24. Frequency response of a causal discretised Butterworth filter*

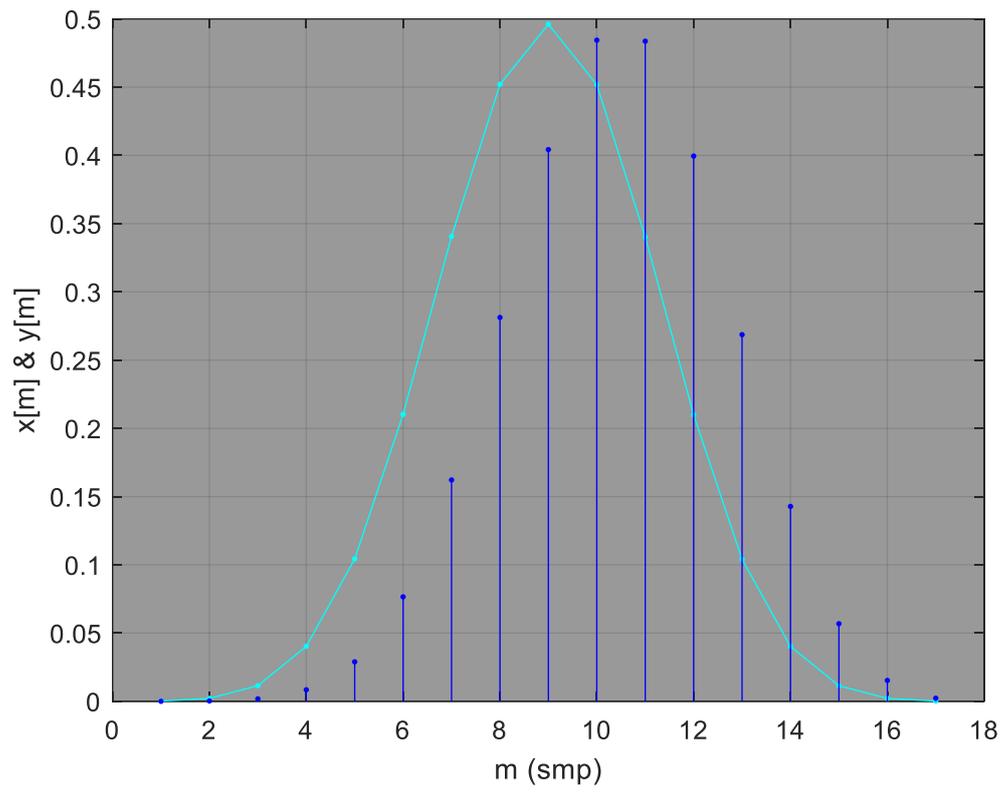

*Figure 25. Slepian pulse response of a causal discretised Butterworth filter*

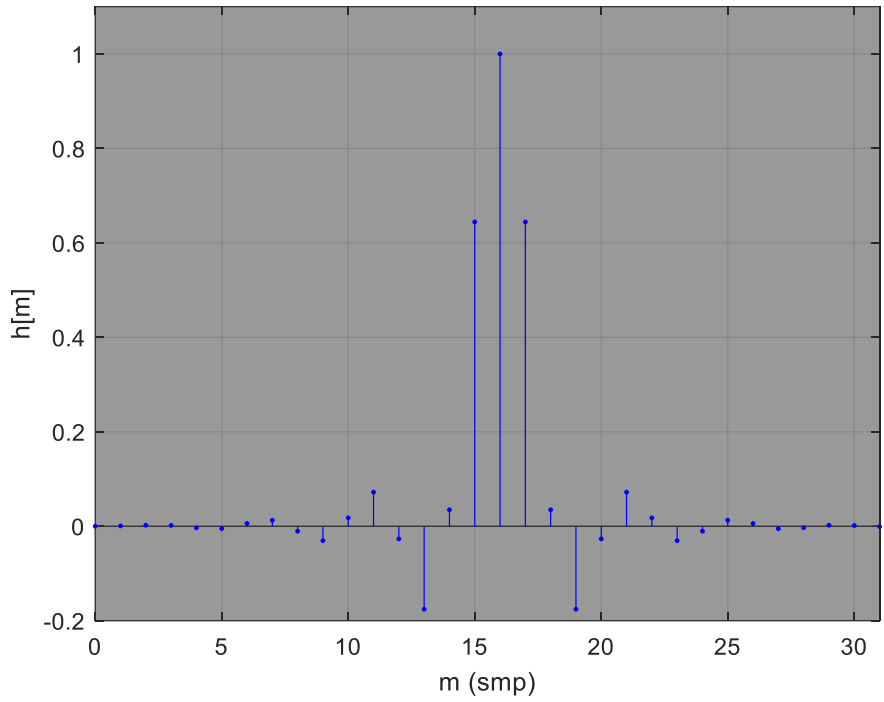

*Figure 26. Impulse response of a discretised Butterworth filter, realized using an FIR filter and a causal IIR filter*

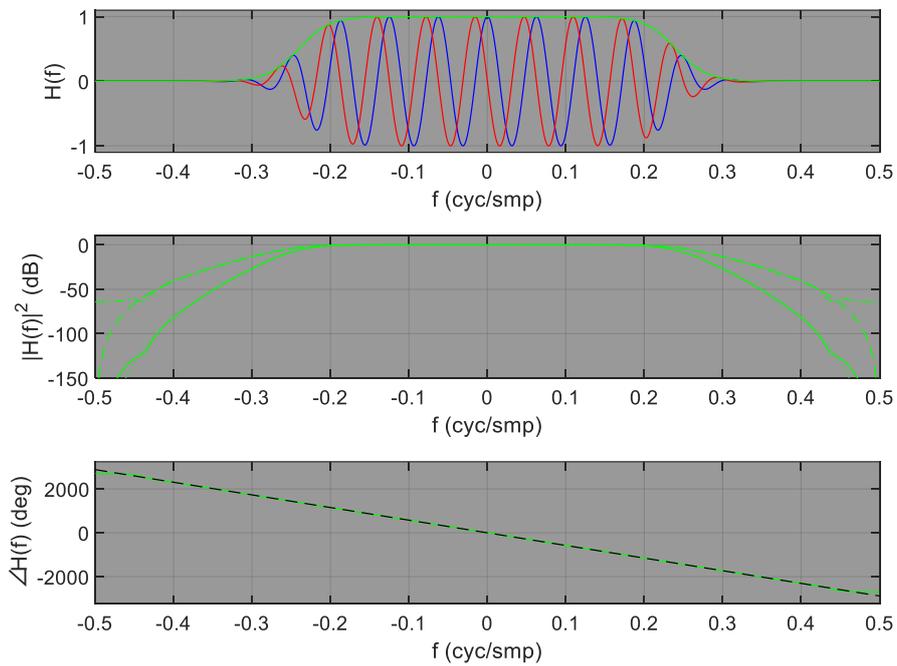

*Figure 27. Frequency response of a discretised Butterworth filter, realized using an FIR filter and a causal IIR filter*

# Part III: Modulation and wireless-communication applications

The examples below consider a noisy (0 dB) channel that only supports the propagation of real waves (i.e. not complex). The channel (i.e. tx & rx hardware and radio-frequency propagation) is otherwise ideal (i.e. unity magnitude scaling and zero phase shift at all frequencies) and there are no other users or jammers to consider. Furthermore, perfect tx-rx synchronization of time, phase and frequency is assumed to have been established, via a known sequence of calibration pulses on link establishment, for instance. Low-pass filters for pulse-modulation schemes employing phase-shift-keying and orthogonal sub-channels are discussed. The purpose of these examples is not to compare one modulation scheme with another; rather, the aim is to show how the performance of a given modulation scheme (i.e. bit rate and symbol resolvability) may be computed from the parameters of the tx & rx low-pass pulse-shaping filters.

For this hypothetical system the one-sided bandwidth of the channel is a free parameter, i.e. from zero to half the sampling rate of the digitizer (samples per second) for a pulse duration of infinity or unity (samples). The channel capacity is

$$\mathcal{C} = f_{\text{chn}} \log_2(1 + \text{SNR}) \tag{45}$$

where

$f_{\text{chn}}$ is the cut-off frequency (i.e. one-sided 'bandwidth') of the channel ($f_{\text{chn}} = 1/2$ cycles per sample)

and

SNR is the ratio (on a linear scale) of the signal power to the noise power (SNR = 1)

therefore

$$\mathcal{C} = 0.5 \log_2(2) = 0.5 \text{ bits per sample.} \tag{46}$$

## Radio-frequency definitions used in Part III

### Examples 1-6

Modem: **Mod**ulation and **dem**odulation.

$f_{\text{chn}}$: The cut-off relative frequency (cycles per sample) i.e. the one-sided 'bandwidth' of a communication channel.

SNR: Signal-to-Noise (power) Ratio.

$\mathcal{C}$: Channel capacity (bits per sample).

$\widetilde{K}$: Number of sub-channel pairs in a multiplexed channel.

$\widetilde{M}$: Odd number of sub-channels in a multiplexed channel, $\widetilde{M} = 2\widetilde{K} + 1$.

$\tilde{k}$: Sub-channel index, $\tilde{k} = -\widetilde{K} \dots \widetilde{K}$.

$K^{\#}$: Number of constellation points in the complex plane, i.e. the number of symbols in an un-multiplexed channel or the number of tokens per sub-channel in a multiplexed channel.

$k^{\#}$: Symbol or token index, $k^{\#} = 0 \dots K^{\#} - 1$.

$\widetilde{K}^{\#}$: Number of symbols in a multiplexed channel, $\widetilde{K}^{\#} = \widetilde{M} K^{\#}$.

$\phi_0$: Phase offset of the constellation.

$\phi_\Delta$: Phase increment of the constellation.

$\phi[k^{\#}]$: Phase of the $k^{\#}$th symbol or token in a constellation, $\phi[k^{\#}] = \phi_0 + \phi_\Delta k^{\#}$.

$\blacksquare^{\downarrow}$: A quantity at the lower rate, i.e. at the symbol or pulse rate.

$\blacksquare^{\uparrow}$: A quantity at the higher rate, i.e. at the sampling rate of the analogue-to-digital or digital-to-analogue converter.

$N^{\downarrow}$: Number of symbols in a message or the number of pulses in a burst.

$n^{\downarrow}$: Index of the $n^{\downarrow}$th symbol or pulse in a message or burst, for $n^{\downarrow} = 0 \dots N^{\downarrow} - 1$.

$\chi_{\text{tx}}[n^{\downarrow}]$: Decimal value of the $n^{\downarrow}$th symbol in a message sent by the transmitter.

$k_n^{\#}$: Symbol or token index of the $n^{\downarrow}$th symbol in a message, $k_n^{\#} = \chi_{\text{tx}}[n^{\downarrow}]$.

$\rho$: Magnitude of the digital tx pulse train.

$\phi_{\text{tx}}[n^{\downarrow}]$: Phase of the $n^{\downarrow}$th transmitted pulse in a message.

$\varphi_{\text{tx}}[n^{\downarrow}]$: Complex value of the $n^{\downarrow}$th symbol in a message sent by the transmitter, $\varphi_{\text{tx}} = \rho e^{i\phi_{\text{tx}}}$.

$N^{\uparrow}$: Number of samples in a message or burst.

$n^{\uparrow}$: Index of the $n^{\uparrow}$th sample in a burst, for $n^{\uparrow} = 0 \dots N^{\uparrow} - 1$.

$h_{\text{tx}}[n^{\uparrow}]$: Impulse response of the low-pass filter used for pulse shaping at the transmitter.

$M_{\text{tx}}^{\uparrow}$: Duration (samples) of each transmitted pulse, i.e. the length of $h_{\text{tx}}[n^{\uparrow}]$, with $M_{\text{tx}}^{\uparrow} = 2K_{\text{tx}}^{\uparrow} + 1$.

$\varphi_{\text{tx}}[n^{\uparrow}]$: The transmitted pulse train (complex).

$\psi_{\text{tx}}[n^{\uparrow}]$: The sampled oscillator or the digital carrier (complex). Is modulated by the pulse train at the transmitter.

$\omega_{\text{tx}}$: Relative angular frequency (radians per sample) of the digital carrier $\psi_{\text{tx}}$.

$f_{\text{tx}}$: Relative frequency (cycles per sample) of the digital carrier $\psi_{\text{tx}}$.

$\tilde{\psi}_{\text{tx}}[n^{\uparrow}]$: Real part of the modulated digital carrier.

$H_{\text{chn}}(\Omega)$: Frequency-response of the channel's continuous-time transfer-function.

$\tilde{\psi}_{\text{rx}}[n^{\uparrow}]$: Sampled waveform (real) at the receiver.

$\psi_{\text{rx}}[n^{\uparrow}]$: The sampled oscillator or the digital carrier (complex). Is mixed with $\tilde{\psi}_{\text{rx}}[n^{\uparrow}]$ at the receiver.

$\tilde{\varphi}_{\text{rx}}[n^{\uparrow}]$: Mixed waveform comprised of sum and difference components.

$\varphi_{\text{rx}}^{+}[n^{\uparrow}]$: Sum component (complex) of the mixed waveform $\tilde{\varphi}_{\text{rx}}$, centred on $\omega = \omega_{\text{tx}} + \omega_{\text{rx}}$ .

$\varphi_{\text{rx}}^{-}[n^{\uparrow}]$: Difference (i.e. base-band) component (complex) of the mixed waveform $\tilde{\varphi}_{\text{rx}}$, centred on $\omega = \omega_{\text{tx}} - \omega_{\text{rx}}$ .

$\Phi_{\text{rx}}^{+}(\omega)$: Power density spectrum of $\varphi_{\text{rx}}^{+}[n^{\uparrow}]$.

$\Phi_{\text{rx}}^{-}(\omega)$: Power density spectrum of $\varphi_{\text{rx}}^{-}[n^{\uparrow}]$.

$h_{\downarrow}[m]$: Impulse response of the low-pass filter used on rx down-conversion to remove the sum component.

$H_{\downarrow}(\omega)$: Frequency response of $h_{\downarrow}[m]$.

$h_{\text{rx}}[n^{\uparrow}]$: Impulse response of the low-pass filter used for pulse-matched filtering at the receiver.

$M_{\text{rx}}^{\uparrow}$: Assumed duration (samples) of each transmitted symbol ($M_{\text{rx}}^{\uparrow} = M_{\text{tx}}^{\uparrow}$) i.e. the length of $h_{\text{rx}}[n^{\uparrow}]$ for an FIR filter, with $M_{\text{tx}}^{\uparrow} = 2K_{\text{tx}}^{\uparrow} + 1$.

$\varphi_{\text{rx}}[n^{\uparrow}]$: The rx base-band pulse-train at the higher sampling rate.

$\varphi_{\text{rx}}[n^{\downarrow}]$: The down-sampled rx base-band pulse-train at the lower sampling rate.

$c_{\text{tx}}$: A factor (real) used to scale the tx pulse magnitudes in the plots to compensate for tx-rx filter loss.

$\arg\min\limits_{k}\{\varphi[k]\}$: Returns the index $k$ for which $\varphi[k]$ is minimized.

$P_{\text{err}}$: Error power.

$\sigma^2$: Variance of additive white noise in the channel (need not be Gaussian).

WNG: White-Noise Gain, also known as the variance reduction factor.

CPP : Cross-Pulse Product. Compensates for the gain of the low-pass modem filters used on tx & rx.

$\Delta_{\rho}$: The distance between adjacent symbols or tokens $\phi[k^{\#}]$ in the complex plane.

$\Delta_{\sigma}$: The expected radial dispersion of $\varphi_{\text{tx}}[n^{\downarrow}]$ around each symbol $\phi[k^{\#}]$ due to random noise-induced errors.

$\Delta^{\#}$: Symbol resolvability, $\Delta^{\#} = \Delta_{\rho}/2\Delta_{\sigma}$. Is an indicator of expected error rates over a link.

$\omega_{\tilde{k}}$: Relative angular frequency (radians per sample) of the $\tilde{k}$th sub-carrier in a multiplexed channel.

$\psi_{\tilde{k}}[m^{\uparrow}]$ or $\psi_{\tilde{k}}[n^{\uparrow}]$: Sampled sub-carrier of the $\tilde{k}$th sub-channel.

$h_{\tilde{k}}[m^{\uparrow}]$: Impulse response of the $\tilde{k}$th sub-channel, $h_{\tilde{k}}[m^{\uparrow}] = h[m^{\uparrow}]\psi_{\tilde{k}}[m^{\uparrow}]$.

$H_{\tilde{k}}(\omega)$: Frequency response of $h_{\tilde{k}}[m^{\uparrow}]$.

$c_{\tilde{k}}$: Equalization factor (complex) to compensate for loss and lags in the $\tilde{k}$th sub-channel.

*Example 7*

FFT$\{\blacksquare\}$: Fast Fourier Transform operator.

IFFT$\{\blacksquare\}$: Inverse Fast Fourier Transform operator.

$\odot$: Convolution operator.

$*$: Multiplication operator.

$\acute{k}$: Block index, within a data stream.

$m$: Sample index, within a block.

$M$: Length (samples) of an FIR filter kernel $h[m]$.

$L$: Length (samples) of a data block.

$B$: Length (samples) of an extended block, e.g. a zero-padded data block, $B = L + M$.

$\acute{h}[m]$: Zero-padded (terminal) extension of FIR filter kernel $h[m]$.

$\acute{x}[m]$: Zero-padded (terminal) extension of data block $x[m]$.

$H[k]$: The $k$th bin of the DFT of $\acute{h}[m]$, for $m = 0 \ldots B - 1$ and $k = 0 \ldots B - 1$.

$X[k]$: The $k$th bin of the DFT of $\acute{x}[m]$, for $m = 0 \ldots B - 1$ and $k = 0 \ldots B - 1$.

$x_k[m]$: The $m$th sample of the $\acute{k}$th data block, for $m = 0 \ldots L - 1$.

$y_k[m]$: The $m$th sample of the $\acute{k}$th output block, with initiation transient, for $m = 0 \ldots L - 1$.

$\acute{x}_k[m]$: The $m$th sample of the $\acute{k}$th extended (zero-padded) data block, for $m = 0 \ldots B - 1$.

$\acute{y}_k[m]$ The $m$th sample of the $\acute{k}$th extended output block, with initiation and termination transients, for $m = 0 \ldots B - 1$.

## Example 1 – Modulation and demodulation

When a real sinusoid is multiplied by a real factor, only its magnitude is scaled. When a complex sinusoid is multiplied by a complex factor (e.g. $c$) its magnitude is scaled (i.e. by $|c|$) *and* its phase is shifted (i.e. $\angle c$). Thus, as illustrated below, complex arithmetic is very convenient for modelling and manipulating waves and pulsed waveforms.

Phase-shift keying is a digital modulation scheme that transfers information by partitioning a radio-frequency carrier-wave into equal time intervals. If there are $K^{\#}$ symbols in the alphabet (for $K^{\#} \geq 2$) then the phase of the carrier over each partition is shifted by one of $K^{\#}$ discrete equally-spaced values using $\phi[k^{\#}] = \phi_0 + \phi_\Delta k^{\#}$ where $\phi_\Delta = 2\pi/K^{\#}$ and $\phi_0 = \phi_\Delta/2$ (for $K^{\#} > 2$) or $\phi_0 = \phi_\Delta/4$ (for $K^{\#} = 2$). The decimal value of the $k^{\#}$th symbol is equal to $k^{\#}$, for $k^{\#} = 0 \dots K^{\#} - 1$. A Morse-code (i.e. binary) alphabet is reached in the limiting case with one bit per partition (thus $K^{\#} = 2$) and it may be appropriate in severely compromised channels. The magnitude of the carrier over each partition is also shaped (i.e. tapered) by a low-pass filter to optimize power concentration over time and frequency. Thus, the carrier wave is modulated (i.e. multiplied) by a train of complex pulses on transmission that shift and scale the carrier in phase and magnitude, respectively. The receiver then determines the phase of each pulse and *identifies* the corresponding symbol value, or *estimates* the most likely value, when noise, interference, multipath, and uncertainty (e.g. errors in the tx-rx synchronization of time, phase and frequency) are present. A simple radio model of a modulator/demodulator (i.e. a modem) is used in all examples considered here and Example 1 is used to outline its operation.

Let the message contain $N^\downarrow$ symbols for a burst of $N^\downarrow$ pulses and let the decimal value of each symbol be $\chi_{\text{tx}}[n^\downarrow]$, where $n^\downarrow$ is the pulse index, for $n^\downarrow = 0 \dots N^\downarrow - 1$. The transmitted waveform therefore begins as a sequence of complex values $\varphi_{\text{tx}}[n^\downarrow]$ with elements $\rho e^{i\phi_{\text{tx}}[n^\downarrow]}$ where $\rho$ is the magnitude of the pulse train and $\phi_{\text{tx}}[n^\downarrow]$ is the phase shift to be applied to the $n^\downarrow$th pulse on transmission, as determined using $\phi_{\text{tx}}[n^\downarrow] = \phi[k_n^{\#}]$ with $k_n^{\#} = \chi_{\text{tx}}[n^\downarrow]$.

The sequence $\varphi_{\text{tx}}[n^\downarrow]$ is then up sampled by inserting $2K_{\text{tx}}^\uparrow$ zeros after each element, with an up-arrow now used to index waveform samples at the higher rate.

The up-sampled sequence is then low-pass filtered by convolving it with $h_{\text{tx}}[n^\uparrow]$ to form a sequence of shaped complex pulses $\varphi_{\text{tx}}[n^\uparrow]$ at the sampling rate of the transmitter, for $n^\uparrow = 0 \dots N^\uparrow - 1$. If $h_{\text{tx}}[n^\uparrow]$, with a frequency response of $H_{\text{tx}}(\omega)$, is finite in duration with length $M_{\text{tx}}^\uparrow = 2K_{\text{tx}}^\uparrow + 1$ then the pulses abut in time. If $h_{\text{tx}}[n^\uparrow]$ is infinite then the pulses overlap somewhat.

The pulse train $\varphi_{\text{tx}}[n^\uparrow]$ modulates a (complex) oscillator $\psi_{\text{tx}}[n^\uparrow]$ with a frequency of $\omega_{\text{tx}}$, i.e. the sampled carrier, to yield the discrete-time sequence $\tilde{\psi}_{\text{tx}}[n^\uparrow] = Re\{\varphi_{\text{tx}}[n^\uparrow]\psi_{\text{tx}}[n^\uparrow]\}$, that is sent to a digital-to-analogue converter, an amplifier, then the transmitter's antenna.

The continuous-time waveform passes through the propagating medium, the receiver's antenna, analogue band-pass filter, amplifier then analogue-to-digital converter (with the same sampling rate as the transmitter) yielding $\tilde{\psi}_{\text{rx}}[n^\uparrow]$. The continuous-time transfer-function of the channel that links the output of the digital-to-analogue converter $\tilde{\psi}_{\text{tx}}(t)$ to the input of the analogue-to-digital converter $\tilde{\psi}_{\text{rx}}(t)$, is assumed in this example to be ideal and free of noise. It has the frequency response $H_{\text{chn}}(\Omega) = 1$. It is also assumed here that the sampling rates at tx and rx are identical and synchronized.

The sampled waveform $\tilde{\psi}_{rx}[n^\uparrow]$ is then multiplied by a (complex) oscillator $\psi_{rx}[n^\uparrow]$, with frequency and phase that are perfectly matched to $\psi_{tx}[n^\uparrow]$, i.e. $\tilde{\varphi}_{rx}[n^\uparrow] = \psi^*_{rx}[n^\uparrow]\tilde{\psi}_{rx}[n^\uparrow]$ with $\psi_{rx}[n^\uparrow] = \psi_{tx}[n^\uparrow]$. If the channel supports the propagation of a complex waveform, then $\tilde{\varphi}_{rx}[n^\uparrow] = \varphi_{rx}[n^\uparrow]$. However, only the real part is transmitted here, thus $\tilde{\varphi}_{rx}[n^\uparrow] = \varphi^-_{rx}[n^\uparrow] + \varphi^+_{rx}[n^\uparrow]$, where $\varphi^+_{rx}[n^\uparrow]$ is a *spurious sum* component centred on $\omega = \omega_{tx} + \omega_{rx} = 2\omega_{tx}$, and $\varphi^-_{rx}[n^\uparrow]$ is the *desired difference* (i.e. *baseband*) component with a complex spectrum centred on $\omega = \omega_{tx} - \omega_{rx} = 0$, with $\varphi^-_{rx}[n^\uparrow] = \varphi_{tx}[n^\uparrow]$.

Therefore $\tilde{\varphi}_{rx}[n^\uparrow]$ is passed through a low-pass filter, with impulse response $h_\downarrow[m]$ and frequency response $H_\downarrow(\omega)$, that is designed to have $\Phi^-_{rx}(\omega)$ in its pass band and $\Phi^+_{rx}(\omega)$ in its stop band, where $\Phi^-_{rx}(\omega)$ and $\Phi^+_{rx}(\omega)$ are the power spectral densities of $\varphi^-_{rx}[n^\uparrow]$ and $\varphi^+_{rx}[n^\uparrow]$, respectively. For an ideal filter and an ideal (noise-free) channel, this convolution removes the $\varphi^+_{rx}[n^\uparrow]$ component and passes only the $\varphi^-_{rx}[n^\uparrow]$ component thus $\varphi_{rx}[n^\uparrow] = \varphi_{tx}[n^\uparrow]$.

The baseband waveform is now low-pass filtered by convolving it with $h_{rx}[n^\uparrow]$, which is matched to the pulse envelope, i.e. $h_{rx}[n^\uparrow] = h_{tx}[n^\uparrow]$, then down sampled to yield $\varphi_{rx}[n^\uparrow]$ by keeping only the complex sample at the centre of each pulse (i.e. every $M^\uparrow_{rx}$th sample, where $M^\uparrow_{rx} = M^\uparrow_{tx}$) assuming perfect tx-rx time synchronization. The modulation/demodulation process is now complete and it is illustrated in Figure 28, Figure 29 & Figure 33.

The complex pulse train $\varphi_{tx}[n^\uparrow]$ is plotted in the first (uppermost) subplot of Figure 28 (real part in blue and imaginary part in red). The real part of $\varphi_{tx}[n^\uparrow]\psi_{tx}[n^\uparrow]$ i.e. the transmitted waveform $\tilde{\psi}_{tx}[n^\uparrow]$ is plotted in the second. The mixed waveform $\tilde{\varphi}_{rx}[n^\uparrow]$, with sum and difference components $\varphi^+_{rx}[n^\uparrow]$ and $\varphi^-_{rx}[n^\uparrow]$ is plotted in the third. Then the low-pass filtered waveform $\varphi_{rx}[n^\uparrow]$ in the fourth (lowermost) subplot.

The output of the pulse-matched filter, i.e. the convolution of $\varphi_{rx}[n^\uparrow]$ with $h_{rx}[n^\uparrow]$ is shown in Figure 29 (real part in blue and imaginary part in red) with the decimal symbol values of each pulse marked along the time axis. The waveform is shifted forward in time in these plots (i.e. to the left) to compensate for the group delay of all modem filters, i.e. $h_{tx}[n^\uparrow]$, $h_\downarrow[m]$ and $h_{rx}[n^\uparrow]$, so that the pulses are aligned with the complex impulses $c_{tx}\varphi_{tx}[n^\downarrow]$ (plotted using stemmed triangle tokens, blue for the real part and red for the imaginary part). The $c_{tx}$ factor is a constant that scales the impulses so that their magnitude is the same as the maxima of the pulses, to aid visualization. It is equal to the cross-pulse product and an analytical expression for its evaluation is described later in Example 2, where an expression for the symbol resolvability is derived.

A Slepian pulse-shaping filter (with $f_c = 4/M$) is used here and in Examples 2-5. It yields very low side-lobes and well-tapered pulses with negligible magnitude at the ends of each time partition (see Sood & Xiao, 2006). The gaps between pulses may help to protect against delay spread due to multipath propagation in a real channel. Pulse-shaping filters are mainly used here to concentrate signal power in the most linear region of a channel. For the ideal channel considered here, the only non-ideal component of the system, that stands between $h_{tx}[m]$ and $h_{rx}[m]$, is the low-pass IIR filter used for down conversion at the receiver (a consequence of real wave propagation) thus the primarily purpose of the pulse shaping filter is to improve the transient response of $h_\downarrow[m]$.

For a rectangular (un-shaped) pulse, long transients result in large bias errors in the matched filter output (see Figure 30 & Figure 31). The Slepian pulse-shaper (with low side-lobes) suppresses the underdamped (ringing) step-response by removing the high-frequency content of the rectangular pulse, which would otherwise 'excite' the filter, i.e. be amplified by the non-negligible stop-band gain of $H_\downarrow(\omega)$. The frequency response of the fourth-order Butterworth filter used on (digital) down-conversion is plotted in Figure 32. Its cut-off frequency $f_c$ is tuned to be $1.5 \times$ the (one-sided) bandwidth of the channel (i.e. $f_{chn}$) which is equal to the cut-off frequency of the pulse-shaping filter (i.e. $f_c$) in Examples 1-4. In all examples considered here, the wavelength of the carrier is 4 samples per cycle, thus $f_{tx} = 1/4$ cycles per sample, where $f_{tx} = \omega_{tx}/2\pi$. For a perfectly tuned oscillator at the receiver, this conveniently places the sum component on down conversion at the Nyquist notch of $H_\downarrow(\omega)$, i.e. at $f = 1/2$.

The constellation diagram for this ideal system, with Slepian and rectangular pulses, is plotted in Figure 33 & Figure 34. The symbol coordinates $\varphi_{tx}[n^\downarrow]$ and the down-sampled pulse coordinates $\varphi_{rx}[n^\downarrow]$ in this complex plane are shown in red and cyan, respectively, with $\varphi_{rx}[n^\uparrow]$ in magenta. The corresponding (decimal) symbol values are also shown.

In this noise-free example for the Slepian pulse there is no dispersion in $\varphi_{rx}[n^\downarrow]$; however, in the general case (i.e. noisy and biased) the value of the symbol that is closest to the pulse in angle is assigned to the pulse, using

$$\chi_{rx}[n^\downarrow] = \arg\min_{k^\#}\left\{\frac{\angle\varphi_{tx}^*[k^\#]\varphi_{rx}[n^\downarrow]}{|\varphi_{rx}[n^\downarrow]|}\right\} \text{ where } \varphi^*[k^\#] = e^{-i\phi[k^\#]} \tag{47}$$

and $\arg\min_{k}\{\varphi[k]\}$ returns the index $k$ that minimizes $\varphi[k]$.

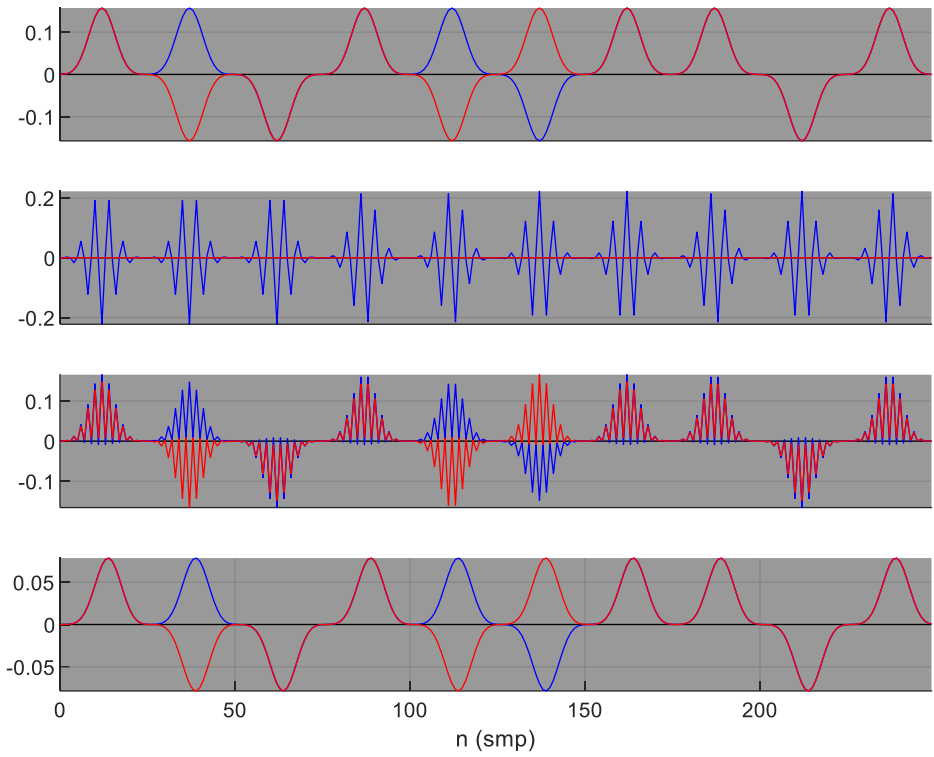

*Figure 28. Modulation and de-modulation of a carrier wave using Slepian pulses*

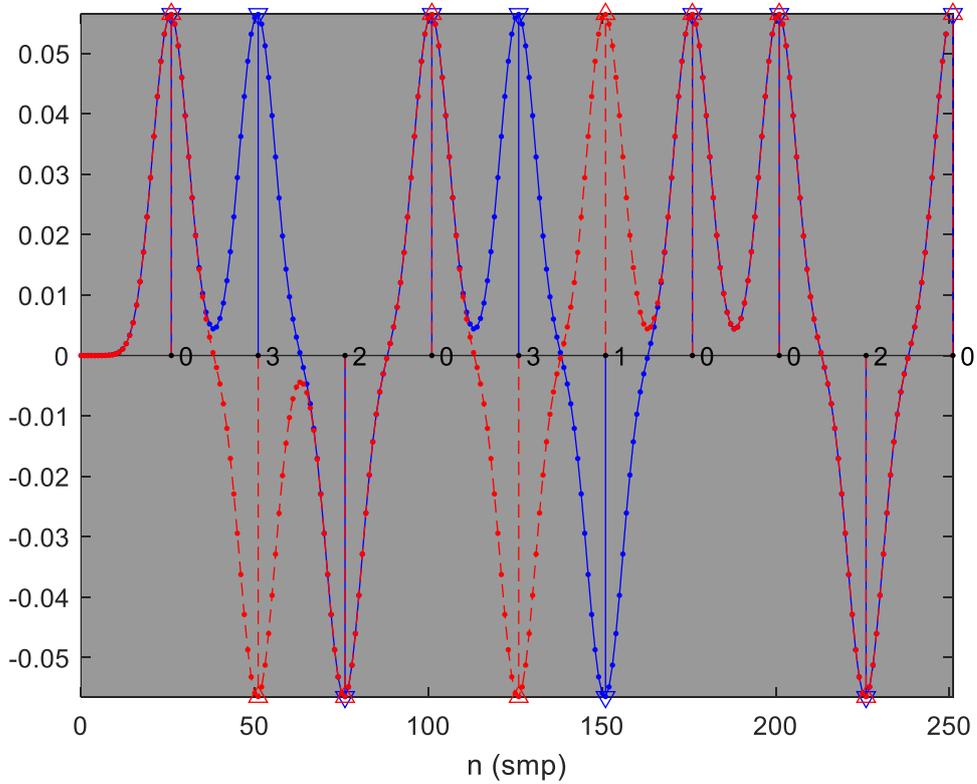

*Figure 29. Complex output of the matched filter at the higher rate, for Slepian pulses, superimposed on the complex symbols at the lower rate*

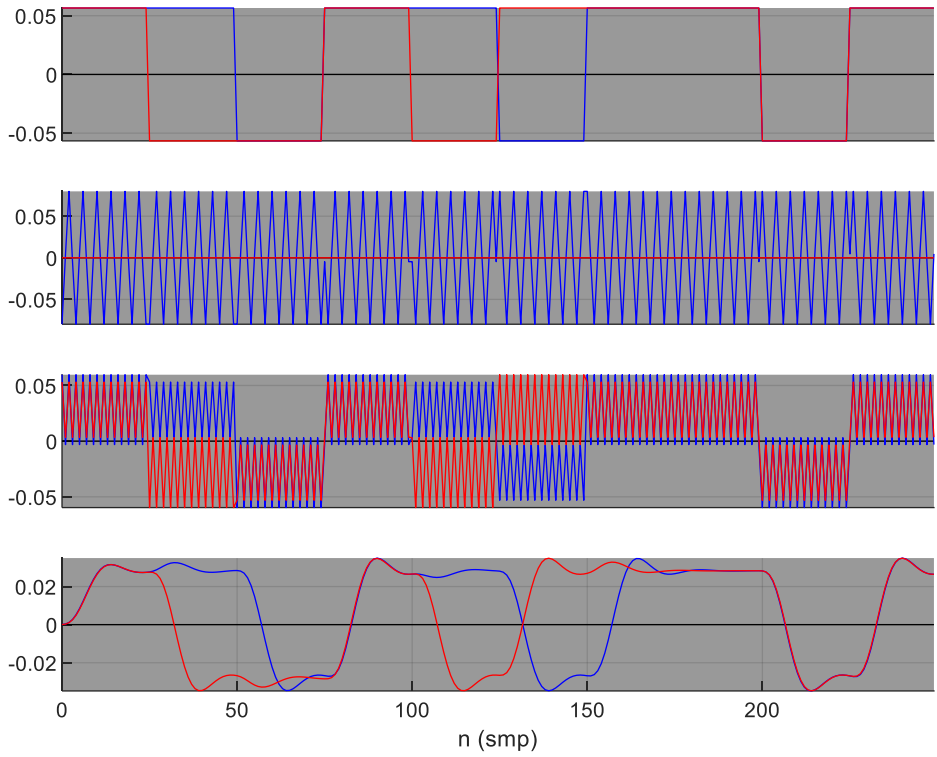

*Figure 30. Modulation and de-modulation of a carrier wave using rectangular pulses*

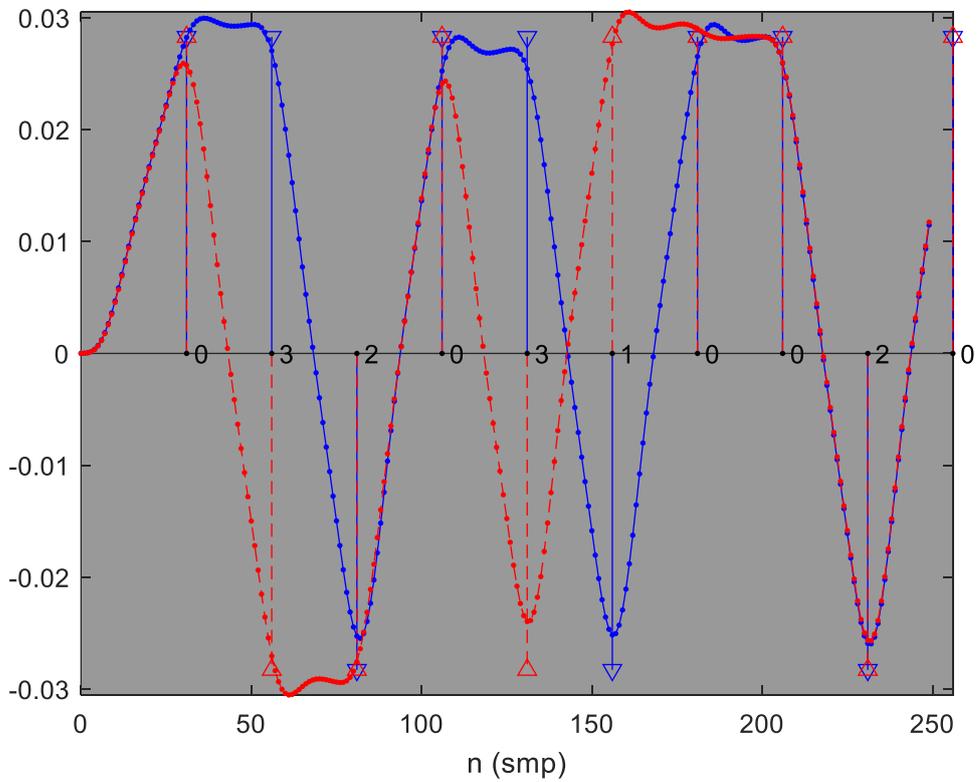

*Figure 31. Complex output of the matched filter at the higher rate, using rectangular pulses, superimposed on the complex symbols at the lower rate*

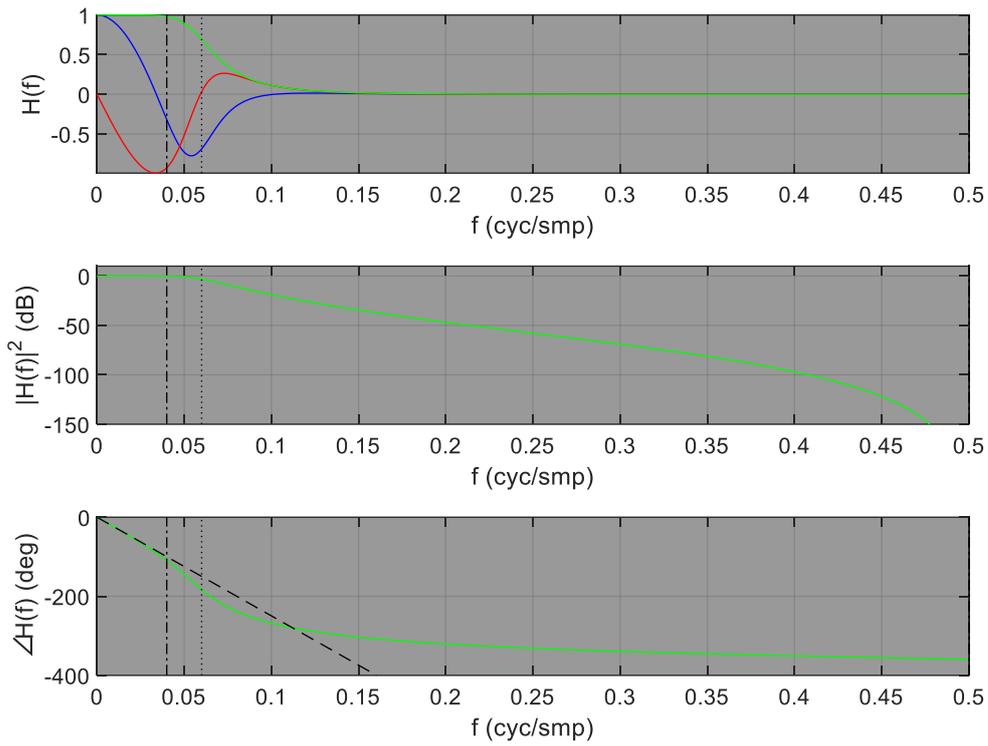

*Figure 32. Frequency response of the digital Butterworth filter used to extract the baseband signal on rx down-conversion*

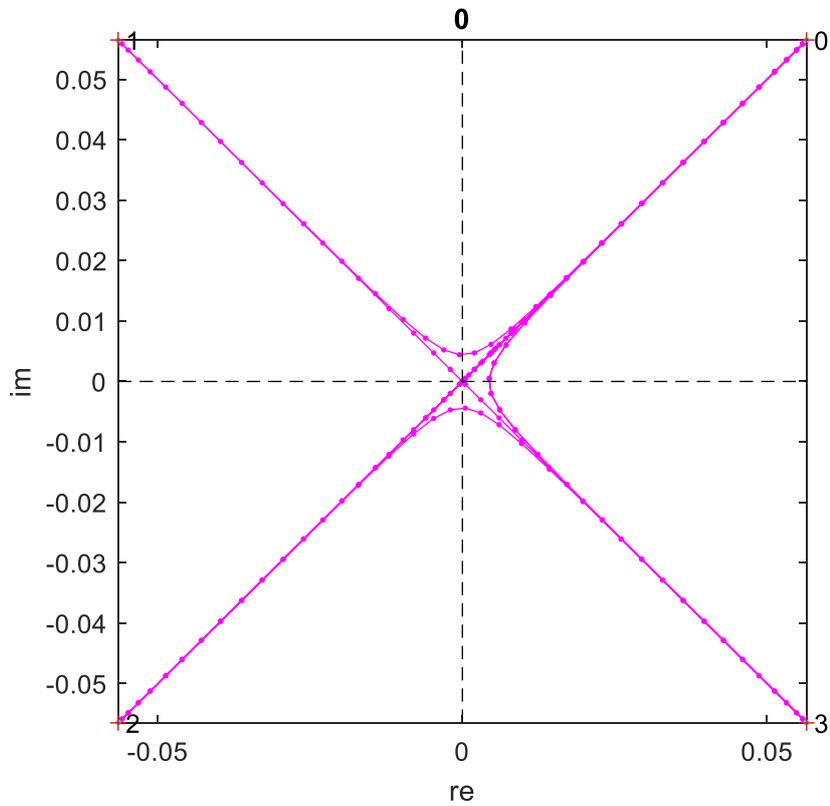

*Figure 33. Constellation diagram for Example 1, with Slepian pulses*

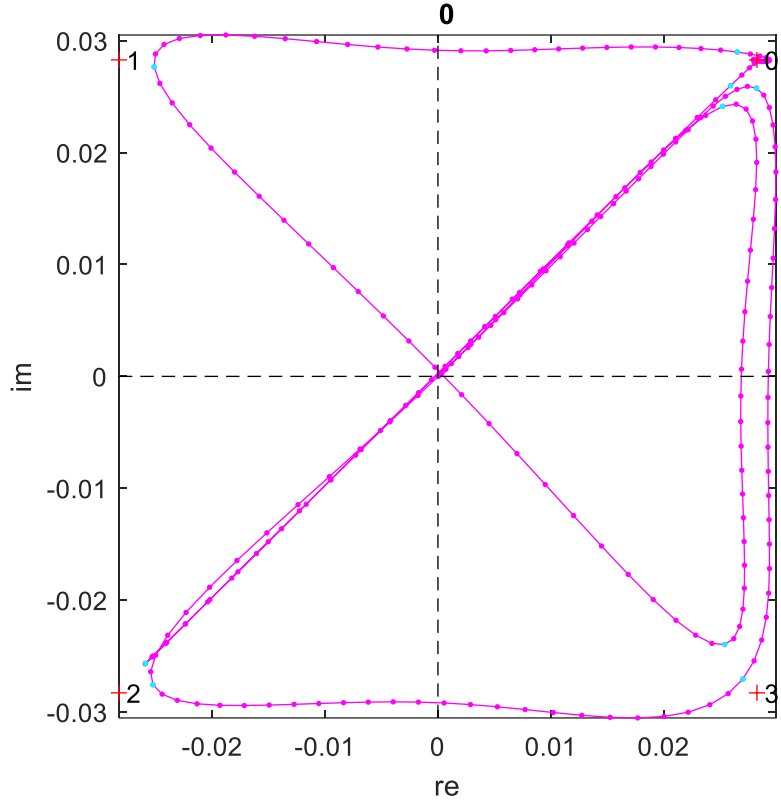

*Figure 34. Constellation diagram for Example 1, with rectangular pulses*

## Example 2 − Short pulses

The example above that is used to describe the ideal system (i.e. modem and channel) is repeated here; however, white Gaussian noise is now added to $\tilde{\psi}_{\text{tx}}[n^\downarrow]$ for a signal-to-noise ratio of zero dB. It is also used in a Monte-Carlo simulation with the number of pulses greatly increased (from 10 to 10,000). The constellation diagram (in a complex plane) for this noisy system is shown in Figure 35 (symbols in red and down-sampled pulses $\varphi_{\text{tx}}[n^\downarrow]$ in cyan). The impact of the channel noise is evident: The complex $\varphi_{\text{rx}}[n^\downarrow]$ values are now scattered around the $\varphi_{\text{tx}}[n^\downarrow]$ points. The overlap between the distributions is not insignificant thus the assignment of symbols, using (47), for some pulses may be ambiguous, resulting in errors. The mean and standard deviation of the distributions for each symbol are plotted in blue. The variance, or error power $P_{\text{err}}$ may also be predicted analytically from the response of the low-pass pulse-shaping filter applied at the receiver using

$$P_{\text{err}} = \text{WNG}\sigma^2 \tag{48}$$

where WNG is the White-Noise Gain of the filter which is defined in the frequency domain as follows:

$$\text{WNG} = \tfrac{1}{2\pi}\int_{-\pi}^{\pi} H_{\text{rx}}^*(\omega)H_{\text{rx}}(\omega)d\omega = \tfrac{1}{2\pi}\int_{-\pi}^{\pi}|H_{\text{rx}}(\omega)|^2 d\omega \ . \tag{49}$$

For FIR filters the integral is readily evaluated using (11), i.e. $\text{WNG} = P_\pi/2\pi$ or

$$\text{WNG} = \sum_{m=0}^{M-1}|h_{\text{rx}}[m]|^2 \ . \tag{50}$$

Fortunately, Parseval's theorem shows that this expression holds for both FIR and stable IIR filters, with $M = \infty$ in the latter case; however, the summation may be terminated after the impulse response has effectively converged on zero. Note that the white-noise gain is approximately proportional to the bandwidth of the filter, if the side-lobe height is negligible.

In Figure 35 the predicted one-sigma limits (i.e. one standard deviation) of the distributions for each symbol, computed using $P_{\text{err}}$, are also plotted in red. The predicted (red solid line) and observed errors (blue dashed line) are clearly in agreement.

The cross-pulse product (CPP) compensates for the gain of the low-pass modem filters used on tx & rx. It is computed using

$$\text{CPP} = \tfrac{1}{4\pi}\int_{-\pi}^{\pi} H_{\text{rx}}^*(\omega)H_{\text{tx}}(\omega)d\omega \text{ or} \tag{51}$$

(for low-pass FIR filters with $M$ odd and a linear phase)

$$\text{CPP} = \tfrac{1}{2}\sum_{m=-K}^{K} h_{\text{rx}}[m]h_{\text{tx}}[m] \text{ where } K = \min(K_{\text{rx}}, K_{\text{tx}}). \tag{52}$$

The CPP of pulse-shaping filters used on tx and rx, is the maximum output of $h_{\text{tx}}[m]$ convolved with $h_{\text{rx}}[m]$, which occurs when both impulse responses are centred in time at $m = 0$. Clearly CPP = WNG/2 when $h_{\text{rx}}[m]$ is matched to $h_{\text{tx}}[m]$. If the pass bands of $H_{\text{tx}}(\omega)$ and $H_{\text{rx}}(\omega)$ are well within the passband of $H_\downarrow(\omega)$, and the passband of $H_\downarrow(\omega)$ is reasonably flat, and the stop-band gains are all negligible, then the response of $h_\downarrow[m]$ does not need to be considered in the CPP. The factor of one half is due to the elimination of the imaginary part on tx.

The CPP and WNG are used here to evaluate the *symbol resolvability*, i.e.

$$\Delta^\# = \Delta_\rho/2\Delta_\sigma \tag{53}$$

where

$\Delta_\rho$ is the distance between adjacent symbols $\phi[k^\#]$ in the complex plane, i.e.

$$\Delta_\rho = \text{CPP}\left|\rho\left(e^{i\phi_\Delta} - 1\right)\right| \text{ and}$$

$\Delta_\sigma$ is the expected radial dispersion of $\varphi_{\text{tx}}[n^\downarrow]$ around each symbol $\phi[k^\#]$ due to noise, i.e.

$$\Delta_\sigma = \sqrt{P_{\text{err}}} \; .$$

In the constellation diagrams shown below (e.g. Figure 35) the radius of the dotted red circle is equal to CPP$\rho$; the radius of the dashed red circle is equal to $\Delta_\rho/2$; the radius of the solid red circle is equal to $\Delta_\sigma$; thus the symbol resolvability ($\Delta^\#$) is the radius of the dashed red circle over the radius of the solid red circle. In all examples, $\rho = 2.0$ is used. The likelihood of symbol-to-pulse assignment errors decreases as symbol resolvability increases. In statistical analysis, events beyond the 4-sigma tails of a normal distribution are considered highly unlikely, and similar values are recommended for low error rates here. The expressions above are for general white-noise processes (i.e. with a flat spectrum) thus they are not restricted to Gaussian distributions. An additional simulation with additive noise generated using a uniform distribution (with mean of zero and a variance of $\sigma^2$) was performed to check this assertion (see Figure 36).

Note that using different pulse-shaping filters on tx and rx does not improve the resolvability. For instance, it may be tempting to use a filter with a narrower bandwidth (thus a longer impulse duration) for a lower white-noise gain; however, the mis-matched tx & rx filters now also have an even lower cross-pulse product, thus symbol resolvability decreases. Therefore, matched filters are used in all examples considered here.

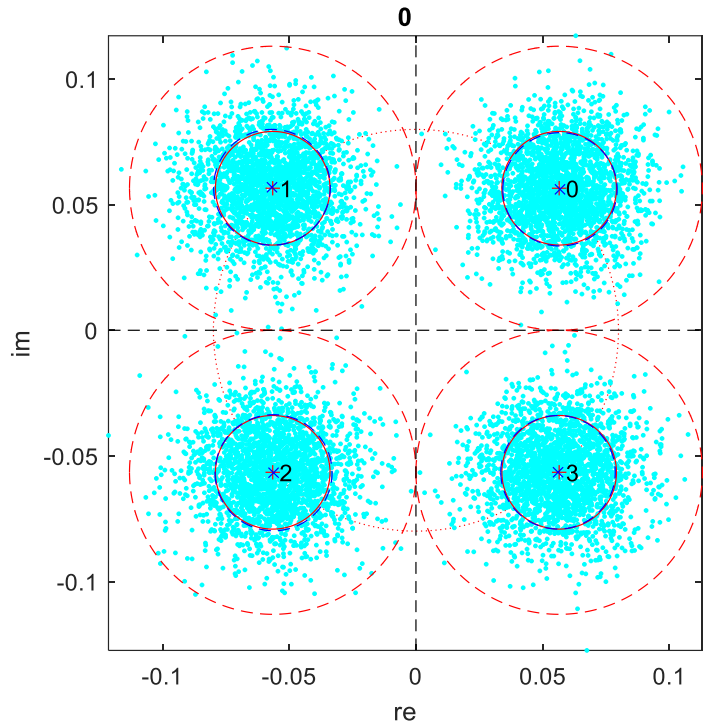

*Figure 35. Constellation diagram for the short pulse in Example 2, with Gaussian noise*

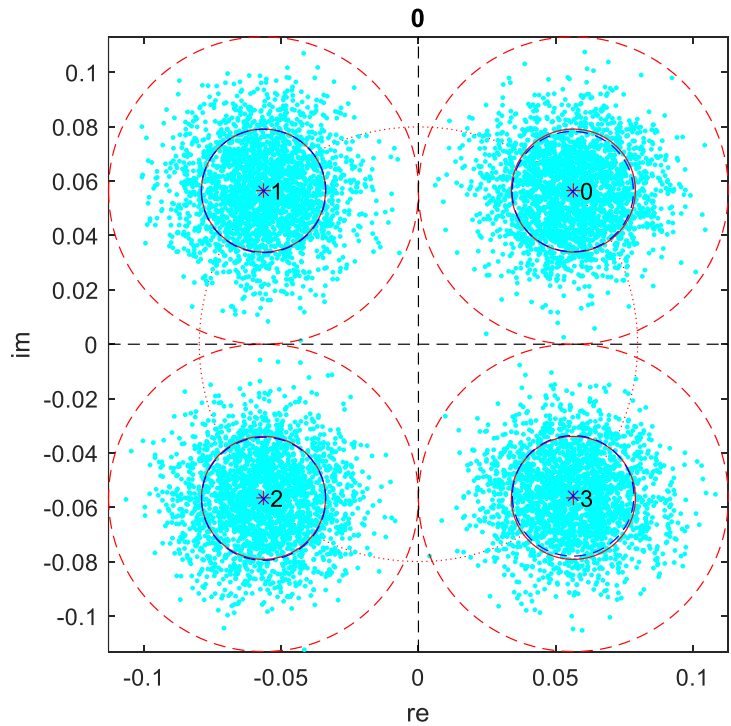

*Figure 36. Constellation diagram for the short pulse in Example 2, with uniform noise*

## Example 3 – Longer pulses

For the simulation in Example 2, the error rate is not negligible as 8 pulses (out of 10,000) are assigned the wrong symbol (for Gaussian noise). The noise immunity of the link may be improved by using longer pulses, thus a narrower bandwidth, for a lower white-noise gain. However, this also decreases the data rate, which is computed using

$$\text{bit rate} = \frac{\log_2 K^{\#}}{M^{\uparrow}} \text{ (bits per sample).} \tag{54}$$

Therefore in this example, the pulse duration is increased (from $K^{\uparrow} = 12$ to $K^{\uparrow} = 36$, with $M^{\uparrow} = 2K^{\uparrow} + 1$). As shown in **Error! Reference source not found.**, this improves the symbol resolvability (from $\Delta^{\#} = 2.500$ to $\Delta^{\#} = 4.2720$) so there are no incorrect assignments of symbols to pulses in the simulation; however, the data rate is lowered (from $0.0800$ to $0.0274$ bits per sample).

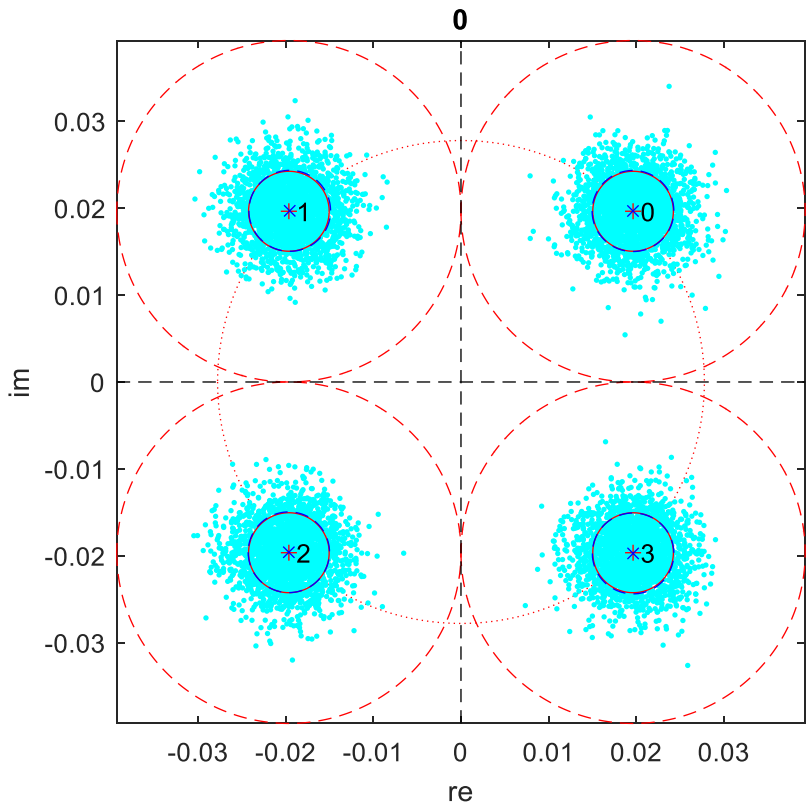

*Figure 37. Constellation diagram for the longer pulse in Example 3.*

## Example 4 – Long pulses with more symbols

The lower data rates of longer pulses may be partially offset using an alphabet with more symbols. In this example, the pulse duration is further increased (from $K^{\uparrow} = 36$ to $K^{\uparrow} = 124$) and the number of symbols is also increased (from $K^{\#} = 4$ to $K^{\#} = 8$ for 2 and 3 bits per channel, respectively). As indicated in the constellation diagram (see Figure 38) this approximately maintains the symbol resolvability (from $\Delta^{\#} = 4.2720$ to $\Delta^{\#} = 4.2700$) so there are no incorrect assignments of symbols to pulses in the simulation; however, the data rate is approximately halved (from $0.0274$ to $0.0120$ bits per sample).

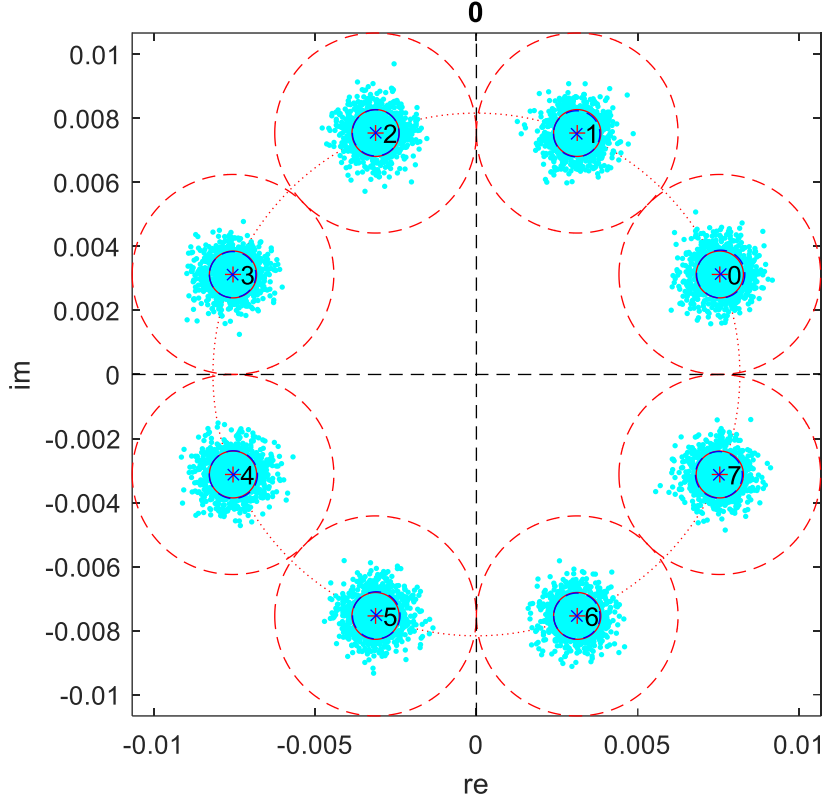

*Figure 38. Constellation diagram for the long pulses in Example 4*

## Example 5 – Long pulses with less symbols and more sub-channels

Pulses with a longer duration have a narrower bandwidth, thus a lower bit rate; however, the bandwidth of a channel with long pulses may be increased if multiple orthogonal pulses are transmitted simultaneously in adjacent sub-channels with closely spaced *digital* sub-carriers. This principle is the foundation of orthogonal frequency-division multiplexing, which is commonly used in wireless digital-communication systems.

The modulation scheme in this and the remaining examples uses an odd number of $\widetilde{M} = 2\widetilde{K} + 1$ sub-channels with sub-carrier frequencies $\omega_{\tilde{k}}$ for $\tilde{k} = -\widetilde{R} \dots \widetilde{R}$. Each sub-channel has $K^{\#}$ *tokens* in the complex plane to shift the phase of its sub-carrier to one of $K^{\#}$ angles. The binary values of the $\widetilde{R}^{\#}$ symbols of the alphabet are formed by appending the binary values of the tokens in each of the $\widetilde{M}$ sub-channels (thus $\widetilde{R}^{\#} = \widetilde{M} K^{\#}$).

If the complex pulses in a channel are orthonormal in time over the symbol interval (of $M^{\uparrow}$ samples) i.e.

$$\sum_{m^{\uparrow}=0}^{M^{\uparrow}-1} h_{\tilde{k}_1}^*[m^{\uparrow}] h_{\tilde{k}_2}[m^{\uparrow}] = \begin{cases} 1, & \tilde{k}_1 = \tilde{k}_2 \\ 0, & \tilde{k}_1 \neq \tilde{k}_2 \end{cases} \tag{55}$$

then the pulses do not 'overlap' in time, even though they 'occupy' the same interval in time, thus they do not 'interfere' with each other. Orthonormality enables the concurrent transmission of pulses in a shared wide-bandwidth channel on a single radio-frequency carrier. Thus, we now seek pulse shapes that have the desired time-frequency profiles *and* satisfy (55) above.

Orthogonalization is very sensitive to rounding errors, thus ill-conditioned problems are readily posed, particularly for large $M^\uparrow$ and $\widetilde{M}$. Fortunately, Parseval's theorem comes to the rescue once again by allowing (55) to be stated in the frequency domain, i.e.

$$\frac{1}{2\pi}\int_{-\pi}^{\pi}H^*_{\tilde{k}_2}(\omega)H_{\tilde{k}_1}(\omega)d\omega = \begin{cases} 1, & \tilde{k}_1 = \tilde{k}_2 \\ 0, & \tilde{k}_1 \neq \tilde{k}_2 \end{cases}. \tag{56}$$

At first glance, this appears to only have complicated matters; however, the creation of the orthonormal family of pulses (in sub-channels) that will share the same carrier (of the channel) is now straightforward. The Slepian pulse $h[m^\uparrow]$ & $H(\omega)$ with a cut-off frequency of $\omega_c$ is used to minimize the overlap between adjacent sub-channels (see Figure 9 to Figure 12). It modulates subcarriers $\psi_{\tilde{k}}[m^\uparrow] = e^{i\omega_{\tilde{k}}m^\uparrow}$, i.e. $h_{\tilde{k}}[m^\uparrow] = h[m^\uparrow]\psi_{\tilde{k}}[m^\uparrow]$ which shifts the centre-frequency of the Slepian $H(\omega)$ from $\omega = 0$ to $\omega = \omega_{\tilde{k}}$ to form sub-channels $H_{\tilde{k}}(\omega)$ that are (maximally) concentrated over $\omega = \omega_{\tilde{k}} \pm \omega_c$ (see Figure 5 to Figure 8). This ensures that requirements (56) thus (55) are approximately satisfied. The cut-off frequency (i.e. the one-sided bandwidth) of the multiplexed channel is $\omega_{chn} = \widetilde{M}\omega_{pls}$ where $\omega_{pls}$ is the cut-off frequency of the pulse-shaping filter (i.e. $\omega_c$) with $f_{chn} = \omega_{chn}/2\pi$.

Down conversion at rx now involves a two-stage mixing process. As before, the sampled waveform $\tilde{\psi}_{rx}[n^\uparrow]$ is multiplied by a (complex) oscillator $\psi_{rx}[n^\uparrow]$, with frequency and phase that are matched to the radio-frequency carrier $\psi_{tx}[n^\uparrow]$, i.e. $\tilde{\varphi}_{rx}[n^\uparrow] = \psi^*_{rx}[n^\uparrow]\tilde{\psi}_{rx}[n^\uparrow]$ with $\psi_{rx}[n^\uparrow] = \psi_{tx}[n^\uparrow]$. It is then low-pass filtered using $h_\downarrow[m]$ to remove the sum component $\varphi^+_{rx}[n^\uparrow]$. The $\tilde{k}$th sub-channel is then extracted and equalized using $c_{\tilde{k}}\psi^*_{\tilde{k}}[n^\uparrow]\varphi^-_{rx}[n^\uparrow]$. For this ideal channel, the equalization factor $c_{\tilde{k}}$ only compensates for the pass-band magnitude error and phase delay of $h_\downarrow[m]$, i.e. $c_{\tilde{k}} = 1/H_\downarrow(\omega_{\tilde{k}})$. The $\tilde{k}$th sub-channel is then match filtered using an impulse response that is matched to the transmitted pulse shape, i.e. $h_{rx}[m^\uparrow] = h_{tx}[m^\uparrow]$, followed by a down-sampling operation.

The constellation diagrams of a multiplexed channel are plotted in Figure 43. Using a long pulse (i.e. $K^\uparrow = 124$) with seven sub-channels (i.e. $\widetilde{K} = 3$) and only one bit per sub-channel (i.e. $K^\# = 2$ for 7 bits per channel) yields a constellation with a resolvability ($\Delta^\# = 4.2173$) that is approximately the same as Example 4 ($\Delta^\# = 4.2700$). However, the bit rate has more than doubled (from 0.0120 to 0.0281 bits per sample) and it is now approximately the same as the bit rate of Example 3 (i.e. 0.0274 bits per sample). Examples 3, 4 & 5 therefore suggest that more symbols (Example 4) are less effective than more sub-channels (Example 5) for restoring the data rates of longer pulses, but this is to be expected because more bandwidth is also utilized ($f_{chn} = 0.0161$ and 0.1124 cycles per sample in Examples 4 & 5, respectively, compared with 0.0548 in example 3). Furthermore, the shorter pulses in Example 3 and the longer multiplexed pulses in Example 5 have nearly equivalent symbol resolutions and error rates. However, Example 3 utilizes the spectrum efficiently because it uses half the bandwidth of Example 5. Orthogonal sub-channels consume more bandwidth because there are large gaps between the sub bands (see Figure 42). These gaps are required for sub-band orthogonality, using realizable pulse-shaping filters, they may also help to protect against Doppler spread in a real channel (guard bands).

Using least-squared-error low-pass filters for pulse shaping with $f_c = 4/M$ and $f_{lo} = f_c/100$ & $f_{hi} = f_c$ and $\tilde{w} = 1.0$ & $\bar{w} = 1000.0$ for a wide transition-band yields a Slepian-like pulse (see Figure 44) with same with the same symbol resolution (i.e. $\Delta^\# = 4.2173$). Using $f_{lo} = f_c/2$ & $f_{hi} = f_c$ and $\tilde{w} = 1.0$ & $\bar{w} = 1000.0$, yields a more oscillatory pulse (see Figure 45) with a somewhat flatter pass-band and the same symbol resolution. Using $\tilde{w} = 100.0$ & $\bar{w} = 1.0$ for a flatter pass-band and a contracted guard-band (see Figure 47) thus an oscillatory impulse-response (see Figure 46) yields a slight improvement in the symbol resolution ($\Delta^{\#}$ = 4.2175).

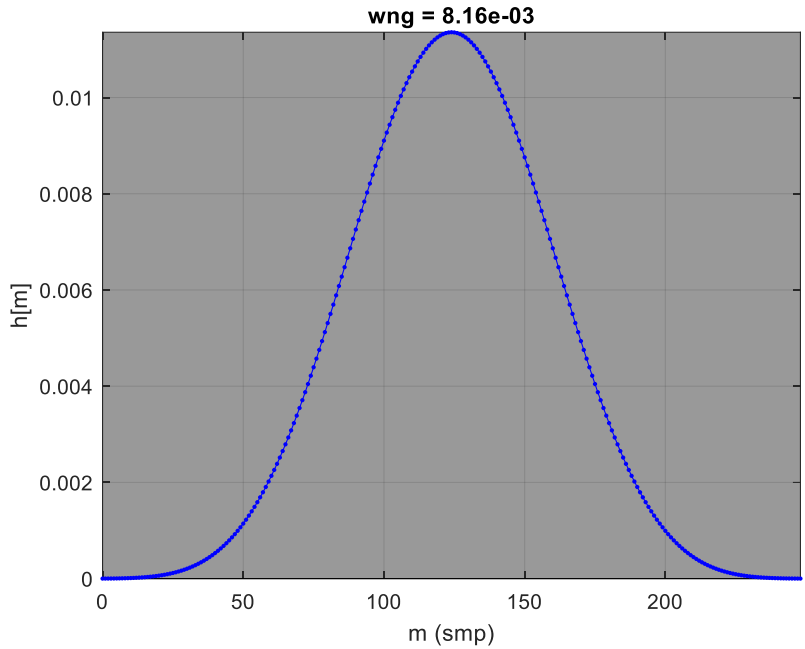

*Figure 39. Impulse response of the Slepian pulse-shaping filter used in Examples 4 & 5*

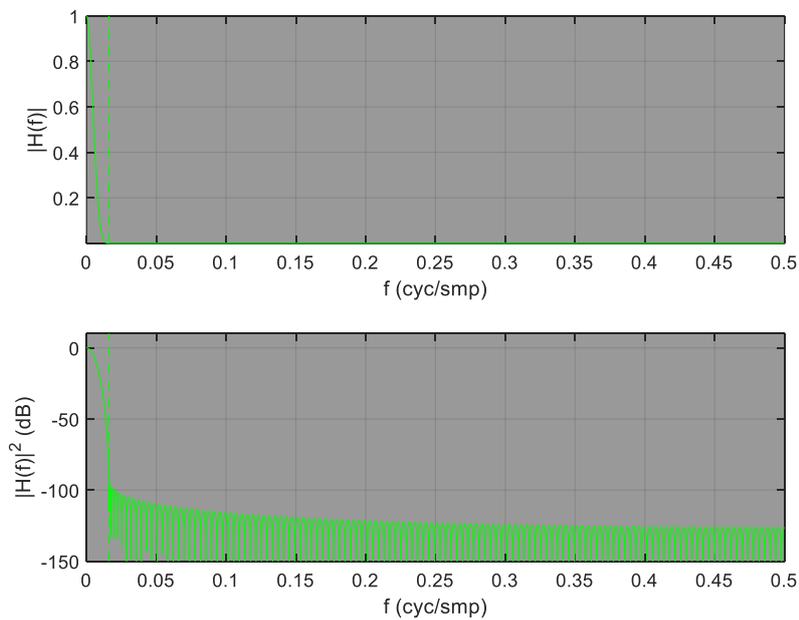

*Figure 40. Frequency response of the Slepian pulse-shaping filter used in Examples 4 & 5*

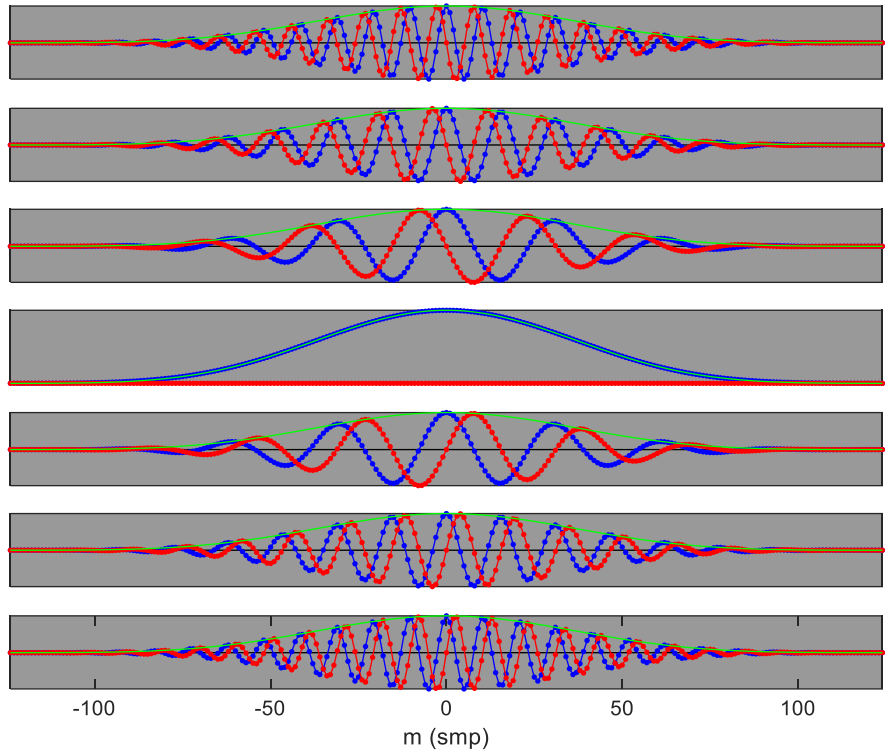

*Figure 41. Impulse response of Slepian pulse-shaped sub-carriers in Example 5, for $\tilde{k} = -\tilde{K} \dots \tilde{K}$ (top to bottom)*

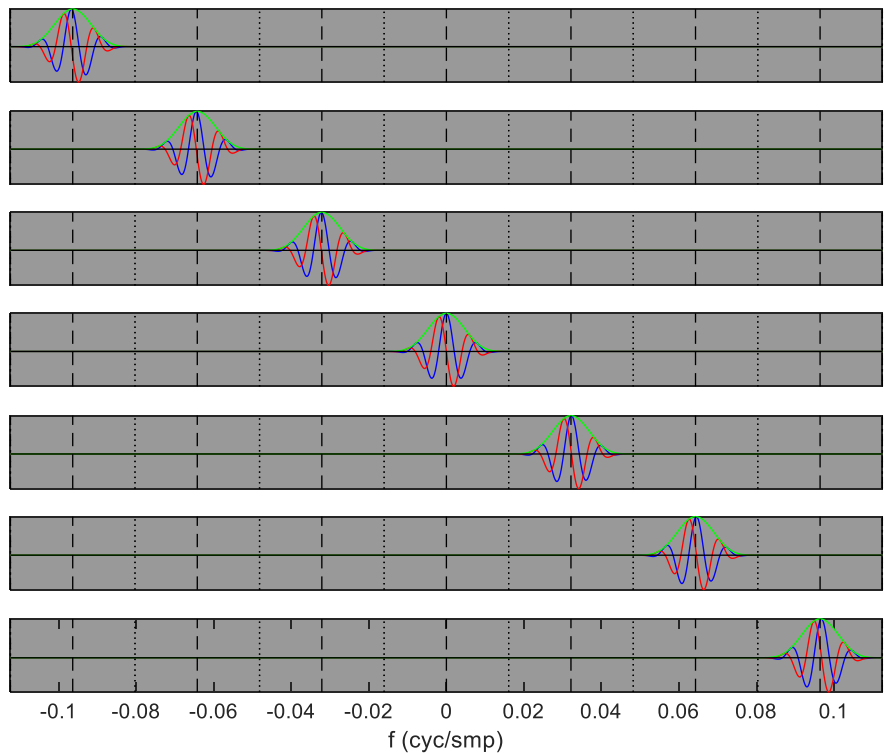

*Figure 42. Frequency response of Slepian pulse-shaped sub-carriers in Example 5, for sub-channels $\tilde{k} = -\tilde{K} \dots \tilde{K}$ (top to bottom)*

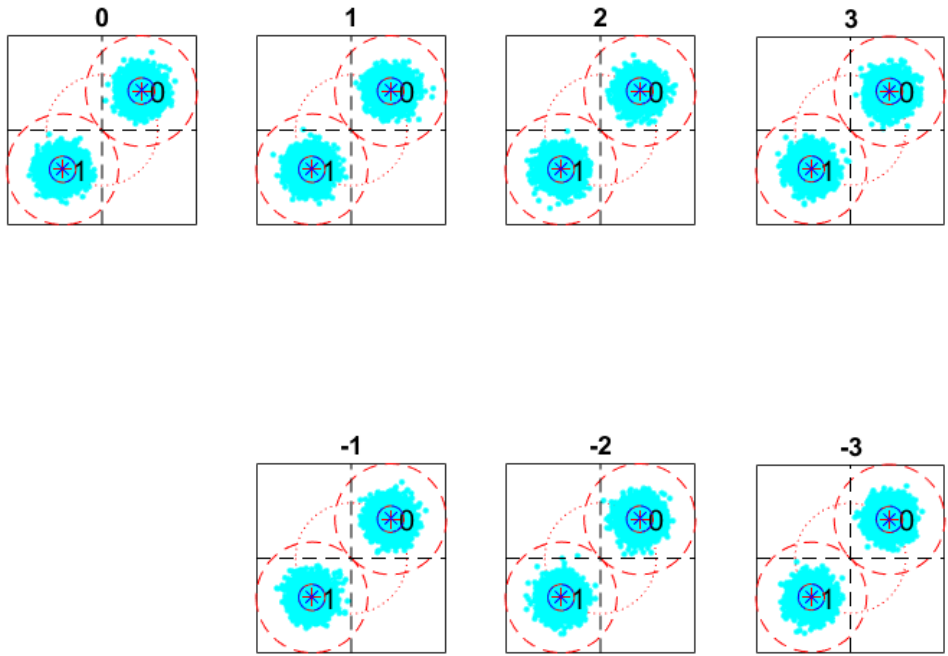

*Figure 43. Constellation diagram for the long pulses in Example 5, for sub-channels $\tilde{k} = -\tilde{R} \dots \tilde{R}$ with $\tilde{R} = 3$*

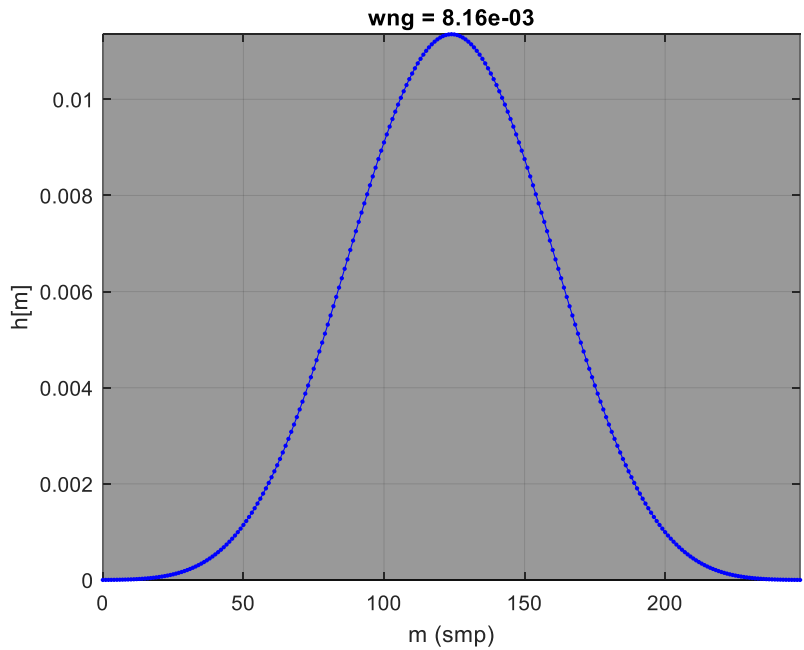

*Figure 44. Impulse response of alternative least-squared-error pulse-shaping filter (with wide transition-band) used in Example 5*

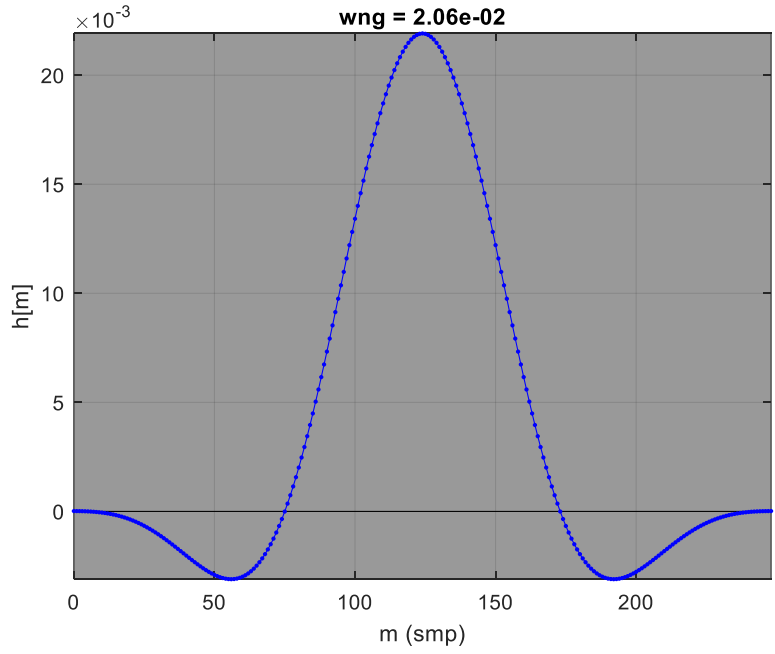

*Figure 45. Impulse response of alternative least-squared-error pulse-shaping filter (with low side-lobes) used in Example 5*

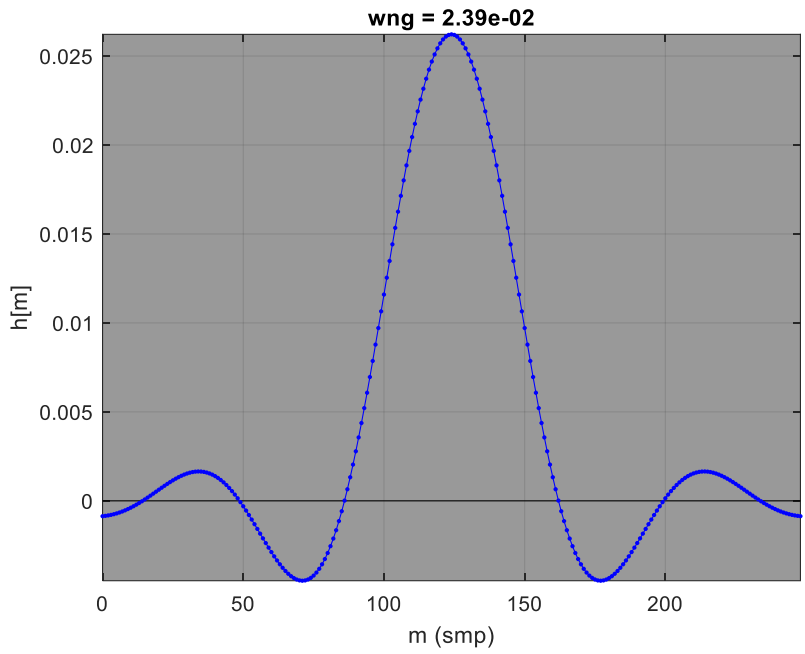

*Figure 46. Impulse response of alternative least-squared-error pulse-shaping filter (with flat pass-band) used in Example 5*

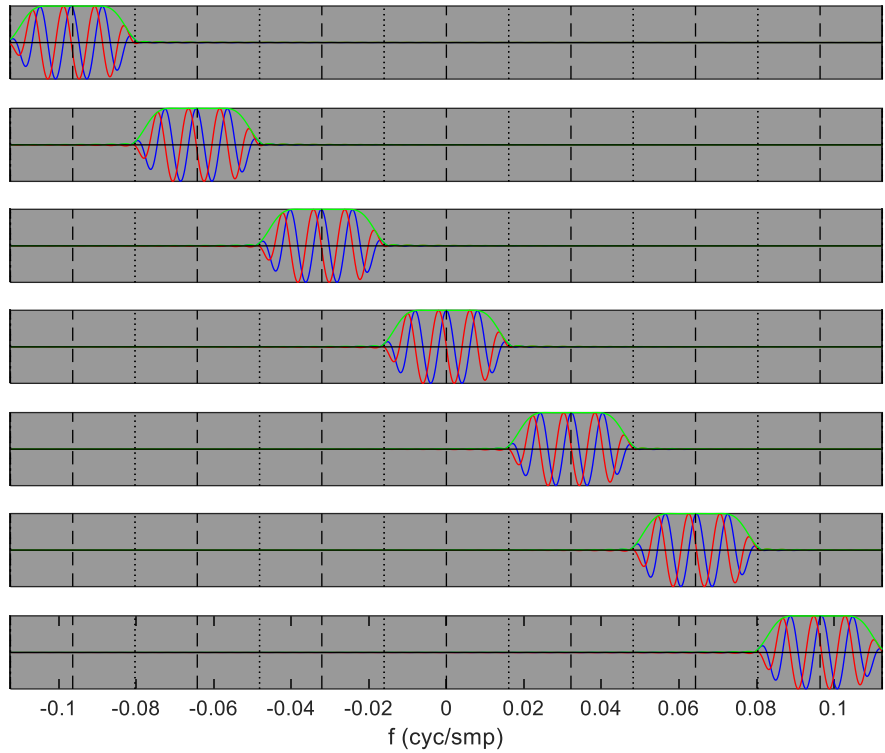

*Figure 47. Frequency response of leased-squared-error pulse-shaped sub-carriers (with flat pass-band) in Example 5, for sub-channels $\widetilde{k} = -\widetilde{K} \dots \widetilde{K}$ (top to bottom)*

## Example 6 – Butterworth pulse shaping for fast recursive realization

During the tx up-sampling process there are $M^{\uparrow}$ outputs at the higher rate for every input at the lower rate. The FIR pulse-shaping filters therefore reduce to an $M^{\uparrow}$-sample multiply and copy operation of $\varphi_{\text{tx}}[n^{\downarrow}]h_{\text{tx}}[n^{\uparrow}]$ into $\varphi_{\text{tx}}[n^{\uparrow}]$ if only non-zero samples are considered. Thus, FIR pulse-shaping need not be an expensive operation. Similarly, during rx down-sampling, there is one output at the lower rate for every $M^{\uparrow}$ inputs at the higher rate, requiring $M^{\uparrow}$ multiply-and-add operations for every $\varphi_{\text{rx}}[n^{\downarrow}]$. Thus, math operations (over the FIR filter kernel and data buffer with samples at the higher rate) are only applied at the sample times of the lower rate and simpler data-transfer and shift operations are applied for every new sample at the higher rate. These reduced complexity multi-rate implementations are readily realized using polyphase filter-banks (see Chapters 6 & 9 in Harris 2021). These structures utilize down-sampling then low-pass filtering (at the lower rate) instead of low-pass filtering (at the higher rate) then down-sampling. However, sliding operations at the higher rate may be necessary if rx-tx time synchronization is absent or approximate.

The sliding time-domain application of FIR pulse-shaping filters in Examples 4 & 5 requires approximately $M = 249$ multiplication and addition operations per output sample. If one output sample is computed for every input sample (at the higher rate) then the convolutions are computationally very expensive. In this example, the modulation scheme employed in Example 5 is repeated here; however, an FIR Butterworth filter is used on tx (with $M^{\uparrow} = 249$) and a 3rd-order (causal) IIR Butterworth filter (i.e. $K = 3$) is used on rx (with $f_c = 2/M^{\uparrow}$). The recursive rx filter requires only 7 multiplication and addition operations per output and the resolution is approximately the same as the previous example ($\Delta^{\#} = 4.2176$, see Figure 48 - Figure 52 for details).

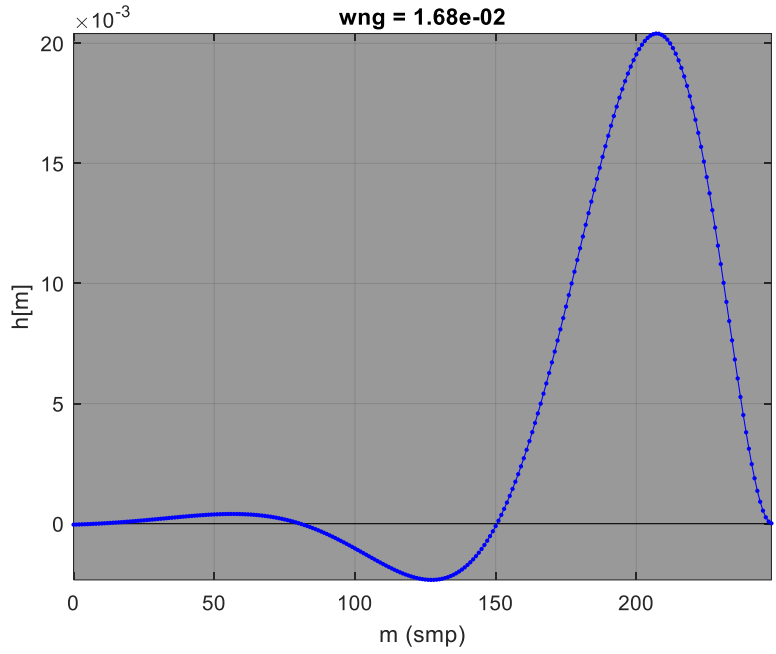

*Figure 48. Impulse response of (non-recursive) FIR Butterworth filter used to shape pulses on tx in example 6. When it is flipped left-right in time it is the truncated impulse-response of the (recursive) IIR Butterworth filter used on rx.*

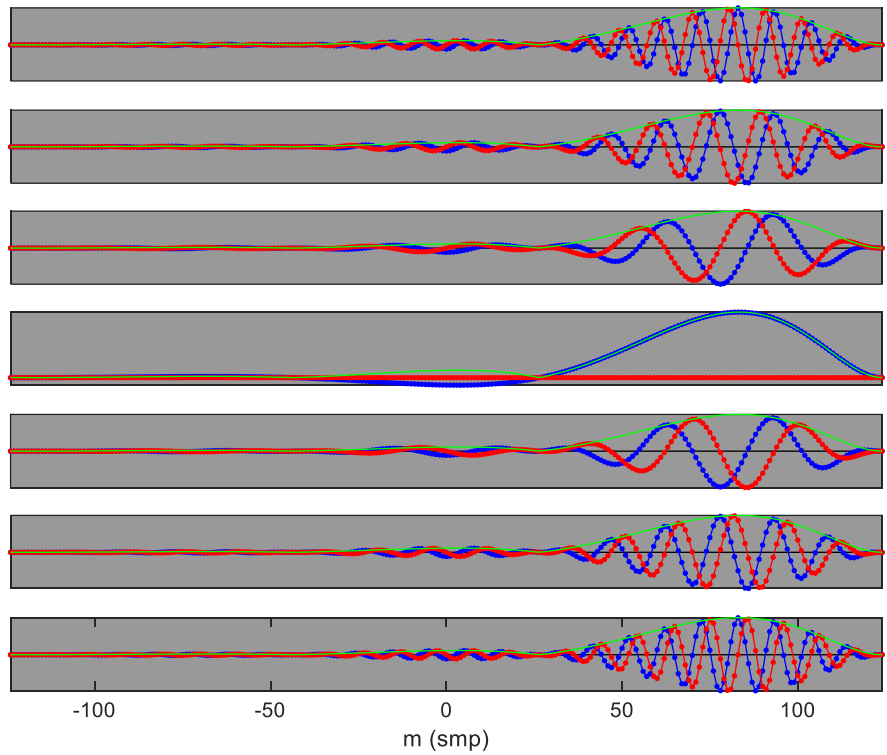

*Figure 49. Impulse response of Butterworth sub-carriers in Example 6 for sub-channels $\tilde{k} = -\tilde{K} \ldots \tilde{K}$ (top to bottom)*

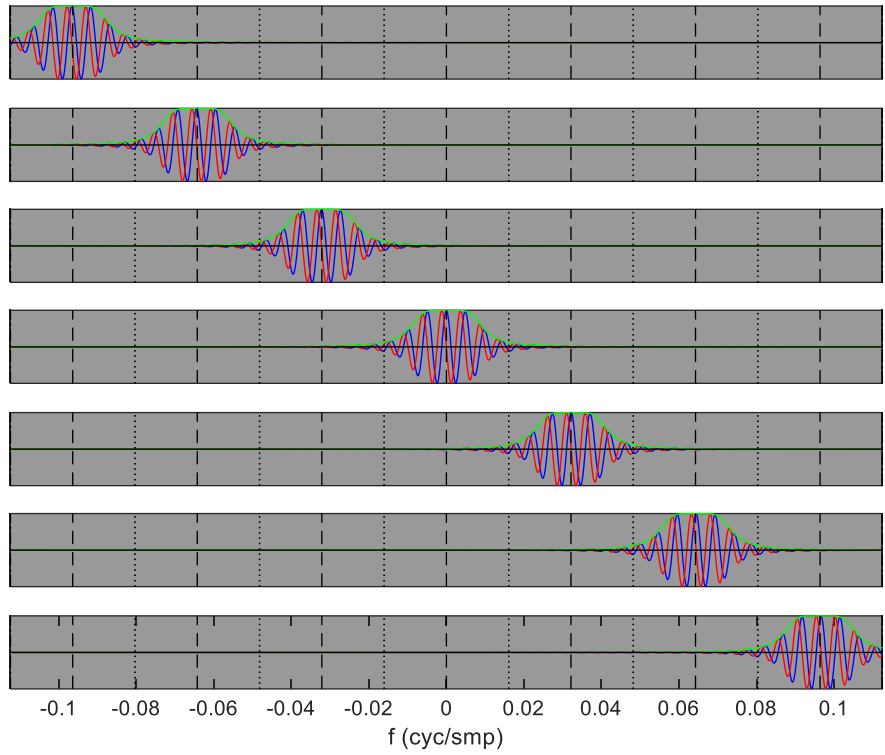

*Figure 50. Frequency response of Butterworth sub-carriers in Example 6 for sub-channels $\tilde{k} = -\tilde{K} \dots \tilde{K}$ (top to bottom)*

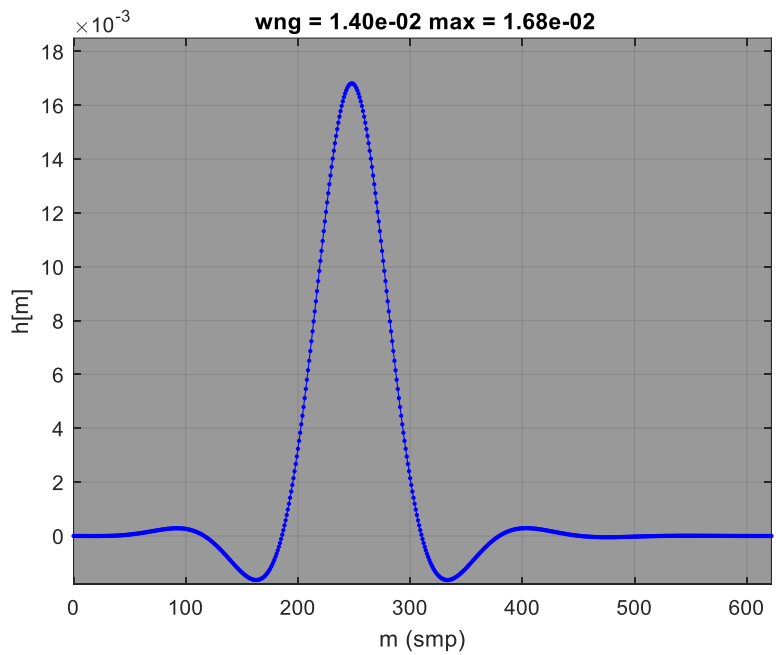

*Figure 51. Impulse response of the FIR and IIR Butterworth filters (at tx and rx) in series*

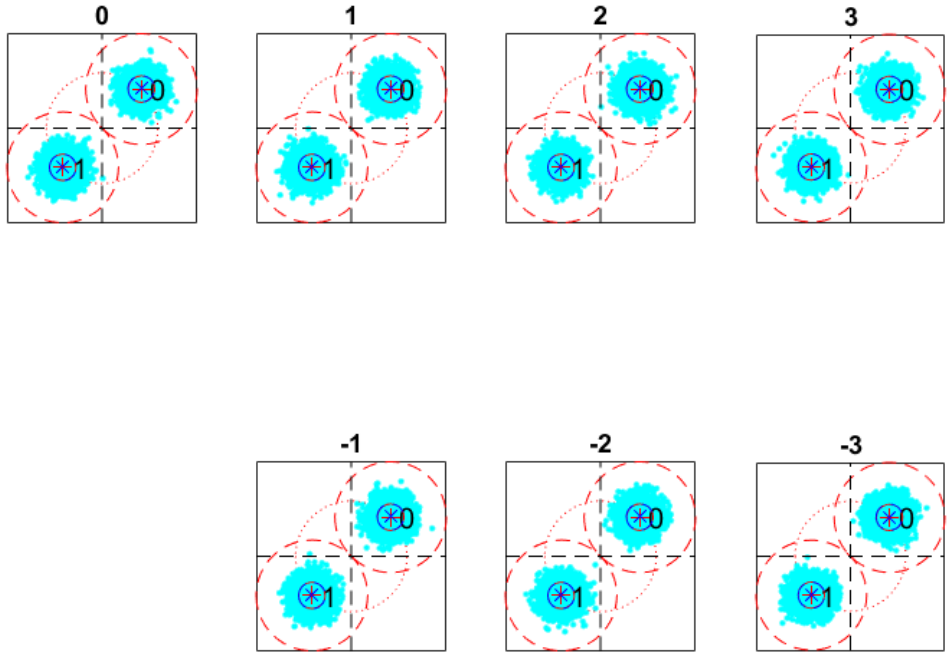

*Figure 52. Constellation diagram for Example 6 for sub-channels $\bar{k} = -\tilde{K} \dots \tilde{K}$ with $\tilde{K} = 3$*

## Example 7 − Fast non-recursive FIR convolution via the FFT

IIR filters utilize feedback for efficient recursive application in the time domain. While it is possible to design recursive FIR filters, such approaches are not recommended because they involve marginally stable poles (i.e. on the unit circle) that are cancelled by zeros. Such filters are therefore susceptible to the gradual accumulation of rounding errors (due to finite numerical precision) over long periods of time, thus they are not discussed further here.

Convolution in the time domain is multiplication in the frequency domain; therefore, the FFT may be used to efficiently apply FIR filters. For an FIR filter with an $M$-sample impulse response $h[m]$, and an $L$-sample input sequence $x[m]$, the output of $x[m]$ convolved with $h[m]$ in (3) i.e.

$$y[m] = h[m] \circledast x[m] \tag{57}$$

is computed using

$\acute{y}[m] = \text{IFFT}\{H[k] * X[k]\}$ where

$H[k] = \text{FFT}\{\acute{h}[m]\}$ and $X[k] = \text{FFT}\{\acute{x}[m]\}$

for $m = 0 \dots B - 1$ and $k = 0 \dots B - 1$.

The FFT and the IFFT are applied to $B$ samples and bins, respectively, where $B$ is the block (or batch) length. The accents above are used to denote extended sequences in the time domain of length $B = L + M$. For $h[m]$ and $x[m]$, with $L > M$, a 'zero-padded' extension is used, i.e.

$$\acute{h}[m] = \begin{bmatrix} \underbrace{h[0] \quad \dots \quad h[M-1]}_{M} & \underbrace{0 \quad \dots \quad 0}_{L} \end{bmatrix} \text{ and} \tag{58}$$

$$\acute{x}[m] = \begin{bmatrix} \underbrace{x[0] \quad \dots \quad x[L-1]}_{L} & \underbrace{0 \quad \dots \quad 0}_{M} \end{bmatrix}. \tag{59}$$

Zero-padding ensures that the FFT arguments have the same length (of $B$). It also ensures that input samples at the end of the block do not affect the output at the start of the block, due to the 'circular' nature of the Fourier transform. The output of the convolution for this block of data, i.e. the filtering of $x[n]$ by $h[m]$, is then extracted using $y[n] = \acute{y}[n]$, for $n = 0 \dots L - 1$. Note that $L$ should be chosen such that $B$ is a power of two, to maximize the efficiency of the FFT. For large $L$ and $M$ the speedup is significant; however in online systems, an $L$-sample latency is introduced because the FFT operation must wait for all $L$ samples to be acquired before the (extended) data block can be processed. In addition to this speedup/latency trade-off, finite random-access memory places an upper practical limit on $L$. Infinitely long sequences in online systems may however be handled by processing contiguous input blocks sequentially then 'splicing' their corresponding outputs together to remove the initiation and termination transients at the beginning and end of each block.

When the $\acute{k}$th block of data $x_k[m]$, is extracted from an 'infinite' stream of input data $x[n]$, then zero-padded and convolved with the zero-padded filter coefficients as per (57)-(59) above in isolation, there is an initiation transient in the output $y_k[m]$ that lasts for $M - 1$ samples, due to the absence of the samples in the preceding block $x_{k-1}[m]$. Similarly, in the extended output $\acute{y}_k[m]$, there is a termination transient that lasts for $M - 1$ samples, due to the absence of the samples in the following block i.e. $x_{k+1}[m]$. In the latter case, this termination transient 'spills' into the extension samples. However when processing contiguous 'data' blocks sequentially (i.e. $x_0[m], x_1[m] \dots x_k[m] \dots x_\infty[m]$) the correct output is simply obtained by adding the termination transient of the preceding (extended) block to the initiation transient of the current block, i.e.

$$y_k[m] = \begin{cases} \acute{y}_{k-1}[L+m] + \acute{y}_k[m], & 0 < m < M \\ \acute{y}_k[m], & M \leq m < L \end{cases} \tag{60}$$

for $\acute{k} > 0$ and $m = 0 \dots L - 1$

where $\acute{k}$ is the index of the block and $m$ is the index of the sample within the block; or

$$y_k[m] = \acute{y}_k[m]$$

for $\acute{k} = 0$ and $m = 0 \dots L - 1$

i.e. for the initial block of the data stream.

Then $y[n] = y_k[m]$ for $n = 0 \dots \infty$, with $n = \acute{k}L + m$.

This principle is the basis of FFT-based filtering algorithms such as the so-called overlap-and-add and the overlap-and-save methods. Note that $H[k]$ need only be computed once whereas $X[k]$ is computed for each new data block.

This block-based processing flow is illustrated in Figure 53 & Figure 54. A synthetic input sequence was generated by adding white Gaussian noise to train of band-limited pulses. The shape of the pulse and the corresponding matched filter (see Figure 53) was designed via the least-squared-error procedure (with $K = 36$, $f_c = 4/M$, $\omega_{lo} = \pi f_c$, $\omega_{hi} = 2\pi f_c$, $\tilde{w} = 1$ & $\bar{w} = 1000$). The pulse sequence (cyan) and input sequence (green) are shown in the bottom subplot of Figure 54. The output of the time-domain convolution is shown in each subplot (red). The input sequence is processed using three blocks, each with $M = 2K + 1 = 73$ and $B = 2^8 = 256$ thus $L = B - M =$

183. The output of each block $\acute{y}_k[m]$, for $\acute{k} = 0 \dots 2$ and $m = 0 \dots B - 1$, is shown in the first three subplots (blue). In these subplots, the initiation and termination transients are clearly visible, particularly in the overlap between the $\acute{k} = 0$ & $\acute{k} = 1$ blocks, because a pulse is centred on the data block boundary. The solid black and dashed black lines are used to mark the ends of the data blocks and extended blocks, respectively. The output $y[n]$, formed by splicing the block outputs together, is shown in the lowermost subplot (blue); block indices are also shown at the start of each contiguous block. In this subplot, the coincidence of the blue and red lines indicates that the block-based (i.e. frequency-domain) and sample-based (i.e. time-domain) implementations are equivalent.

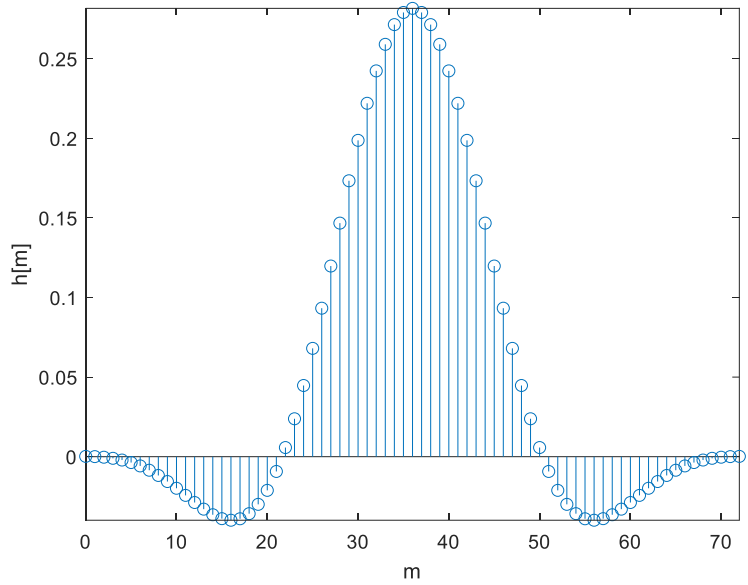

*Figure 53. Pulse shape and impulse response of the corresponding matched filter*

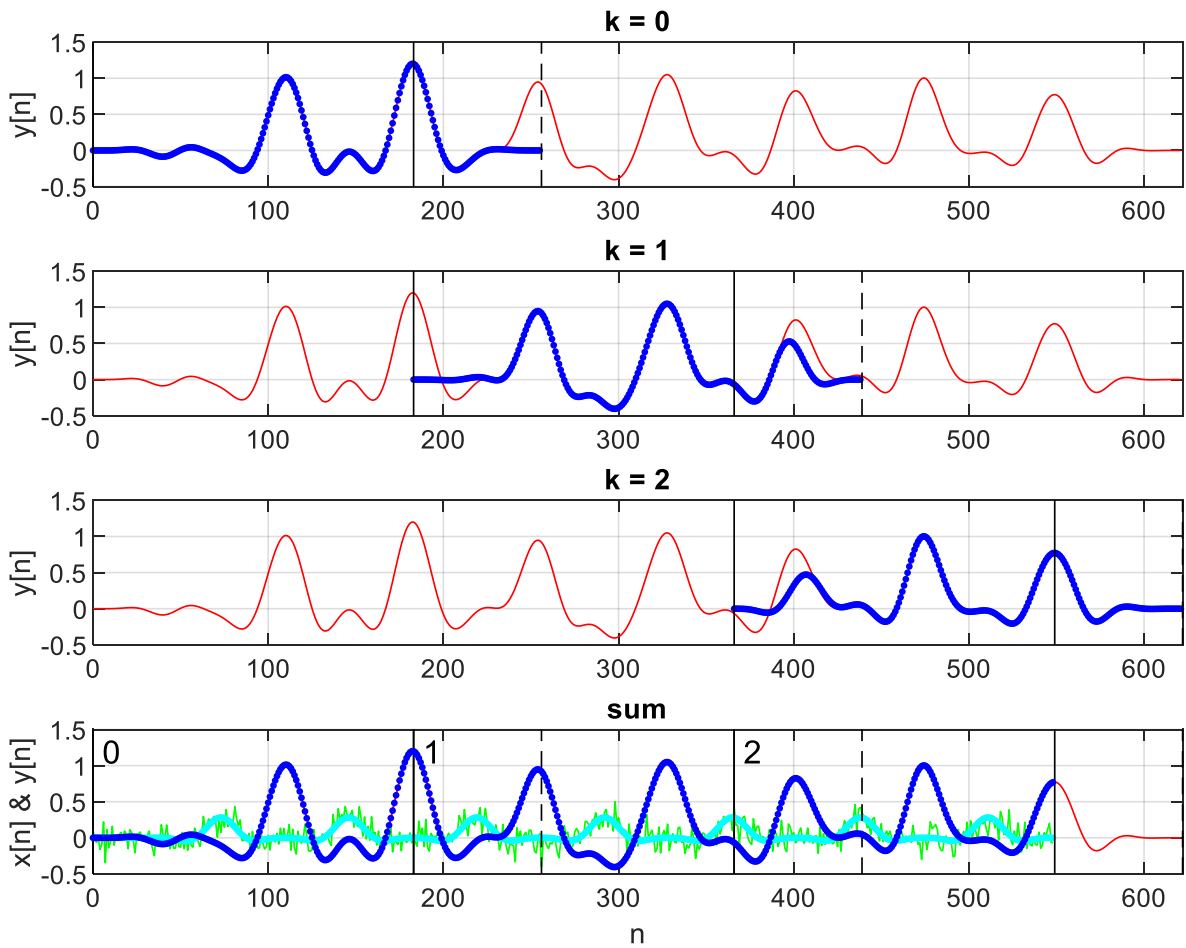

*Figure 54. Illustration of block-based frequency-domain convolution via the fast Fourier transform*

## Closing Remarks

In engineering degrees where signals and systems electives are available, discrete-time systems are only presented after the continuous-time foundations have been established and when time is limited, they may only receive a cursory treatment. These lecture notes may help recent graduates who then find themselves working on digital systems (hardware, firmware and software) gain a better understanding of digital signals-and-systems theory. Wireless communication in the radio-spectrum is used here to illustrate the basic concepts that may also be applied in other digital electronic systems that transmit and or receive pulse-modulated waveforms.

As the material here is relevant at all frequencies in the spectrum and timescales, relative frequency and time units are preferred (i.e. relative to the sampling rate $F_{\mathrm{smp}}$ and the sampling period $T_{\mathrm{smp}}$) so that propagation frequencies (e.g. MHz to THz) and sampling rates (e.g. MHz to GHz) are dimensionless relative quantities with no units.

In Parts I and II, the various trade-offs associated with the design of low-pass filters are discussed. The low-pass digital filter is used here to illustrate the way in which pulses are shaped in time and frequency. In Part III the pros and cons of alternative waveform designs for wireless communication are examined in an ideal scenario. The various waveform designs were deliberately chosen to highlight their approximate equivalence in a noisy channel and the material may suggest that waveform design is fundamentally a zero-sum game such that a gain in system performance in one area is approximately counter-balanced by equal performance loss in another. However, non-ideal hardware, non-ideal propagation, and the presence of other users and interferers, will make some waveforms more attractive than others in some circumstances.

## Suggested reading